\documentclass[a4paper]{article}

\usepackage{a4wide}
\usepackage[UKenglish]{babel}
\usepackage{graphicx,subfigure,pgfplots}
	\graphicspath{{./}{figure/}{figure/boltzmann_loc/}{figure/boltzmann_loc/asym/}{figure/hydro/}}
	\pgfplotsset{compat=newest}
\usepackage[colorlinks=true,linkcolor=blue,citecolor=red]{hyperref}
\usepackage{amsfonts,amsthm,empheq,bbold}
\usepackage[inline]{enumitem}
\usepackage{subfigure}

\newtheorem{theorem}{Theorem}[section]
\newtheorem{proposition}[theorem]{Proposition}
\theoremstyle{remark}\newtheorem{remark}[theorem]{Remark}

\newcommand{\abs}[1]{\left\lvert#1\right\rvert}
\newcommand{\ave}[1]{\langle#1\rangle}
\newcommand{\cB}{\mathcal{B}}
\newcommand{\Beta}{\operatorname{B}}
\newcommand{\cC}{\mathcal{C}}

\newcommand{\cF}{\mathcal{F}}

\newcommand{\pr}[1]{{}^\prime\!#1}
\newcommand{\R}{\mathbb{R}}
\newcommand{\sgn}{\operatorname{sgn}}
\newcommand{\tr}{\operatorname{tr}}

\allowdisplaybreaks

\begin{document}
\title{Hydrodynamic models of preference formation \\ in multi-agent societies}

\author{Lorenzo Pareschi\thanks{Department of Mathematics and Computer Sciences, University of Ferrara, Via Machiavelli 35, 44121 Ferrara, Italy
            (\texttt{lorenzo.pareschi@unife.it})} \and
        Giuseppe Toscani\thanks{Department of Mathematics ``F. Casorati'', University of Pavia, Via Ferrata 1, 27100 Pavia, Italy
            (\texttt{giuseppe.toscani@unipv.it})} \and
        Andrea Tosin\thanks{Department of Mathematical Sciences ``G. L. Lagrange'', Politecnico di Torino, Corso
        		Duca degli Abruzzi 24, 10129 Torino, Italy
            (\texttt{andrea.tosin@polito.it})} \and
        Mattia Zanella\thanks{Department of Mathematical Sciences ``G. L. Lagrange'', Politecnico di Torino, Corso
        		Duca degli Abruzzi 24, 10129 Torino, Italy
        		(\texttt{mattia.zanella@polito.it})}}
\date{}

\maketitle

\begin{abstract}
In this paper, we discuss the passage to hydrodynamic equations for kinetic models of opinion formation. The considered kinetic models feature an opinion density depending on an additional microscopic variable, identified with the personal preference. This variable describes an opinion-driven polarisation process, leading finally to a choice among some possible options, as it happens e.g. in referendums or elections. Like in the kinetic theory of rarefied gases, the derivation of hydrodynamic equations is essentially based on the computation of the local equilibrium distribution of the opinions from the underlying kinetic model. Several numerical examples validate the resulting model, shedding light on the crucial role played by the distinction between opinion and preference formation on the choice processes in multi-agent societies.

\medskip


\end{abstract}

\section{Introduction}
\label{sect:intro}
The mathematical modelling of opinion formation in multi-agent societies has enjoyed in recent years a growing attention~\cite{BN1,BN2,BN3,galam1997PHYSA,MG,SW}. In particular, owing to their cooperative nature, the dynamics of opinion formation have been often dealt with resorting to methods typical of statistical mechanics~\cite{CFL,Cha}. Among other approaches, kinetic theory served as a powerful basis to model fundamental interactions among the so-called agents~\cite{Bou2,Bou1,Bou,CDT,gal11,toscani2006CMS} and to provide a sound structure for related applications~\cite{albi2016BOOKCH,tosin2018CMS}. In kinetic models, analogously to the kinetic theory of rarefied gases, the mechanism leading to the opinion variation is given by binary, i.e. pairwise, interactions among the agents. Then, depending on the parameters of such microscopic rules, the society develops a certain macroscopic equilibrium distribution~\cite{motsch2014SIREV,pareschi2013BOOK}, which describes the formation of a relative consensus about certain opinions.

Two main aspects are usually taken into account in designing the elementary binary interactions. The first is the \textit{compromise}~\cite{deffuant2002JASSS,Wei}, namely the tendency of the individuals to reduce the distance between their respective opinions after the interaction. The second is the \textit{self-thinking}~\cite{toscani2006CMS}, i.e. an erratic individual change of opinion inducing unpredictable deviations from the prescribed deterministic dynamics of the interactions.

Recently, many efforts have been devoted to include further details in the opinion formation models, so as to capture more and more realistic phenomena. The usual strategy consists in taking additional behavioural aspects into account, such as the \textit{stubbornness} of the agents~\cite{lallouache2010PRE,stella2013CDC}, the emergence of \textit{opinion leaders}~\cite{DMPW,Due-Wol}, the influence of \textit{social networks}~\cite{albi2017KRM,toscani2018PRE}, the \textit{expertise} in decision making tasks~\cite{pareschi2017PHYSA}, the \textit{personal conviction}~\cite{Cro,Bis,Bru-Tos,Cro1}. Generally, the aim of such additional parameters is to model on one hand the resistance of the agents to change opinion and, on the other hand, the prominent role played by some individuals in attracting others towards their opinions. In all these contributions, the additional variables act as modifiers of the microscopic interactions. This means that they affect the process of opinion formation but are not affected in turn by the evolving opinions.
 
In the present paper we aim instead at modelling a parallel process to opinion formation, namely the formation of \textit{preferences}. The preference, which is driven by, but need not coincide with, the opinion, is here understood as representative of the \textit{choice} that an agent makes among some possible options, such as e.g. some candidates in an election or yes/no in a referendum~\cite{cristiani2018MMS}. As such, and unlike the opinion, the preference evolves  towards necessarily \textit{polarised} states, which reflect the available options.

To pursue this goal, we consider a novel class of inhomogeneous kinetic models for the joint distribution function $f(t,\,\xi,\,w)$, where $\xi$ is the variable describing the preference and $w$ the one describing the opinion. In particular, $f(t,\,\xi,\,w)d\xi dw$ is the proportion of agents who, at time $t\geq 0$, express a preference in the interval $[\xi,\,\xi+d\xi]$ and simultaneously an opinion in the interval $[w,\,w+dw]$. As far as the modelling of the interactions leading to the evolution of the opinion is concerned, we take advantage of the well consolidated background recalled before. Conversely, concerning the evolution of the preference, we assume transport-type dynamics of the form
$$ \frac{d\xi}{dt}=(w-\alpha)\Phi(\xi). $$
Here, in analogy with the classical kinematics, $w$ plays morally the role of the velocity, i.e. it drags the preference in time, however biased by a \textit{perceived social opinion} $\alpha$, which accounts for the predominant social feeling. Moreover, the zeros of the function $\Phi$ define the options where the preference may polarise.

These ingredients lead us to an inhomogeneous Boltzmann-type kinetic equation of the form
$$ \partial_tf+(w-\alpha)\partial_\xi(\Phi(\xi)f)=Q(f,\,f), $$
where $Q$ is a Boltzmann-type collision operator encoding the opinion formation interaction dynamics. From here, analogously to the classical kinetic theory of rarefied gases, we derive macroscopic equations for the density $\rho=\rho(t,\,\xi)$ and the mean opinion $m=m(t,\,\xi)$ of the agents with preference $\xi$ at time $t$ by means of a local equilibrium closure based on the identification of the local equilibrium distribution of the opinions -- the equivalent of a local ``Maxwellian''. The precise type of hydrodynamic equations that we obtain in this way depends on whether the mean opinion of the agents is or is not conserved in time by the microscopic dynamics of opinion formation. For instance, if it is conserved we get the following system of \textit{conservation laws}:
$$ 	\begin{cases}
		\partial_t\rho+\partial_\xi\left(\Phi(\xi)\rho(m-\alpha)\right)=0,\\
		\partial_t(\rho m)+\partial_\xi\left(\Phi(\xi)(M_{2,\infty}-\alpha\rho m)\right)=0,
	\end{cases} $$
where $M_{2,\infty}$ denotes the energy of the local Maxwellian. For special classes of local equilibrium distributions, that we compute from the collisional kinetic equation by an asymptotic procedure reminiscent of the \textit{grazing collision limit} of the classical kinetic theory~\cite{toscani2006CMS,villani1998ARMA}, we can express such an energy analytically in terms of the hydrodynamic parameters $\rho$, $m$. We recover therefore a self-consistent macroscopic model, which we show to be \textit{hyperbolic} for all the physically admissible values of $(\rho,\,m)$ and able to reproduce the preference polarisations, viz. the choices, discussed above.

In more detail, the paper is organised as follows. In Section~\ref{sect:micro_xi_w} we give preliminary microscopic insights into the joint process of opinion and preference formation, stressing in particular the role of the perceived social opinion $\alpha$. In Section~\ref{sect:kinetic_opinion} we move to an aggregate analysis of the opinion formation by means of Boltzmann-type kinetic models, studying in particular their steady states which, as set forth above, pave the way for the identification of the local equilibrium distributions needed in the passage to the hydrodynamic equations. In Section~\ref{sect:macro} we derive various types of macroscopic models of preference formation out of the aforementioned inhomogeneous Boltzmann-type equation and we link them precisely to key features of the microscopic interactions among the agents. In Section~\ref{sect:num} we present several numerical tests, both at the kinetic and at the macroscopic scales, which exemplify the distinction between the preference formation and the opinion formation processes and show how much such a distinction enhances the interpretation of the social dynamics. Finally, in Section~\ref{sect:summary} we discuss further developments and research prospects.

\section{A microscopic look at the opinion-preference interplay}
\label{sect:micro_xi_w}
The first mathematical models of consensus formation in opinion dynamics were proposed in the form of systems of ordinary differential equations (ODEs) describing the behaviour of a finite number of agents. After the pioneering works~\cite{degroot1974JASA,french1956PR}, which introduced simple agent-based models to understand the effects of the influence among connected individuals, many research efforts have been devoted to the construction of sophisticated differential models of opinion formation. Besides those dealing with simple consensus dynamics, in recent years new models of more realistic social phenomena have been proposed to capture additional aspects. Without intending to review all the literature, we give here some references on certain classes of models for finite systems: the celebrated Hegselmann-Krause model~\cite{hegselmann2002JASSS}, which considers bounded-confidence-type interactions to stress the impact of homophily in learning processes, see also~\cite{ceragioli2012NARWA,lorenz2007IJMPC}; models incorporating leader-follower effects~\cite{ni2010SCL}; models of social interactions on realistic networks~\cite{watts2007JCR}; models of opinion control~\cite{canuto2012SICON}. 

An aspect which, to our knowledge, has been so far basically disregarded in mathematical models of opinion dynamics, in spite of its realism, is the distinction between the \textit{opinion} in the strict sense of the individuals about single issues and their overall \textit{preference}, which, in some cases, is actually mainly responsible for their choices. For instance, in case of referendums or political elections, the preference of a voter can be identified with his/her \textit{voting intention}, which may not always coincide with his/her opinion on every topic debated during the election campaign. Obviously, the preference evolves in time with the opinion, in such a way that if a certain opinion persists for a sufficiently long time it can affect the preference considerably. For example, a voter with a voting intention biased towards right (left, respectively) parties, who however finds him/herself frequently in agreement with the positions taken by left (right, respectively) parties on key topics of the election campaign, may end up with a final vote opposite to his/her original intention.

As a foreword to the subsequent contents of the paper, in this section we present some preliminary considerations about the effect of the interplay between opinion and preference, taking advantage of a deterministic microscopic model for a finite number of agents. Let us consider then a system composed by $N$ agents with microscopic state given by a pair $(\xi_i,\,w_i)\in [-1,\,1]^2$, where $w_i$ is the opinion of the $i$th agent, $\xi_i$ is his/her preference and $i=1,\,\dots,\,N$. Sticking to a standard custom in the literature of models of opinion dynamics, we describe mathematically the opinion of an agent as a bounded scalar variable $w_i$ conventionally taken in the interval $[-1,\,1]$. In particular, we understand the values $w_i=\pm 1$ as the two extreme opinions, while $w_i=0$ as the maximum of indecisiveness. Since, from the physical point of view, the preference is commensurable with an opinion, we adopt the same mathematical conventions also for the variable $\xi_i$.

In order to highlight the different roles of the opinion and the preference variables, we rely on the analogy with the classical kinetic theory of rarefied gases. There, a molecule moving on a line is characterised by its position $x\in\R$ and its velocity $v\in\R$. In the absence of external forces, the velocity remains constant, whereas the position varies according to the kinematic law
\begin{equation}
	\frac{dx}{dt}=v.
	\label{eq:x_v.kinematic}
\end{equation}
In practice, a particle with positive velocity will move rightwards, while a particle with negative velocity will move leftwards. In first approximation, it seems natural to assume that the opinion plays the role of the velocity and the preference that of the position. Indeed, at least in the case in which an agent has to end up with one of the two preferences $\pm 1$, like e.g. in a referendum, one can assume that a large part of the agents with positive opinion will move their preferences rightwards, while agents with negative opinion will move their preferences leftwards. Clearly, one cannot resort to a law like~\eqref{eq:x_v.kinematic}, which allows the position to increase or decrease indefinitely in time: a correction is required in order to maintain the preference variable in the allowed interval $[-1,\,1]$.

A primary example of the analogy just discussed is provided below, where the time evolution of the opinions and the preferences of the agents is modelled by the ODE system
\begin{empheq}[left=\empheqlbrace]{align}
	& \dfrac{d\xi_i}{dt}=(w_i-\alpha)\Phi(\xi_i) \label{eq:micro_xi} \\
	& \dfrac{dw_i}{dt}=\dfrac{1}{N}\displaystyle{\sum_{j=1}^N}P(w_i,\,w_j)(w_j-w_i) \label{eq:micro_w}
\end{empheq}
for $i=1,\,\dots,\,N$, supplemented with initial conditions $(\xi_i(0),\,w_i(0))=(\xi_{0,i},\,w_{0,i})\in [-1,\,1]^2$. The second equation describes standard alignment dynamics among the opinions of the agents, i.e. \textit{consensus}, driven by the interaction/compromise function $0\leq P(\cdot,\,\cdot)\leq 1$, see e.g.,~\cite{pareschi2013BOOK,toscani2006CMS}. The first equation describes instead the evolution of the preference of the $i$th agent based on the signed distance between its true opinion $w_i$ and a reference opinion $\alpha\in [-1,\,1]$ perceived in the society, which we will refer to as the \textit{perceived social opinion}. The function $\Phi:[-1,\,1]\to [0,\,1]$ has to be primarily chosen so as to guarantee that $\xi_i(t)\in [-1,\,1]$ for all $t>0$. However, as we will see in a moment, this function will be also useful to take into account meaningful \textit{polarisations} of the preferences.

In model~\eqref{eq:micro_xi}-\eqref{eq:micro_w}, the coupling between opinion and preference is actually one-directional, indeed the evolution of $\xi_i$ depends on that of the $w_i$'s but not vice versa. In particular, the system~\eqref{eq:micro_w} for the $w_i$'s can be solved \textit{a priori}, before analysing the dynamics~\eqref{eq:micro_xi} of the $\xi_i$'s. Let us consider, in particular, the case of interactions with bounded confidence, which are described by taking
\begin{equation}
	P(w_i,w_j)=\chi(\abs{w_j-w_i}\leq\Delta),
	\label{eq:P_BC}
\end{equation}
where $\chi$ denotes the characteristic function and $\Delta\in [0,\,2]$ is a given confidence threshold, above which agents do not interact because their opinions are too far away from each other. If $\Delta=0$ then only agents with the very same opinion interact, whereas if $\Delta=2$ then we speak of all-to-all interactions, considering that $\abs{w_j-w_i}\leq 2$ for all $w_i,\,w_j\in [-1,\,1]$. The latter case is actually equivalent to choosing $P\equiv 1$.

\begin{figure}[!t]
\centering
\subfigure[$\Delta = 1$]{\includegraphics[scale=0.32]{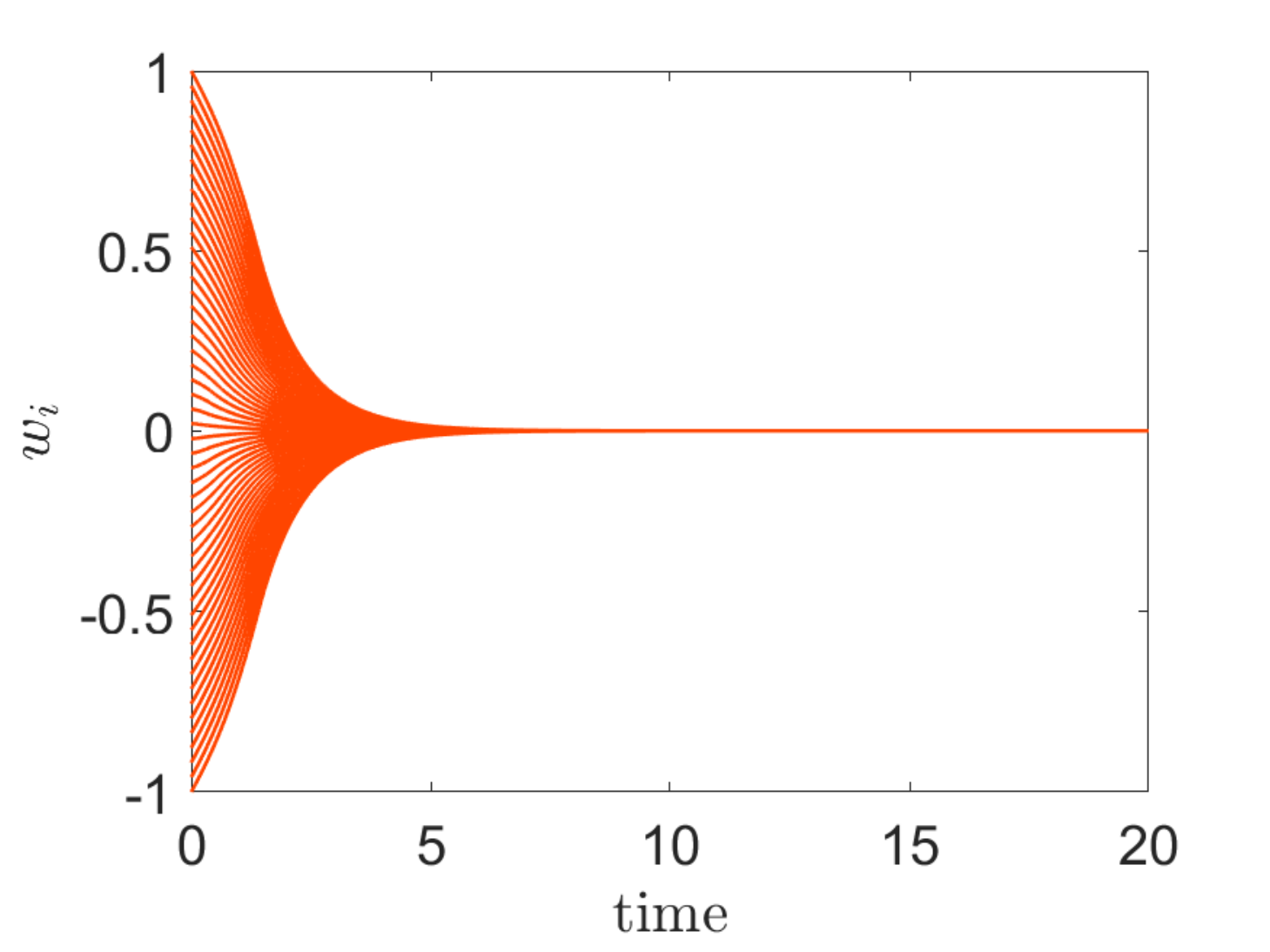}}
\subfigure[$\Delta = 0.4$]{\includegraphics[scale=0.32]{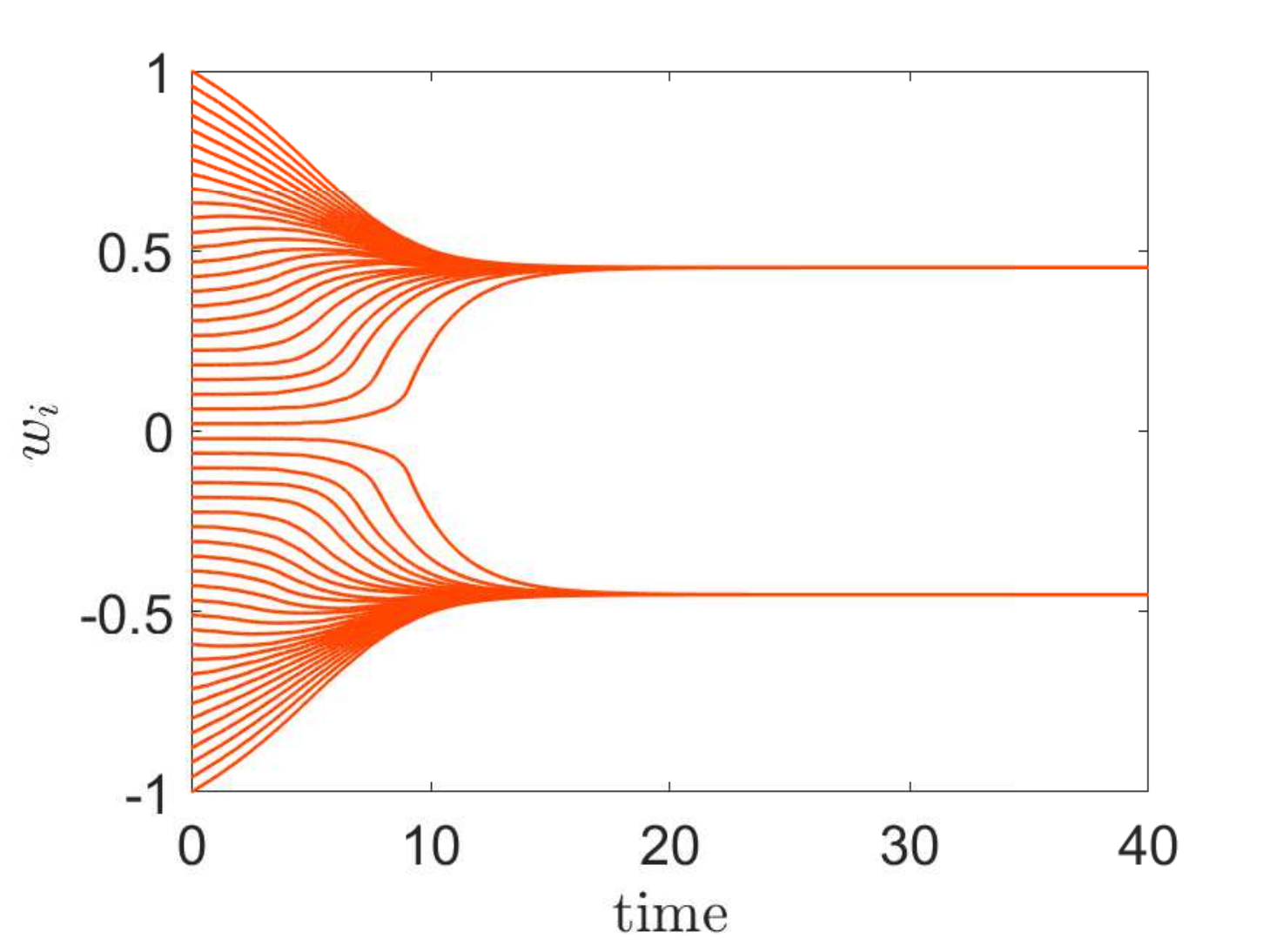}}
\subfigure[$\Delta = 0.2$]{\includegraphics[scale=0.32]{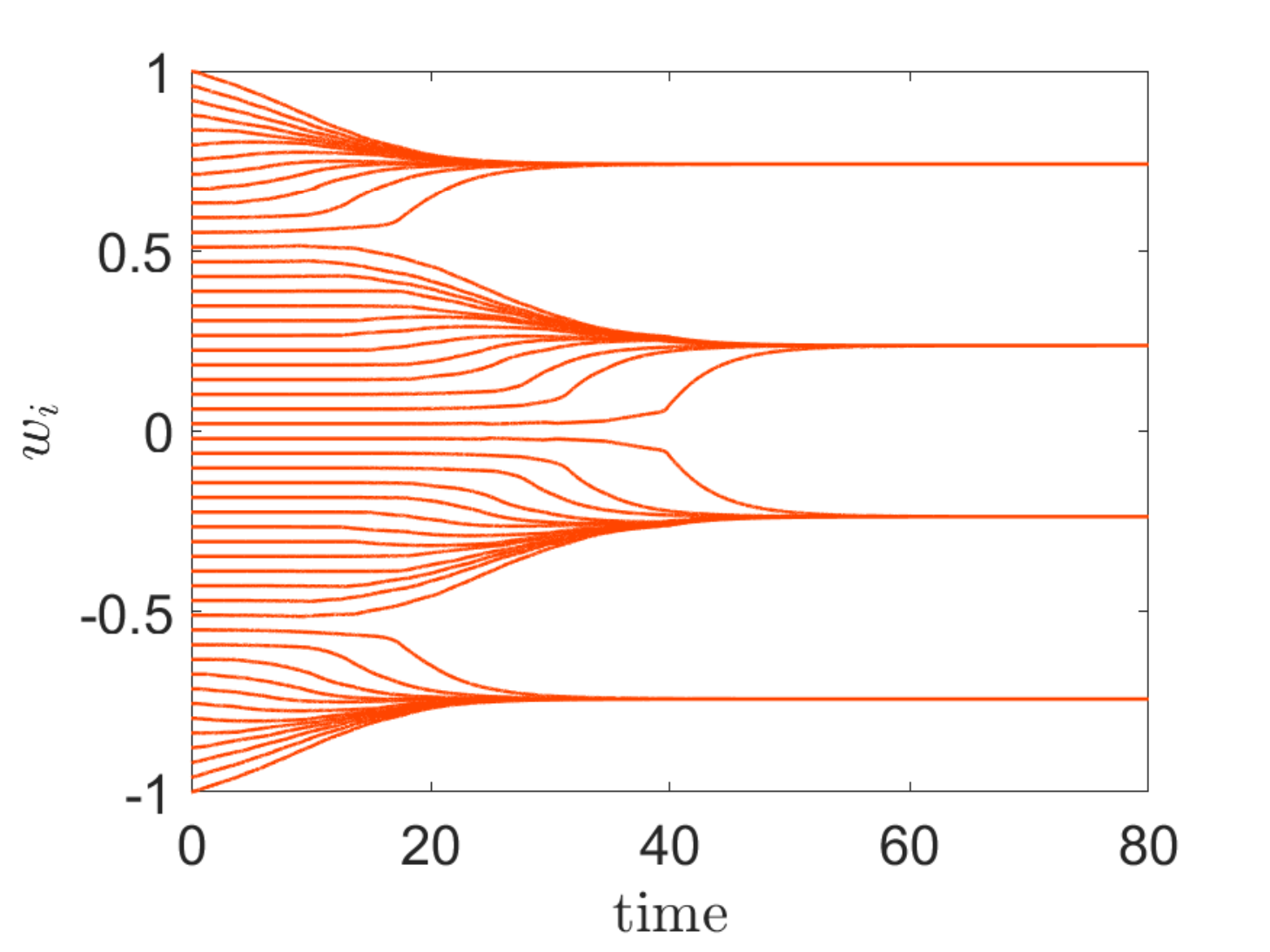}}
\caption{Solution of~\eqref{eq:micro_w} with $N=50$ agents and $P$ given by~\eqref{eq:P_BC} for decreasing values of the confidence threshold $\Delta$. The initial opinions $w_{0,i}$ have been sampled uniformly in $[-1,\,1]$. The ODE system has been integrated numerically via a standard fourth order Runge-Kutta method.}
\label{fig:BCmicro}
\end{figure}

Depending on the value of $\Delta$, one can observe a loss of global consensus. Asymptotically, the opinions may form several clusters, whose number is dictated by $\Delta$ and by the initial conditions $w_{0,1},\,\dots,\,w_{0,N}$, see Figure~\ref{fig:BCmicro}. However, since the function $P$ given in~\eqref{eq:P_BC} is symmetric, i.e. $P(w_i,\,w_j)=P(w_j,\,w_i)$ for all $i,\,j=1,\,\dots,\,N$, the mean opinion $\frac{1}{N}\sum_{i=1}^Nw_i$ is conserved in time and for all $t>0$ coincides, in particular, with the mean opinion at $t=0$.

\medskip

Now we give some insights into the preference dynamics modelled by~\eqref{eq:micro_xi}, at least under special forms of the function $\Phi$. In order to avoid that the preference $\xi_i$ leaves the interval $[-1,\,1]$, a very natural condition is $\Phi(\pm 1)=0$. This implies that the constant functions $\xi_i(t)=-1$ and $\xi_i(t)=1$ are indeed stationary solutions to~\eqref{eq:micro_xi} and may therefore represent attractive or repulsive equilibria of the system, depending on the sign of $w_i-\alpha$.

For instance, we may choose
$$ \Phi(\xi)=1-\abs{\xi}. $$
Taking for granted from~\eqref{eq:micro_w} that $w_i(t)\leq 1$ for all $t>0$, if we integrate~\eqref{eq:micro_xi} starting from an initial condition $\xi_{0,i}\in [0,\,1]$ then for all times $t>0$ in which $\xi_i$ remains non-negative we find
$$ \xi_i(t)\leq 1-(1-\xi_{0,i})e^{-(1-\alpha)t}\leq 1. $$
Likewise, taking for granted from~\eqref{eq:micro_w} that $w_i(t)\geq -1$ for all $t>0$, if we start from an initial condition $\xi_{0,i}\in [-1,\,0]$ then for all times $t>0$ in which $\xi_i$ remains non-positive we deduce
$$ \xi_i(t)\geq -1+(1+\xi_{0,i})e^{-(1+\alpha)t}\geq -1. $$
This argument, applied to the various time intervals in which $\xi_i$ has constant sign, shows indeed that $\xi_i(t)\in [-1,\,1]$ for all $t\geq 0$. Nevertheless, we cannot solve~\eqref{eq:micro_xi} exactly, because from~\eqref{eq:micro_w} we cannot calculate exactly the function $t\mapsto w_i(t)$. On the other hand, we can get a useful idea at least of the large time dynamics of~\eqref{eq:micro_xi} by fixing $w_i$ to its asymptotic value, say $w_{\infty,i}\in [-1,\,1]$, and considering the equation
$$ \frac{d\xi_i}{dt}=\left(w_{\infty,i}-\alpha\right)(1-\abs{\xi_i}), $$
whose solution reads
$$ \xi_i(t)=
	\begin{cases}
		-1+(1+\xi_{0,i})e^{\left(w_{\infty,i}-\alpha\right)t} & \text{if } \xi_i(t)\leq 0 \\
		1-(1-\xi_{0,i})e^{-\left(w_{\infty,i}-\alpha\right)t} & \text{if } \xi_i(t)\geq 0.
	\end{cases} $$
From here we easily deduce that:
\begin{itemize}
\item if $\xi_{0,i}<0$ and $w_{\infty,i}<\alpha$ then $\xi_i\to -1$ for $t\to +\infty$;
\item if $\xi_{0,i}>0$ and $w_{\infty,i}>\alpha$ then $\xi_i\to 1$ for $t\to +\infty$.
\end{itemize}
In both cases, the final preference confirms and consolidates the initial one, because $w_{\infty,i}-\alpha$ has the same sign as $\xi_{0,i}$. Conversely,
\begin{itemize}
\item if $\xi_{0,i}<0$ but $w_{\infty,i}>\alpha$ then $\xi_i\to 1$ for $t\to +\infty$;
\item if $\xi_{0,i}>0$ but $w_{\infty,i}<\alpha$ then $\xi_i\to -1$ for $t\to +\infty$.
\end{itemize}
In these cases, the final preference reverses the initial one, because $w_{\infty,i}-\alpha$ has opposite sign with respect to $\xi_{0,i}$.

Another possible choice of the function $\Phi$ is:
\begin{equation}
	\Phi(\xi)=\abs{\xi}\left(1-\xi^2\right),
	\label{eq:Phi_3poles}
\end{equation}
which vanishes also at $\xi=0$. Thus we are led to consider the equation
$$ \frac{d\xi_i}{dt}=\left(w_{\infty,i}-\alpha\right)\abs{\xi_i}\left(1-\xi_i^2\right), $$
whose solution reads
$$ \xi_i(t)=\frac{\xi_{0,i}}{\sqrt{\xi_{0,i}^2+\left(1-\xi_{0,i}^2\right)e^{-2\sgn{\xi_{0,i}}\left(w_{\infty,i}-\alpha\right)t}}}. $$
Now the asymptotic trend of the preference can be summarised as follows.
\begin{itemize}
\item For $w_{\infty,i}<\alpha$:
\begin{itemize}
\item if $\xi_{0,i}<0$ then $\xi_i\to -1$ for $t\to +\infty$;
\item if $\xi_{0,i}>0$ then $\xi_i\to 0$ for $t\to +\infty$.
\end{itemize}
\item For $w_{\infty,i}>\alpha$:
\begin{itemize}
\item if $\xi_{0,i}<0$ then $\xi_i\to 0$ for $t\to +\infty$;
\item if $\xi_{0,i}>0$ then $\xi_i\to 1$ for $t\to +\infty$.
\end{itemize}
\end{itemize}
We observe that the large time behaviour of the preference is again a polarisation in poles coinciding with the zeroes of the function $\Phi$. Unlike the previous case, however, the presence of an intermediate pole at $\xi=0$ prevents a complete reversal of the initial preference when the latter has opposite sign with respect to $w_{\infty,i}-\alpha$. In such a situation, the agents simply become indecisive, their preference tending indeed to zero.

In order to illustrate the actual coupled dynamics of~\eqref{eq:micro_xi}-\eqref{eq:micro_w}, we solve numerically the coupled system of equations with $N=50$ agents and with the functions $P$, $\Phi$ given in~\eqref{eq:P_BC},~\eqref{eq:Phi_3poles}, respectively.

\begin{figure}[!t]
\centering
\subfigure[$\Delta=1,\,\alpha=-0.3$]{\includegraphics[scale=0.5]{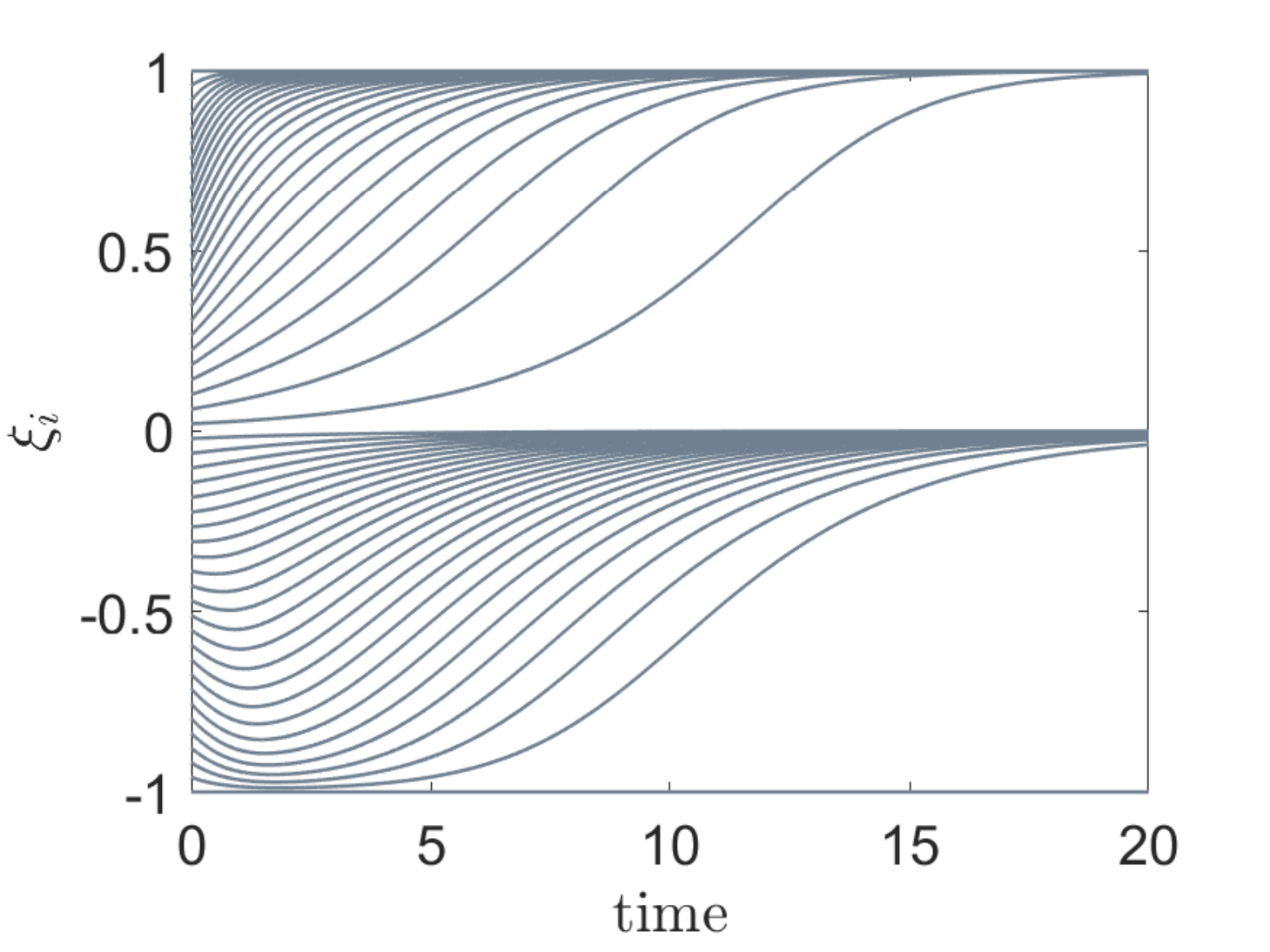}}
\subfigure[$\Delta=1,\,\alpha=0.3$]{\includegraphics[scale=0.5]{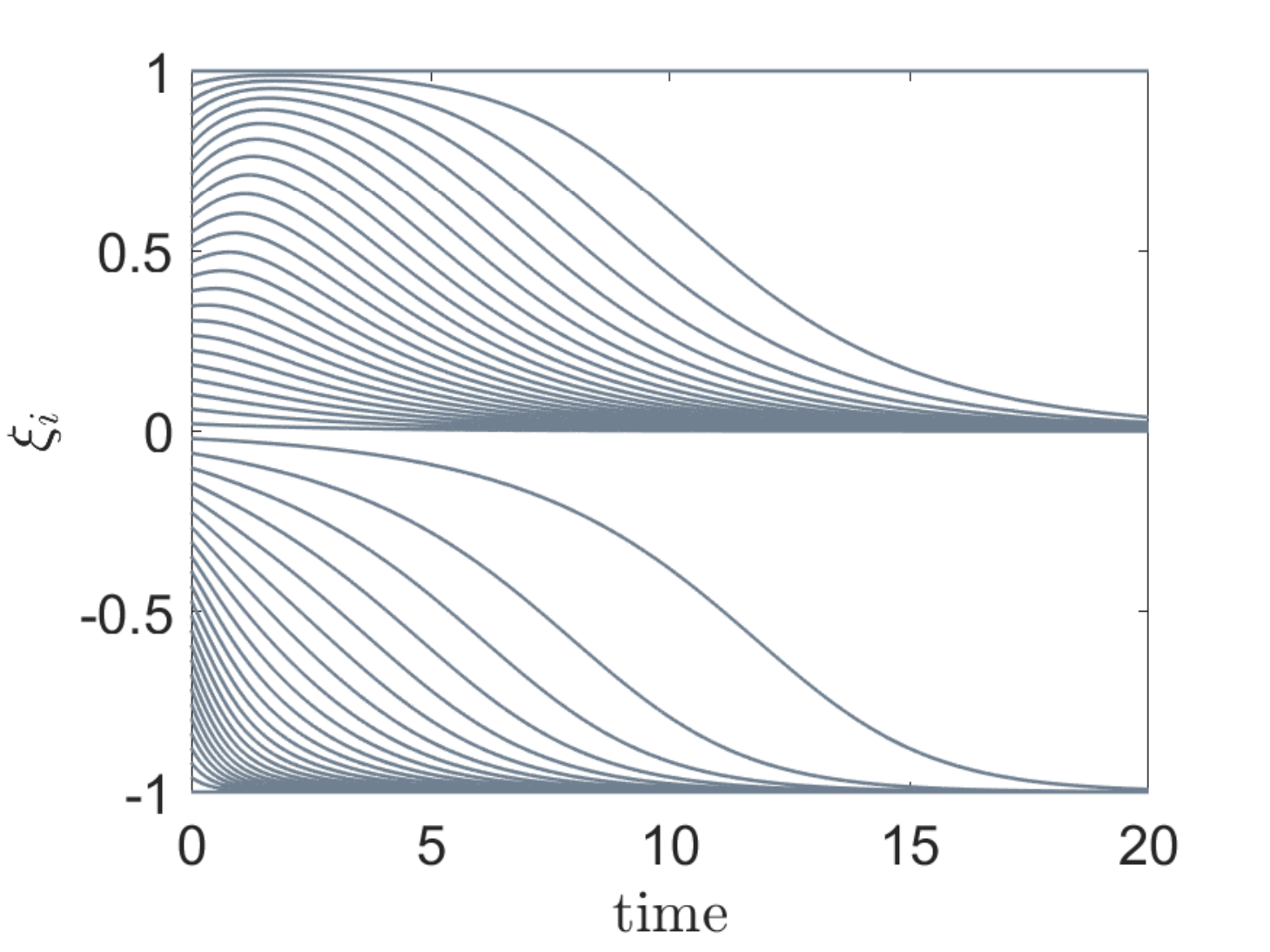}}
\caption{The curves $t\mapsto\xi_i(t)$ generated by the coupled system~\eqref{eq:micro_xi}-\eqref{eq:micro_w} with $N=50$ agents and $P$, $\Phi$ like in~\eqref{eq:P_BC},~\eqref{eq:Phi_3poles} with $\Delta=1$ and $\alpha=\pm 0.3$. The initial values $w_{0,i}$, $\xi_{0,i}$ have been sampled uniformly in the interval $[-1,\,1]$. The ODE system has been integrated numerically via a standard fourth order Runge-Kutta method.}
\label{fig:xi1}
\end{figure}

In Figure~\ref{fig:xi1} we present the curves $t\mapsto\xi_i(t)$ in the case $\Delta=1$ and for $\alpha=\pm 0.3$. Since the agents reach a global consensus around the centrist opinion $w=0$, cf. Figure~\ref{fig:BCmicro}(a), with a leftward-biased perceived social opinion $\alpha<0$ we observe the polarisation of the preferences either towards the indecisiveness $\xi=0$, if the initial preference was in turn leftward-biased, i.e. $\xi_{0,i}<0$, or in $\xi=1$, if the initial preference was rightward-biased, i.e. $\xi_{0,i}>0$, cf. Figure~\ref{fig:xi1}(a). Conversely, with a rightward-biased perceived social opinion $\alpha>0$ we observe indecisiveness if $\xi_{0,i}>0$ and consolidation in $\xi=-1$ if $\xi_{0,i}<0$, cf. Figure~\ref{fig:xi1}(b).

\begin{figure}[!t]
\centering
\subfigure[$\Delta=0.4,\,\alpha=-0.3$]{\includegraphics[scale=0.5]{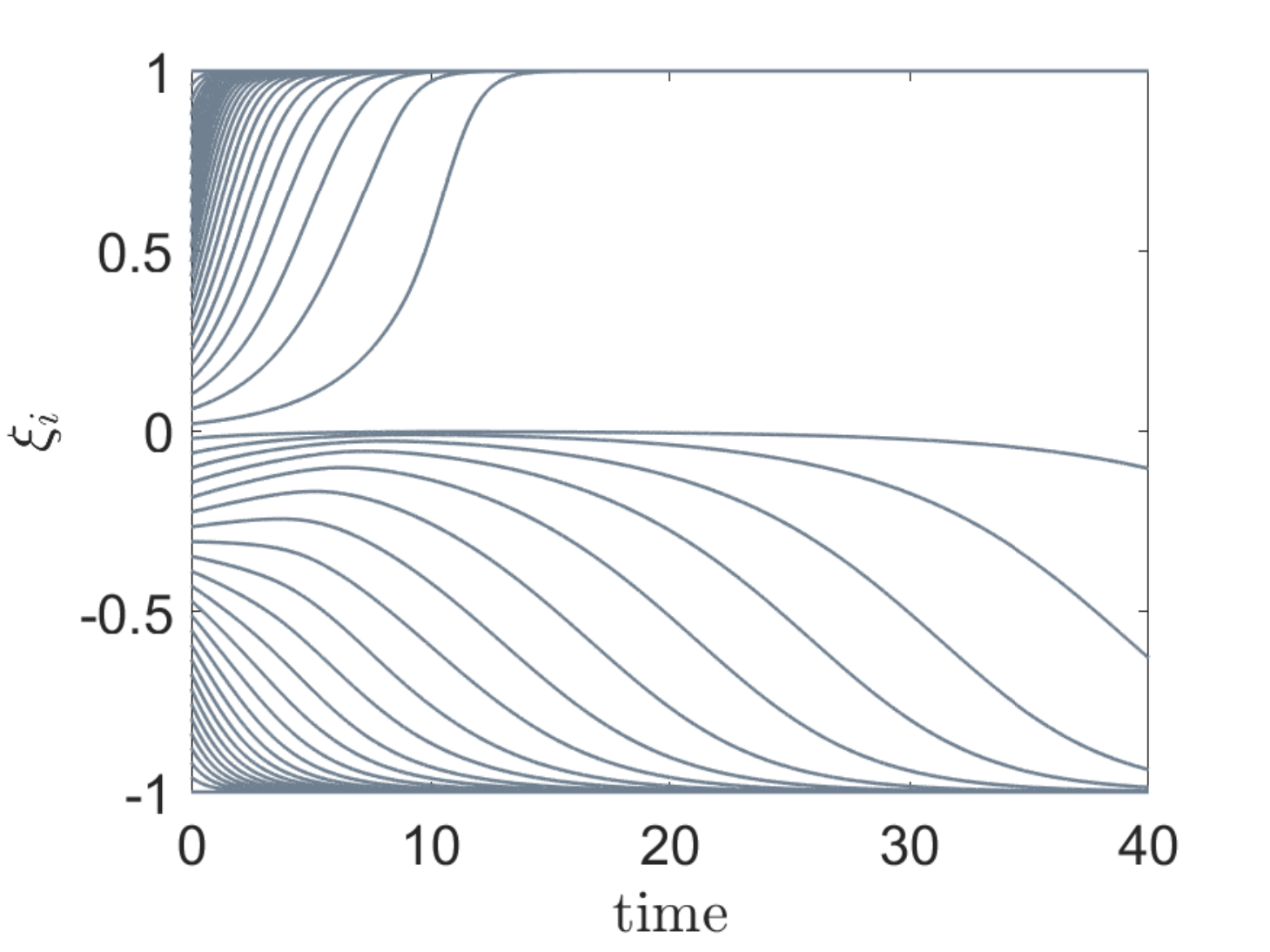}}
\subfigure[$\Delta=0.4,\,\alpha=0.3$]{\includegraphics[scale=0.5]{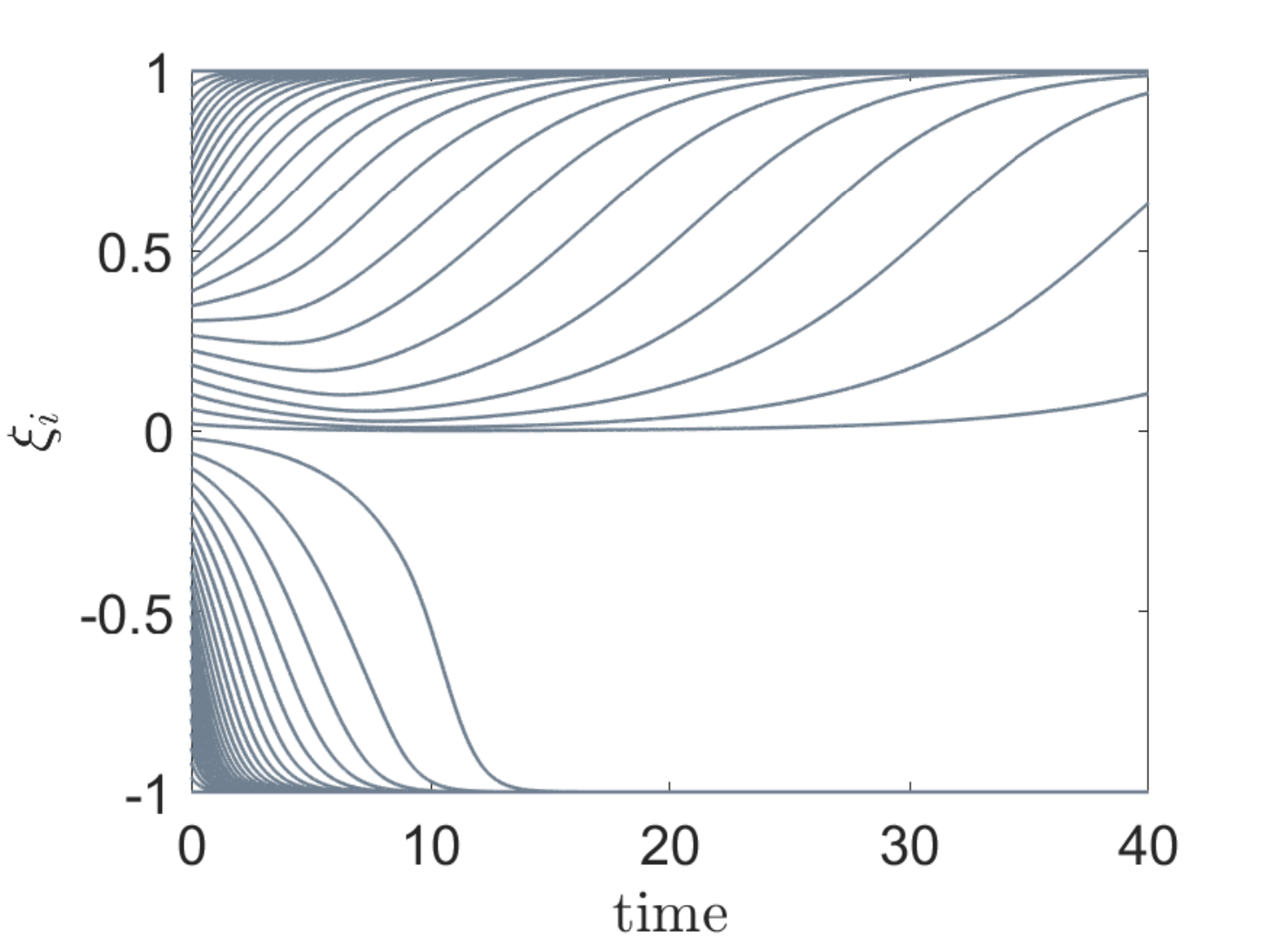}} \\
\subfigure[$\Delta=0.4,\,\alpha=-0.6$]{\includegraphics[scale=0.5]{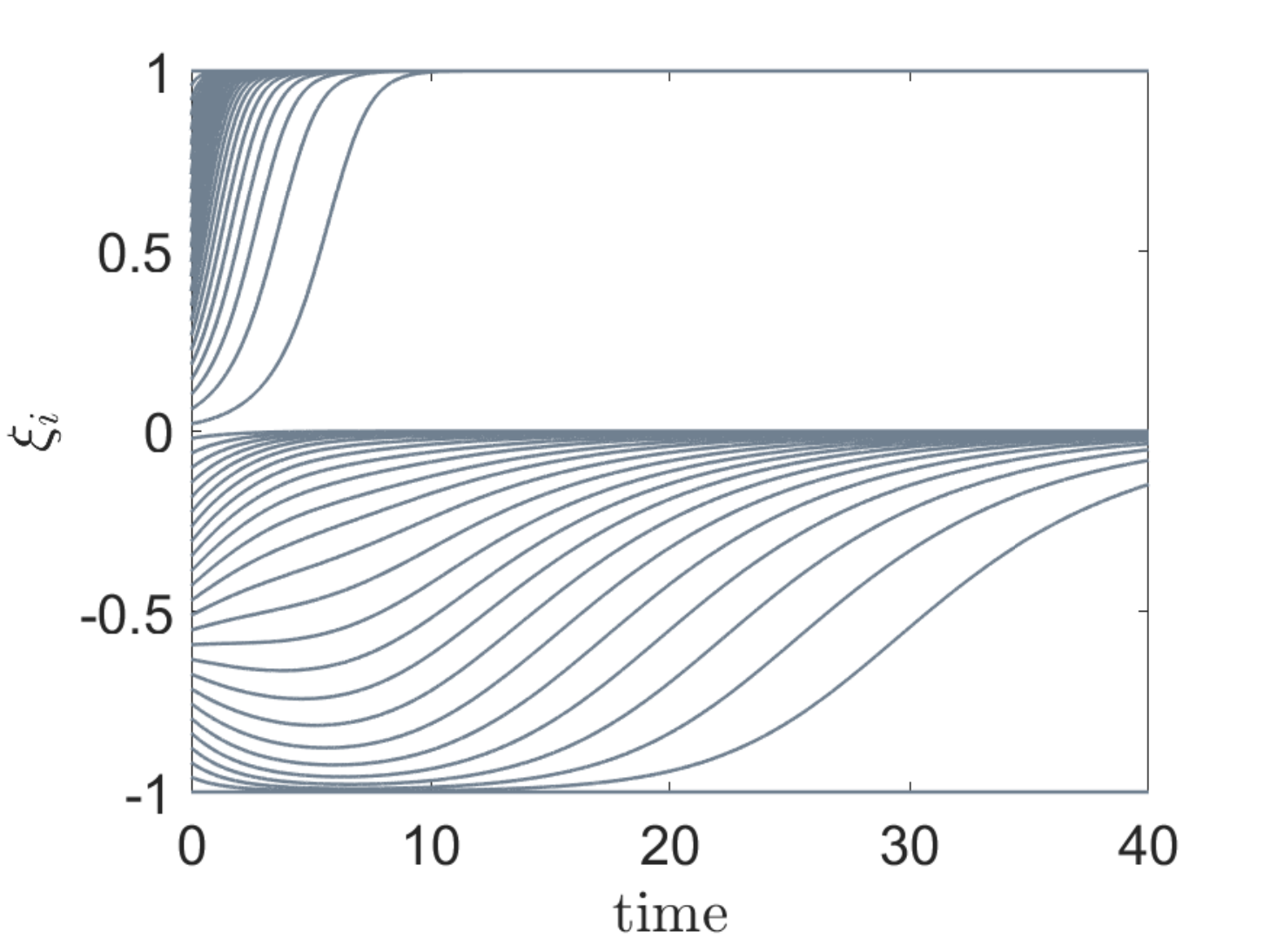}}
\subfigure[$\Delta=0.4,\,\alpha=0.6$]{\includegraphics[scale=0.5]{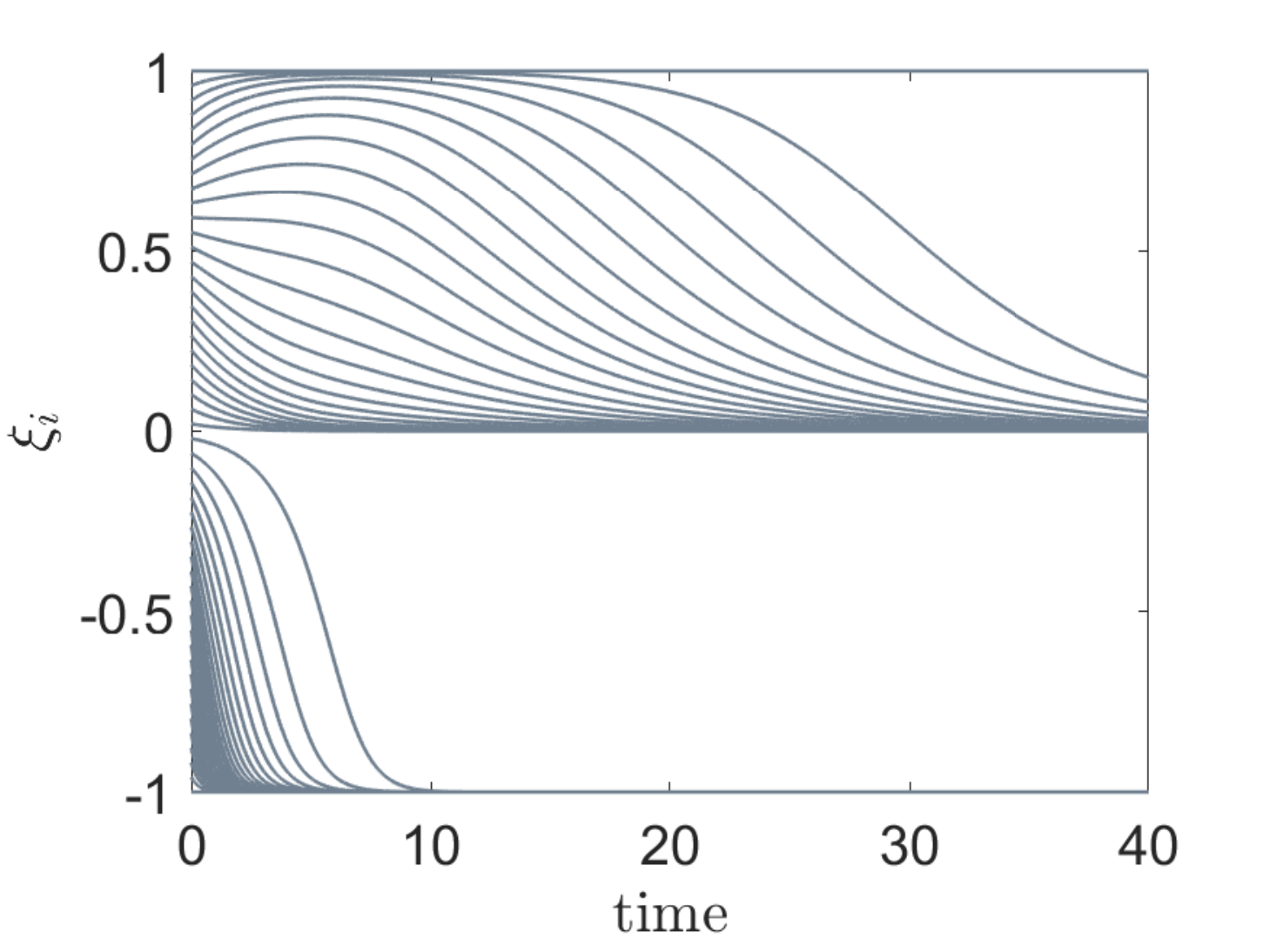}}
\caption{The same as in Figure~\ref{fig:xi1} but with $\Delta=0.4$, cf. Figure~\ref{fig:BCmicro}(b), and $\alpha=\pm 0.3$ (top row), $\alpha=\pm 0.6$ (bottom row).}
\label{fig:xi2}
\end{figure}

Such rather simple dynamics may become more complex under the formation of multiple opinion clusters. To exemplify this case, we consider now $\Delta=0.4$, like in Figure~\ref{fig:BCmicro}(b), and again the two cases $\alpha=\pm 0.3$, cf. Figure~\ref{fig:xi2}(a, b) along with also $\alpha=\pm 0.6$, cf. Figure~\ref{fig:xi2}(c, d). In this case, simultaneous polarisations in $\xi=\pm 1$ can also be observed, depending on the distribution of the pairs $(\xi_{0,i},\,w_{0,i})$ at the initial time.

\section{Aggregate analysis of opinion dynamics}
\label{sect:kinetic_opinion}
The discussion set forth in the previous section shows that it is in general quite hard to analyse exactly the interplay between opinions and preferences from a strictly microscopic point of view. Due to the severe dependence of the microscopic system on the particular initial state and trajectory of each agent, the main difficulty is, as usual, to grasp the essential facts able to explain the big picture, namely to depict the collective behaviour. For this reason, from this section we move to a more aggregate analysis, which, starting from a description of opinion dynamics by methods of statistical physics and kinetic theory, will finally lead us to macroscopic equations for the preference dynamics written in terms of hydrodynamic parameters such as the density of the agents and their mean opinion.

\subsection{Microscopic binary interactions}
\label{sect:microscopic}
In order to approach the opinion dynamics~\eqref{eq:micro_w} from the point of view of kinetic theory, we need to set up a consistent scheme of binary, i.e. pairwise, interactions among the agents. To this purpose, inspired by~\cite{carrillo2010SIMA}, we consider~\eqref{eq:micro_w} for just two agents, say $i$, $j$, and we discretise the differential equation with the forward Euler formula during a small time step $0<\gamma<1$. Setting
$$ w:=w_i(t),\quad w_\ast:=w_j(t),\quad w':=w_i(t+\gamma),\quad w_\ast':=w_j(t+\gamma) $$
we obtain the binary rules
\begin{equation}
\begin{split}
    w'&=w+\gamma P(w,\,w_\ast)(w_\ast-w)+D(w)\eta, \\
    w_\ast'&=w_\ast+\gamma P(w_\ast,\,w)(w-w_\ast)+D(w_\ast)\eta,
    \label{eq:binary}
\end{split}
\end{equation}
where we have also added a random contribution, given by a centred random variable $\eta$, modelling stochastic fluctuations induced by the self-thinking of the agents. Here, $D(\cdot)\geq 0$ is an opinion-dependent diffusion coefficient modulating the amplitude of the stochastic fluctuations, that is the variance of $\eta$.

In general, the binary interactions~\eqref{eq:binary} are such that
\begin{equation}
    \ave{w'+w_\ast'}=w+w_\ast+\gamma\left(P(w,\,w_\ast)-P(w_\ast,\,w)\right)(w_\ast-w),
    \label{eq:bin.ave}
\end{equation}
where $\ave{\cdot}$ denotes the expectation with respect to the distribution of $\eta$. Hence the mean opinion is in general not conserved on average in a single binary interaction unless $P$ is symmetric, i.e. $P(w,\,w_\ast)=P(w_\ast,\,w)$ for all $w,\,w_\ast\in [-1,\,1]$.
Furthermore, at leading order for $\gamma$ small enough we have
\begin{align}
    \begin{aligned}[b]
        \ave{(w')^2+(w_\ast')^2} &= w^2+w_\ast^2+2\gamma\left(wP(w,\,w_\ast)-w_\ast P(w_\ast,\,w)\right)(w_\ast-w) \\
        &\phantom{=} +\left(D^2(w)+D^2(w_\ast)\right)\sigma^2+o(\gamma),
    \end{aligned}
    \label{eq:bin.energy}
\end{align}
where $\sigma^2>0$ denotes the variance of $\eta$. Therefore, in general, also the energy is not conserved on average in a single binary interaction, not even for a symmetric function $P$.

Equations~\eqref{eq:bin.ave},~\eqref{eq:bin.energy} show that a particularly interesting case is when $P$ is constant, for then from~\eqref{eq:bin.ave} we deduce that the mean opinion is conserved in each binary interaction, while from~\eqref{eq:bin.energy} we see that, at least in the absence of stochastic fluctuations (i.e. formally for $\sigma^2=0$), the average energy is dissipated:
$$ \ave{(w')^2+(w_\ast')^2}=w^2+w_\ast^2-2\gamma(w_\ast-w)^2+o(\gamma)\leq w^2+w_\ast^2+o(\gamma). $$

In order to be physically admissible, the interaction rules~\eqref{eq:binary} have to be such that $\abs{w'},\,\abs{w_\ast'}\leq 1$ for $\abs{w},\,\abs{w_\ast}\leq 1$. Observing that
\begin{align*}
	\abs{w'} &= \abs{(1-\gamma P(w,\,w_\ast))w+\gamma P(w,\,w_\ast)w_\ast+D(w)\eta} \\
	&\leq (1-\gamma P(w,\,w_\ast))\abs{w}+\gamma P(w,\,w_\ast)+D(w)\abs{\eta},
\end{align*}
where we have used the fact that $\abs{w_\ast}\leq 1$, we see that a sufficient condition for $\abs{w'}\leq 1$ is
$$ D(w)\abs{\eta}\leq(1-\gamma P(w,\,w_\ast))(1-\abs{w}), $$
which is satisfied if there exists a constant $c>0$ such that
\begin{equation}
    \begin{cases}
        \abs{\eta}\leq c(1-\gamma P(w,\,w_\ast)) \\[1mm]
        cD(w)\leq 1-\abs{w},
    \end{cases}
    \quad \forall\,w,\,w_\ast\in [-1,\,1].
    \label{eq:eta_D}
\end{equation}
Considering that $P(w,\,w_\ast)\leq 1$ by assumption, the first condition can be further enforced by requiring $\abs{\eta}\leq c(1-\gamma)$, which implies that $\eta$ has to be chosen as a compactly supported random variable. The second condition forces instead $D(\pm 1)=0$. Taking inspiration from~\cite{toscani2006CMS}, possible choices are: $D(w)=1-\abs{w}$ and $c=1$, which produces $\abs{\eta}\leq 1-\gamma$; or $D(w)=1-w^2$ and $c=\frac{1}{2}$, which yields $\abs{\eta}\leq\frac{1}{2}(1-\gamma)$. Another less obvious option is
\begin{equation}
    D(w)=\sqrt{\left(1-(1+\gamma^s)w^2\right)_+} \quad \text{and} \quad c=\frac{\gamma^{s/2}}{\sqrt{1+\gamma^s}},
        \quad s>0,
    \label{eq:D_beta}
\end{equation}
where $(\cdot)_+:=\max\{0,\,\cdot\}$ denotes the positive part, which produces $\abs{\eta}\leq\frac{\gamma^{s/2}(1-\gamma)}{\sqrt{1+\gamma^s}}$. This function $D$ converges uniformly to $\sqrt{1-w^2}$ in $[-1,\,1]$ when $\gamma\to 0^+$. Notice, however, that such a uniform limit does not comply with~\eqref{eq:D_beta} regardless of choice of $c>0$, because of the infinite derivative at $w=\pm 1$.

Exactly the same considerations hold true for the second interaction rule in~\eqref{eq:binary}.

\subsection{Kinetic description and steady states}
\label{sect:boltzmann}
Introducing the distribution function $f=f(t,\,w):\R_+\times [-1,\,1]\to\R_+$, such that $f(t,\,w)dw$ is the fraction of agents with opinion in $[w,\,w+dw]$ at time $t$, the binary rules~\eqref{eq:binary} can be encoded in a Boltzmann-type kinetic equation, which, in weak form, writes:
\begin{multline}
	\frac{d}{dt}\int_{-1}^1\varphi(w)f(t,\,w)\,dw \\
	=\frac{1}{2}\int_{-1}^1\int_{-1}^1\ave{\varphi(w')+\varphi(w_\ast^\prime)-\varphi(w)-\varphi(w_\ast)}f(t,\,w)f(t,\,w_\ast)\,dw\,dw_\ast,
    \label{eq:boltzmann.f}
\end{multline}
where $\varphi:[-1,\,1]\to\R$ is an arbitrary test function, i.e. any observable quantity depending on the microscopic state of the agents. Choosing $\varphi(w)=1$, we obtain that the integral of $f$ with respect to $w$ is constant in time, i.e. that the total number of agents is conserved. This also implies that, up to normalisation at the initial time, $f$ can be thought of as a probability density for every $t>0$. Choosing instead $\varphi(w)=w$ we discover
\begin{equation}
	\frac{d}{dt}\int_{-1}^1 wf(t,\,w)\,dw=
		\frac{\gamma}{2}\int_{-1}^1\int_{-1}^1(P(w,\,w_\ast)-P(w_\ast,\,w))(w_\ast-w)f(t,\,w)f(t,\,w_\ast)\,dw\,dw_\ast,
	\label{eq:mean}
\end{equation}
therefore the mean opinion $M_1:=\int_{-1}^1 wf(t,\,w)\,dw$ is either conserved in time, if $P$ is symmetric so that the right-hand side of the previous equation vanishes, or not conserved, if $P$ is non-symmetric. This difference has important consequences on the steady distributions of~\eqref{eq:boltzmann.f}, which in turn will impact considerably on the equations describing the formation of the preferences. Therefore, in what follows we investigate it in some detail.

\subsubsection{Symmetric~\texorpdfstring{$\boldsymbol{P}$}{}}
\label{sect:P.symm}
The prototype of a symmetric $P$ is the constant function $P\equiv 1$. In this case, from~\eqref{eq:boltzmann.f} we can recover an explicit expression of the asymptotic distribution function at least in the so-called \textit{quasi-invariant regime}, i.e. the one in which the variation of the opinion in each binary interaction is small. To describe such a regime, we scale the parameters $\gamma$, $\sigma^2$ in~\eqref{eq:binary} as
\begin{equation}
	\gamma\to\epsilon\gamma, \qquad \sigma^2\to\epsilon\sigma^2,
	\label{eq:scaling}
\end{equation}
where $\epsilon>0$ is an arbitrarily small scaling coefficient. Parallelly, in order to study the large time behaviour of the system, we introduce the new time scale $\tau:=\epsilon t$ and we scale the distribution function as $g(\tau,\,w):=f(\frac{\tau}{\epsilon},\,w)$. In this way, it is clear that, at every fixed $\tau>0$ and in the limit $\epsilon\to 0^+$, $g$ describes the large time trend of $f$. Since $\partial_\tau g=\frac{1}{\epsilon}\partial_t f$, substituting in~\eqref{eq:boltzmann.f} and using the symmetry of the interactions~\eqref{eq:binary} with $P\equiv 1$ we see that the equation satisfied by $g$ is
\begin{equation}
	\frac{d}{d\tau}\int_{-1}^1\varphi(w)g(\tau,\,w)\,dw
		=\frac{1}{\epsilon}\int_{-1}^1\int_{-1}^1\ave{\varphi(w')-\varphi(w)}g(\tau,\,w)g(\tau,\,w_\ast)\,dw\,dw_\ast.
    \label{eq:boltzmann.g}
\end{equation}

Now, because of the scaling~\eqref{eq:scaling}, if $\varphi$ is sufficiently smooth then the difference $\ave{\varphi(w')-\varphi(w)}$ is small and can be expanded about $w$ to give:
$$ \ave{\varphi(w)-\varphi(w')}=\varphi'(w)\ave{w'-w}+\frac{1}{2}\varphi''(w)\ave{\left(w'-w\right)^2}
	+\frac{1}{6}\varphi'''(\bar{w})\ave{(w'-w)^3}, $$
where $\min\{w,\,w'\}<\bar{w}<\max\{w,\,w'\}$. Plugging into~\eqref{eq:boltzmann.g} this produces
\begin{align*}
    \frac{d}{d\tau}\int_{-1}^1\varphi(w)g(\tau,\,w)\,dw &= \gamma\int_{-1}^1\varphi'(w)(m-w)g(\tau,\,w)\,dw \\
	&\phantom{=} +\frac{\sigma^2}{2}\int_{-1}^1\varphi''(w)D^2(w)g(\tau,\,w)\,dw+R_\varphi(g,\,g),
\end{align*}
where we have denoted by $m\in [-1,\,1]$ the constant mean opinion and where $R_\varphi(g,\,g)$ is a reminder such that $\abs{R_\varphi(g,\,g)}=O(\sqrt{\epsilon})$ under the assumption that $\eta$ has finite third order moment, i.e. $\ave{\abs{\eta}^3}<+\infty$, cf.~\cite{toscani2006CMS} for details. Hence for $\epsilon\to 0^+$ it results $R_\varphi(g,\,g)\to 0$ and we get
$$ \frac{d}{d\tau}\int_{-1}^1\varphi(w)g(\tau,\,w)\,dw=
		\gamma\int_{-1}^1\varphi'(w)(m-w)g(\tau,\,w)\,dw+\frac{\sigma^2}{2}\int_{-1}^1\varphi''(w)D^2(w)g(\tau,\,w)\,dw. $$
Integrating back by parts the terms on the right-hand side and assuming $\varphi(\pm 1)=\varphi'(\pm 1)=0$, due to the arbitrariness of $\varphi$ this can be recognised as a weak form of the \textit{Fokker-Planck equation}
\begin{equation}
    \partial_\tau g=\frac{\sigma^2}{2}\partial^2_w\left(D^2(w)g\right)+\gamma\partial_w((w-m)g).
    \label{eq:FP}
\end{equation}
Fixing\footnote{In view of the scaling~\eqref{eq:scaling}, as $\epsilon\to 0^+$ the function~\eqref{eq:D_beta} converges uniformly to $\sqrt{1-w^2}$, which can therefore be chosen as diffusion coefficient in the Fokker-Planck equation~\eqref{eq:FP} \textit{after} performing the quasi-invariant limit.} $D(w)=\sqrt{1-w^2}$, the unique asymptotic ($\tau\to +\infty$) solution with unitary mass, say $g_\infty(w)$, to~\eqref{eq:FP} reads
\begin{equation}
    g_\infty(w)=\frac{(1+w)^{\frac{1+m}{\lambda}-1}(1-w)^{\frac{1-m}{\lambda}-1}}{2^{\frac{2}{\lambda}-1}
    		\Beta\left(\frac{1+m}{\lambda},\,\frac{1-m}{\lambda}\right)},
    			\qquad \lambda:=\frac{\sigma^2}{\gamma},
    \label{eq:ginf}
\end{equation}
where $\Beta(\cdot,\,\cdot)$ denotes the Beta function. Notice that such a $g_\infty$ is a Beta probability density function on the interval $[-1,\,1]$. Using the known formulas for the moments of Beta random variables, we easily check that its mean is indeed $m$ and we compute its energy as
\begin{equation}
	M_{2,\infty}:=\int_{-1}^1 w^2g_\infty(w)\,dw=\frac{2m^2+\lambda}{2+\lambda}.
	\label{eq:M2}
\end{equation}

\begin{figure}[!t]
\centering
\includegraphics[width=0.5\textwidth]{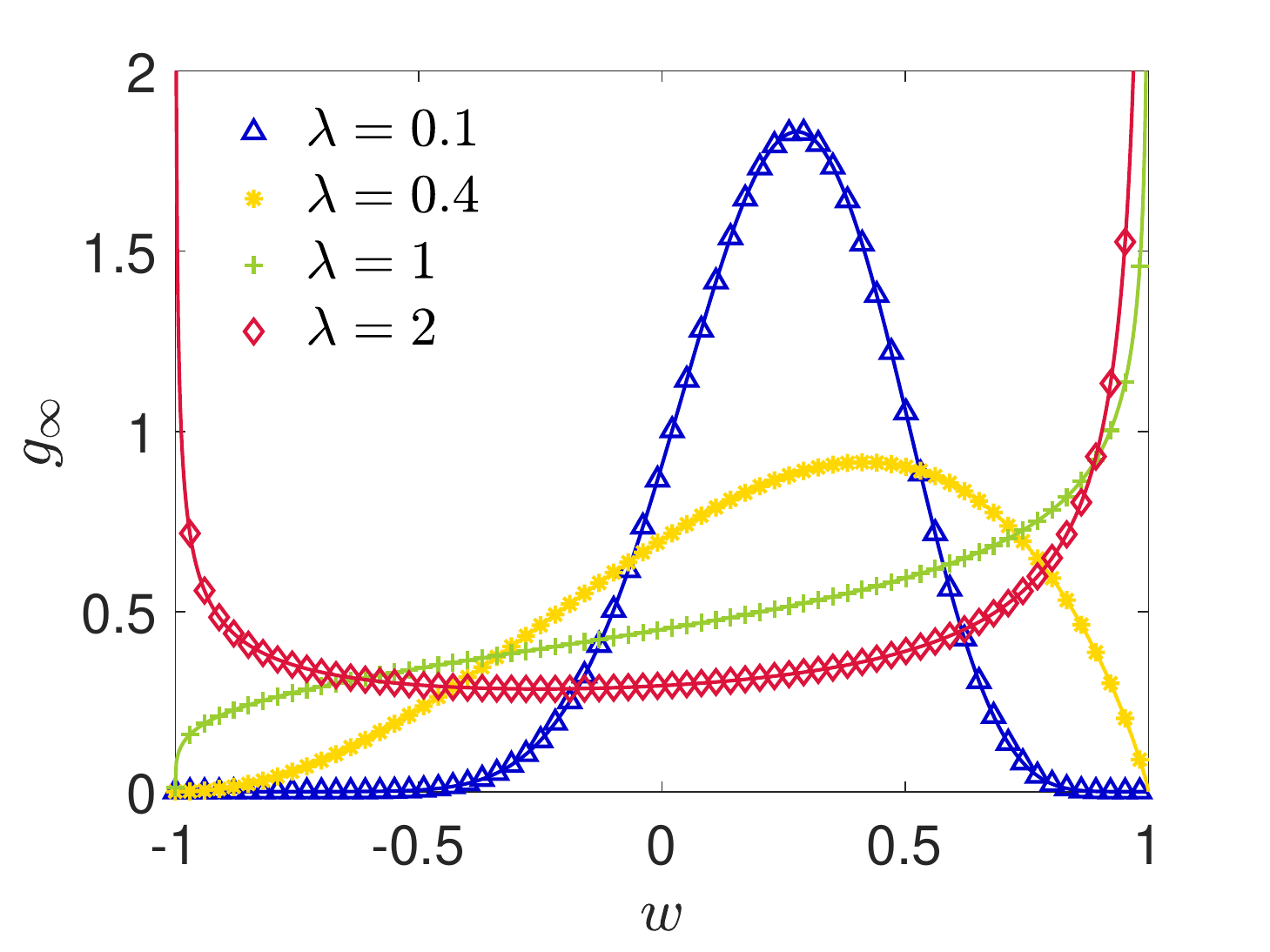}
\caption{Asymptotic opinion distribution~\eqref{eq:ginf} with mean $m=0.25$ and four different values of the parameter $\lambda$.}
\label{fig:ginf.beta}
\end{figure}

In Figure~\ref{fig:ginf.beta} we illustrate some typical trends of the distribution function~\eqref{eq:ginf} with positive mean, $m=0.25$ in this example. We observe that, depending on the value of $\lambda$, such a distribution may depict a transition from a strong consensus around the mean ($\lambda=0.1$) to a milder consensus ($\lambda=0.4$) and further to a radicalisation in the extreme opinion $w=1$ ($\lambda=1$) up to the appearance of a double radicalisation in the two opposite extreme opinions $w=\pm 1$ ($\lambda=2$).

\subsubsection{Non-symmetric~\texorpdfstring{$\boldsymbol{P}$}{}}
\label{sect:P.nonsymm}
A natural prototype of a non-symmetric function $P$ is a linear perturbation of a constant $P$ depending on only one of the two variables $w$, $w_\ast$. More specifically, we consider
\begin{equation}
	P(w,\,w_\ast)=P(w_\ast)=pw_\ast+q,
	\label{eq:P_nonsymm}
\end{equation}
where $p,\,q\in\R$ have to be chosen in such a way that $pw_\ast+q\in [0,\,1]$ for all $w_\ast\in [-1,\,1]$. This is obtained if
$$ 0\leq q\leq 1, \qquad \abs{p}\leq\min\{q,\,1-q\}. $$
With respect to model~\eqref{eq:binary}, such a function $P$ describes a situation in which agents with opinion $w_\ast>0$ are more persuasive than agents with opinion $w_\ast<0$ if $p>0$ and vice versa if $p<0$.

Using~\eqref{eq:P_nonsymm} in~\eqref{eq:mean} we obtain that the evolution of the mean opinion $M_1=M_1(t)$ is ruled by
$$ \frac{dM_1}{dt}=\frac{p\gamma}{2}\int_{-1}^1\int_{-1}^1{(w_\ast-w)}^2f(t,\,w)f(t,\,w_\ast)\,dw\,dw_\ast, $$
whence we see that the sign of the time derivative $\frac{dM_1}{dt}$ coincides with that of $p$. Thus, if $p>0$ the mean opinion is non-decreasing, while if $p<0$ the mean opinion is non-increasing. Continuing the previous calculation, we further find:
$$ \frac{dM_1}{dt}=\frac{p\gamma}{2}\left(M_2-M_1^2\right), $$
which indicates that at the steady state ($t\to +\infty$) it results invariably $M_{2,\infty}=M_{1,\infty}^2$. This implies that the asymptotic distribution has zero variance, thus that it is necessarily a Dirac delta centred in the asymptotic mean opinion, i.e. $f_\infty(w)=\delta(w-M_{1,\infty})$. Plugging this into~\eqref{eq:boltzmann.f} we discover
$$ \ave{\varphi(M_{1,\infty}+D(M_{1,\infty})\eta)}-\varphi(M_{1,\infty})=0, $$
which has to hold for every test function $\varphi$. As a consequence, we deduce $D(M_{1,\infty})=0$, whence $M_{1,\infty}=\pm 1$ if the only zeroes of the diffusion coefficient are $w=\pm 1$ like in the examples considered in Section~\ref{sect:microscopic}.

In conclusion, with the non-symmetric function $P$ given by~\eqref{eq:P_nonsymm} we fully characterise the asymptotic distribution function as:
\begin{itemize}
\item $f_\infty(w)=\delta(w+1)$ if $p<0$, the mean opinion decreasing from its initial value to $M_{1,\infty}=-1$;
\item $f_\infty(w)=\delta(w-1)$ if $p>0$, the mean opinion increasing from its initial value to $M_{1,\infty}=1$.
\end{itemize}

The considerations above can be generalised to the following function $P$:
\begin{equation}
	P(w,\,w_\ast)=rw+pw_\ast+q,
	\label{eq:P_nonsymm.gen}
\end{equation}
where $p\neq r$, so that $P$ is non-symmetric, and where the coefficients $p,\,q,\,r\in\R$ have to be chosen in such a way that $rw+pw_\ast+q\in [0,\,1]$ for all $(w,\,w_\ast)\in [-1,\,1]^2$. Repeating the previous calculations, we conclude that:
\begin{itemize}
\item $f_\infty(w)=\delta(w+1)$ if $p-r<0$; in this case, the mean opinion decreases from its initial value to $M_{1,\infty}=-1$;
\item $f_\infty(w)=\delta(w-1)$ if $p-r>0$; in this case, the mean opinion increases from its initial value to $M_{1,\infty}=1$.
\end{itemize}

From the modelling point of view, we may interpret the difference $p-r$ as a balance between the persuasion ability of the agents, expressed by $p$, and their tendency to be persuaded, expressed by $r$. Notice indeed that for $p=0$ and $r\neq 0$ we obtain the mirror case of~\eqref{eq:P_nonsymm}, in which agents with opinion $w>0$ are more inclined to change their opinion than agents with opinion $w<0$ if $r>0$ and vice versa if $r<0$.

\bigskip

The discussion above clearly shows that an arbitrarily small perturbation of a constant $P$, by destroying the conservation of the mean opinion, may drag the system towards asymptotic configurations much less variegated than~\eqref{eq:ginf} independently of the parameters $\gamma$, $\sigma^2$ of the interactions.

\section{Macroscopic description of preference formation}
\label{sect:macro}
According to model~\eqref{eq:micro_xi}-\eqref{eq:micro_w}, the opinions of the agents evolve through mutual interactions independent of the preferences; on the other hand, the preference of each agent is transported in time by his/her opinion. This suggests that a proper way to account for the interplay between opinion and preference in an aggregate manner is by means of an inhomogeneous Boltzmann-type kinetic equation, whose transport term describes the evolution of the preference and whose ``collisional'' term accounts simultaneously for the changes in the opinions.

\subsection{Inhomogeneous Boltzmann-type description and hydrodynamics}
\label{sect:boltzmann.inhomog}
A Boltzmann-type description of the opinion dynamics in the form of binary interactions~\eqref{eq:binary} coupled to the transport of the preference~\eqref{eq:micro_xi} is obtained by introducing the kinetic distribution function
$$ f=f(t,\,\xi,\,w):\R_+\times [-1,\,1]\times [-1,\,1]\to\R_+, $$
such that $f(t,\,\xi,\,w)d\xi\,dw$ is the proportion of agents that at time $t$ have a preference in $[\xi,\,\xi+d\xi]$ and an opinion in $[w,\,w+dw]$. The distribution function $f$ satisfies the following weak Boltzmann-type equation:
\begin{multline}
	\partial_t\int_{-1}^1\varphi(w)f(t,\,\xi,\,w)\,dw
		+\partial_\xi\left(\Phi(\xi)\int_{-1}^1(w-\alpha)\varphi(w)f(t,\,\xi,\,w)\,dw\right) \\
	=\frac{1}{2}\int_{-1}^1\int_{-1}^1\ave{\varphi(w')+\varphi(w_\ast')-\varphi(w)-\varphi(w_\ast)}
		f(t,\,\xi,\,w)f(t,\,\xi,\,w_\ast)\,dw\,dw_\ast,
	\label{eq:boltzmann.inhomog.f}
\end{multline}
where the transport term (second term on the left-hand side) has been written taking into account that, according to~\eqref{eq:micro_xi}, the transport velocity of the preference $\xi$ is $(w-\alpha)\Phi(\xi)$ and where $w'$, $w_\ast'$ on the right-hand side are given by~\eqref{eq:binary}.

From the distribution function $f$, by integration with respect to the opinion $w$, we can compute macroscopic quantities in the space of the preferences, such as the \textit{density} of the agents with preference $\xi$ at time $t$:
$$ \rho(t,\,\xi):=\int_{-1}^1f(t,\,\xi,\,w)\,dw $$
and the \textit{mean opinion} of the agents with preference $\xi$ at time $t$:
$$ m(t,\,\xi):=\frac{1}{\rho(t,\,\xi)}\int_{-1}^1wf(t,\,\xi,\,w)\,dw. $$

The interest in~\eqref{eq:boltzmann.inhomog.f} is that it allows one to obtain evolution equations directly for the quantities $\rho$, $m$, provided one is able to characterise the large time statistical trends of the opinions, like in Section~\ref{sect:kinetic_opinion}. The underlying key idea is to consider a so-called \textit{hydrodynamic regime}, in which the opinions reach a local equilibrium much more quickly than the preferences, pretty much in the spirit of the microscopic investigations performed in Section~\ref{sect:micro_xi_w}.

Let $0<\delta\ll 1$ be a small parameter, which we use to define a macroscopic time scale $\tau:=\delta t$, i.e. the time scale of the evolution of the preferences, which then turns out to be much larger, viz. slower, than the characteristic one of the binary interactions among the agents. If we want that, on this new scale, the preference dynamics remain the same, from~\eqref{eq:micro_xi} we see that we need to scale simultaneously the transport speed of the preference by letting $\Phi(\xi)\to\delta\Phi(\xi)$.

Let $g(\tau,\,\xi,\,w):=f(\frac{\tau}{\delta},\,\xi,\,w)$, whence $\partial_\tau g=\frac{1}{\delta}\partial_t f$. Plugging into~\eqref{eq:boltzmann.inhomog.f} we find that $g$ satisfies the equation
\begin{multline}
	\partial_\tau\int_{-1}^1\varphi(w)g(\tau,\,\xi,\,w)\,dw
		+\partial_\xi\left(\Phi(\xi)\int_{-1}^1(w-\alpha)\varphi(w)g(\tau,\,\xi,\,w)\,dw\right) \\
	=\frac{1}{2\delta}\int_{-1}^1\int_{-1}^1\ave{\varphi(w')+\varphi(w_\ast')-\varphi(w)-\varphi(w_\ast)}
		g(\tau,\,\xi,\,w)g(\tau,\,\xi,\,w_\ast)\,dw\,dw_\ast.
	\label{eq:boltzmann.inhomog.g}
\end{multline}

Basically, the aforesaid scaling produces the coefficient $1/\delta$ in front of the interaction term, hence $\delta$ is analogous to the Knudsen number in classical fluid dynamics. Since we are assuming that $\delta$ is small, a hydrodynamic regime is justified and, in particular, it can be described by a splitting of~\eqref{eq:boltzmann.inhomog.g}, cf.~\cite{Dur-Tos}, totally analogous to the one often adopted in the numerical solution of the inhomogeneous Boltzmann equation, see e.g.~\cite{dimarco2018JCP,dimarco2014AN,pareschi2001SISC}. One first solves the fast interactions:
\begin{multline}
	\partial_\tau\int_{-1}^1\varphi(w)g(\tau,\,\xi,\,w)\,dw \\
	=\frac{1}{2\delta}\int_{-1}^1\int_{-1}^1\ave{\varphi(w')+\varphi(w_\ast')-\varphi(w)-\varphi(w_\ast)}
		g(\tau,\,\xi,\,w)g(\tau,\,\xi,\,w_\ast)\,dw\,dw_\ast,
	\label{eq:splitting.int}
\end{multline}
which, owing to the high frequency $1/\delta$, reach quickly an equilibrium described by a local (in $\xi$ and $\tau$) asymptotic distribution function playing morally the role of a local Maxwellian. Notice indeed that~\eqref{eq:splitting.int} is actually an equation on the time scale of the microscopic interactions, because $\tau$ can be scaled back to $t$ using the factor $1/\delta$. Next, one transports such a local equilibrium distribution according to the remaining terms of~\eqref{eq:boltzmann.inhomog.g} on the slower hydrodynamic scale:
\begin{equation}
	\partial_\tau\int_{-1}^1\varphi(w)g(\tau,\,\xi,\,w)\,dw
		+\partial_\xi\left(\Phi(\xi)\int_{-1}^1(w-\alpha)\varphi(w)g(\tau,\,\xi,\,w)\,dw\right)=0.
	\label{eq:splitting.transp}
\end{equation}

Due to~\eqref{eq:splitting.int}, and taking the definition of $\rho$ into account, the local ``Maxwellian'' can be given the form $g(\tau,\,\xi,\,w)=\rho(\tau,\,\xi)g_\infty(w)$, where $g_\infty$ is one of the asymptotic opinion distribution functions found in Section~\ref{sect:kinetic_opinion}. This is the distribution transported by~\eqref{eq:splitting.transp}, hence we finally obtain
\begin{equation}
	\partial_\tau\left(\rho\int_{-1}^1\varphi(w)g_\infty(w)\,dw\right)
		+\partial_\xi\left(\Phi(\xi)\rho\int_{-1}^1(w-\alpha)\varphi(w)g_\infty(w)\,dw\right)=0
	\label{eq:kinetic.hydro}
\end{equation}
and we can use the knowledge of $g_\infty$ to compute explicitly the remaining integral terms.

\subsection{First order hydrodynamic models}
\label{sect:first_order.hydro}
Let us consider at first the case of the non-symmetric functions $P$~\eqref{eq:P_nonsymm},~\eqref{eq:P_nonsymm.gen} discussed in Section~\ref{sect:P.nonsymm}. The asymptotic opinion distribution is either $g_\infty(w)=\delta(w+1)$ or $g_\infty(w)=\delta(w-1)$, depending on the asymmetry of $P$. Plugging into~\eqref{eq:kinetic.hydro} along with the choice $\varphi(w)=1$ we find therefore either
\begin{equation}
	\partial_\tau\rho-(1+\alpha)\partial_\xi\left(\Phi(\xi)\rho\right)=0
	\label{eq:FO.-1}
\end{equation}
or
\begin{equation}
	\partial_\tau\rho+(1-\alpha)\partial_\xi\left(\Phi(\xi)\rho\right)=0.
	\label{eq:FO.1}
\end{equation}
In both cases, we get a self-consistent equation for the sole density $\rho$ and we speak thus of \textit{first order} hydrodynamic model.

Unlike typical conservation laws, in~\eqref{eq:FO.-1} and~\eqref{eq:FO.1} the flux does not only depend on the variable $\xi$ through the conserved quantity $\rho$ but also explicitly through the function $\Phi$. An analogous characteristic is found, for instance, in conservation-law-based macroscopic models of vehicular traffic featuring different flux functions in different roads, see~\cite{garavelloNHM2007}.

We observe that both~\eqref{eq:FO.-1} and~\eqref{eq:FO.1} admit the family of stationary distributional solutions
\begin{equation}
	\rho_\infty(\xi)=\sum_{k=1}^M\rho_k\delta(\xi-\xi_k), \qquad \rho_k\geq 0,
	\label{eq:rhoinf}
\end{equation}
where the $\xi_k$'s are the zeroes of the function $\Phi$. This indicates that models~\eqref{eq:FO.-1} and~\eqref{eq:FO.1} reproduce the asymptotic polarisation of the agents in the preference poles individuated by the points where $\Phi$ vanishes. The coefficients $\rho_k$ represent the masses concentrating in each pole. Furthermore,~\eqref{eq:FO.-1} describes invariably a leftward transport of $\rho$ in the space of the preferences, because $-(1+\alpha)<0$ for all $\alpha\in (-1,\,1]$ (if $\alpha=-1$ the density is simply not transported). Conversely,~\eqref{eq:FO.1} describes invariably a rightward transport of $\rho$, since $1-\alpha>0$ for all $\alpha\in [-1,\,1)$ (now the density is not transported if $\alpha=1$).

\subsection{Second order hydrodynamic model}
\label{sect:second_order.hydro}
We now consider the symmetric case $P\equiv 1$ discussed in Section~\ref{sect:P.symm}, which produces the asymptotic opinion distribution $g_\infty$ given by~\eqref{eq:ginf}. Notice that this distribution is parametrised by the (local) mean opinion $m=m(\tau,\,\xi)$, because the latter is conserved by the opinion dynamics. This implies that, if we plug such a $g_\infty$ into~\eqref{eq:kinetic.hydro} together with the choice $\varphi(w)=1$, we do not get a self-consistent equation for the density $\rho$. In fact, we find:
$$ \partial_\tau\rho+\partial_\xi\left(\Phi(\xi)\rho(m-\alpha)\right)=0, $$
with both hydrodynamic parameters $\rho$, $m$ unknown. In order to close the macroscopic equations, we need a further equation relating $\rho$ and $m$, which we can obtain from~\eqref{eq:kinetic.hydro} with $\varphi(w)=w$ and recalling also~\eqref{eq:M2}:
$$ \partial_\tau(\rho m)+\partial_\xi\left(\Phi(\xi)\rho\left(\frac{2m^2+\lambda}{2+\lambda}-\alpha m\right)\right)=0. $$

On the whole, we get the \textit{second order} (i.e., composed of a self-consistent pair of equations) hydrodynamic model
\begin{equation}
	\begin{cases}
		\partial_\tau\rho+\partial_\xi\left(\Phi(\xi)\rho(m-\alpha)\right)=0 \\[2mm]
		\partial_\tau(\rho m)+\partial_\xi\left(\Phi(\xi)\rho\left(\dfrac{2m^2+\lambda}{2+\lambda}-\alpha m\right)\right)=0,
	\end{cases}
	\label{eq:SO}
\end{equation}
where the parameter $\lambda=\sigma^2/\gamma$, which here enters the game through the energy of the stationary opinion distribution~\eqref{eq:ginf}, is reminiscent of the self-thinking (diffusion) of the agents.

Also in this case,~\eqref{eq:rhoinf} is a family of admissible stationary distributional solutions. Hence model~\eqref{eq:SO} can in turn reproduce the asymptotic polarisation of the preferences already observed in the microscopic model.

It is useful to ascertain under which conditions system~\eqref{eq:SO} is hyperbolic in the natural state space $\{(\rho,\,m)\in\R_+\times [-1,\,1]\}$. To this purpose, we rewrite it in the quasilinear matrix form
$$ \partial_\tau U+\Phi(\xi)A(U)\partial_\xi U+\Phi'(\xi)F(U)=0, $$
where $U:=(\rho,\,m)^T$, $A(U)$ is the matrix
\begin{equation}
	A(U):=
	\begin{pmatrix}
		m-\alpha & \rho \\[1mm]
		\frac{\lambda\left(1-m^2\right)}{\rho(2+\lambda)} & \frac{(2-\lambda)m}{2+\lambda}-\alpha
	\end{pmatrix}
	\label{eq:A}
\end{equation}
and $F(U)$ denotes lower order terms, which is not important to write explicitly. Since $\Phi$ is real-valued, system~\eqref{eq:SO} is hyperbolic if both eigenvalues of $A(U)$ are real. To check this, we compute the discriminant $\Delta(U)$ of the characteristic polynomial of $A(U)$:
$$ \Delta(U):=\tr^2{A(U)}-4\det{A(U)}=\frac{4\lambda}{\lambda+2}\left(1-\frac{2m^2}{\lambda+2}\right). $$
Since $m\in [-1,\,1]$, thus $m^2\in [0,\,1]$, and $\lambda\geq 0$, we easily see that $\Delta(U)$ is always non-negative. Therefore, we conclude:
\begin{proposition}
System~\eqref{eq:SO} is hyperbolic in the whole state space $\{(\rho,\,m)\in\R_+\times [-1,\,1]\}$ for every choice of the parameters $\alpha\in [-1,\,1]$, $\lambda\geq 0$ and for every function $\Phi:[-1,\,1]\to [0,\,1]$.
\end{proposition}

\subsection{General first and second order hydrodynamic models}
\label{sect:general.hydro}
If the asymptotic opinion distribution $g_\infty$ is not known analytically, like e.g. in the significant case~\eqref{eq:P_BC}, the hydrodynamic models can still be written from~\eqref{eq:kinetic.hydro}, although only in a semi-analytical form.

Assume that the microscopic dynamics~\eqref{eq:binary} do not conserve the mean opinion. Then the sole conserved quantity is the mass of the agents and from~\eqref{eq:kinetic.hydro} with $\varphi(w)=1$ we obtain the first order model
$$ \partial_\tau\rho+(M_{1,\infty}-\alpha)\partial_\xi(\Phi(\xi)\rho)=0 $$
in the unknown $\rho=\rho(\tau,\,\xi)$, where $M_{1,\infty}:=\int_{-1}^1wg_\infty(w)\,dw$ is the asymptotic mean opinion. The latter may be computed e.g. from ~\eqref{eq:mean}, by means of an appropriate numerical approach.

Conversely, if the microscopic dynamics~\eqref{eq:binary} conserve the mean opinion then $g_\infty$ is parametrised by $m$ and from~\eqref{eq:kinetic.hydro} with $\varphi(w)=1,\,w$ we obtain the second order model
\begin{equation}
	\begin{cases}
		\partial_\tau\rho+\partial_\xi\left(\Phi(\xi)\rho(m-\alpha)\right)=0 \\[2mm]
		\partial_\tau(\rho m)+\partial_\xi(\Phi(\xi)\rho\left(M_{2,\infty}(m)-\alpha m\right))=0
	\end{cases}
	\label{eq:SO.gen}
\end{equation}
in the unknowns $\rho=\rho(\tau,\,\xi)$, $m=m(\tau,\,\xi)$. Here, $M_{2,\infty}(m):=\int_{-1}^1w^2g_\infty(w)\,dw$ is the energy of the asymptotic opinion distribution, expressed as a function of the conserved quantity $m$.

The precise calculation of $M_{2,\infty}$ requires, in general, an accurate numerical reconstruction of $g_\infty$. The latter is a stationary solution to the Fokker-Planck equation
\begin{equation}
	\partial_\tau g=\frac{\sigma^2}{2}\partial_w^2\left(D^2(w)g\right)+\gamma\partial_w(\cB[g]g)
	\label{eq:FP.gen}
\end{equation}
with
\begin{equation}
	\cB[g](\tau,\,w):=\int_{-1}^1P(w,\,w_\ast)(w-w_\ast)g(\tau,\,w_\ast)\,dw_\ast,
	\label{eq:FP.gen-B}
\end{equation}
which is obtained in the quasi-invariant regime starting from the binary interactions~\eqref{eq:binary} with a symmetric but not necessarily constant compromise function $P$. In particular, the following implicit representation of $g_\infty$ can be given:
\begin{equation}
	g_\infty(w)=\frac{C}{D^2(w)}\exp{\left(-\frac{2}{\lambda}\int\frac{\cB[g_\infty](w)}{D^2(w)}\,dw\right)},
	\label{eq:ginf.gen}
\end{equation}
where $C>0$ is a normalisation constant and the integral on the right-hand side denotes any antiderivative of the function $w\mapsto\cB[g_\infty](w)/D^2(w)$. For instance, if $P$ is the function~\eqref{eq:P_nonsymm.gen} with
$$ r=p, \qquad 0\leq q\leq 1, \qquad \abs{p}\leq\frac{1}{2}\min\{q,\,1-q\}, $$
so that $P$ is symmetric and $P(w,\,w_\ast)\in [0,\,1]$ for all $(w,\,w_\ast)\in [-1,\,1]^2$, then from~\eqref{eq:ginf.gen} we find the semi-explicit expression
$$ g_\infty(w)=Ce^{\frac{2p}{\lambda}w}
	{(1+w)}^{\frac{q(1+m)-p\left(1-M_{2,\infty}\right)}{\lambda}-1}(1-w)^{\frac{q(1-m)+p\left(1-M_{2,\infty}\right)}{\lambda}-1}, $$
which brings the calculation of $M_{2,\infty}$ back to the numerical solution of the non-linear system of equations
\begin{equation*}
	\begin{cases}
		\int_{-1}^1g_\infty(w)\,dw=1 \\[2mm]
		\int_{-1}^1w^2g_\infty(w)\,dw=M_{2,\infty}
	\end{cases}
\end{equation*}
parametrised by $m$.

In general, however, we observe that the type of dependence of $M_{2,\infty}$ on $m$ valid for $P\equiv 1$, cf.~\eqref{eq:M2}, is somewhat paradigmatic. In fact, let us consider~\eqref{eq:boltzmann.g} in the quasi-invariant limit $\epsilon\to 0^+$ for binary interactions~\eqref{eq:binary} with a symmetric $P$. Fixing $D(w)=\sqrt{1-w^2}$, we obtain the following equation for $M_2$:
$$ \frac{dM_2}{d\tau}=2\gamma\int_{-1}^1\int_{-1}^1w(w_\ast-w)P(w,\,w_\ast)g(\tau,\,w)g(\tau,\,w_\ast)\,dw\,dw_\ast+\sigma^2(1-M_2). $$
Set $a:=\inf_{w,w_\ast\in[-1,\,1]}P(w,\,w_\ast)$, $0\leq a\leq 1$. Then
$$ -(2+\lambda)M_2+2am^2+\lambda\leq\frac{1}{\gamma}\cdot\frac{dM_2}{d\tau}\leq
	-(2a+\lambda)M_2+2m^2+\lambda, $$
which produces asymptotically
$$ \frac{2am^2+\lambda}{2+\lambda}\leq M_{2,\infty}\leq\frac{2m^2+\lambda}{2a+\lambda}. $$
This suggests that a perhaps rough but possibly useful approximation of $M_{2,\infty}$, to be used in~\eqref{eq:SO.gen}, is the average of these lower and upper bounds, i.e.:
$$ M_{2,\infty}\approx\left(\frac{a}{\lambda+2}+\frac{1}{2a+\lambda}\right)m^2+\frac{\lambda(1+a+\lambda)}{(\lambda+2)(2a+\lambda)}, $$
which for $a=0$, like e.g. in case~\eqref{eq:P_BC}, yields $M_{2,\infty}\approx\frac{m^2}{\lambda}+\frac{\lambda+1}{\lambda+2}$.

\section{Numerical tests}
\label{sect:num}
In this section we exemplify, by means of several numerical tests, the main features of the formation of preferences at the kinetic and hydrodynamic scales as described by the models presented in the previous sections.

The numerical approach is essential, in particular, to investigate the cases in which the compromise function $P$ does not allow for an explicit computation of the asymptotic opinion distribution $g_\infty$. Therefore, first we will briefly review \textit{Structure Preserving} (SP) numerical methods, which are able to capture the large time solution to possibly non-local Fokker-Planck equations with non-constant diffusion, such as those introduced in Sections~\ref{sect:P.symm} and~\ref{sect:general.hydro}, see~\cite{pareschi2018BOOKCH,pareschi2018JSC}. Next, we will compare the large time distributions so computed with those obtained from the numerical solution of the original Boltzmann-type equation~\eqref{eq:boltzmann.g} in the quasi-invariant limit ($\epsilon\ll 1$) by means of classical \textit{Monte Carlo} (MC) methods for kinetic equations~\cite{dimarco2014AN,pareschi2013BOOK}. After validating in this way the accuracy of the numerical solver for the sole opinion dynamics (homogeneous kinetic model), we will investigate the inhomogeneous kinetic model~\eqref{eq:boltzmann.inhomog.f} as well as the hydrodynamic models derived therefrom.

\subsection{MC and SP methods for the homogeneous kinetic equation~\texorpdfstring{\eqref{eq:boltzmann.g}}{}}
\label{sect:MC-SP}
We begin we rewriting the Boltzmann-type equation~\eqref{eq:boltzmann.g} in strong form:
\begin{equation}
	\partial_\tau g=\dfrac{1}{\epsilon}\left(Q^+(g,\,g)-g\right),
	\label{eq:boltzmann_strong}
\end{equation}
where $Q^+$ is the gain part of the kinetic collision operator:
$$ Q^+(g,\,g)(\tau,\,w):=\left\langle\int_{-1}^1\frac{1}{\pr{J}}g(\tau,\,\pr{w})g(\tau,\,\pr{w}_\ast)\,dw_\ast\right\rangle. $$
Here, $(\pr{w},\,\pr{w}_\ast)$ are the pre-interaction opinions generating the post-interaction opinions $(w,\,w_\ast)$ according to the binary interaction rule~\eqref{eq:binary} and $\pr{J}$ is the Jacobian of the transformation from the former to the latter.

To compute the solution of~\eqref{eq:boltzmann_strong}, we adopt a direct MC scheme based on the Nanbu algorithm for Maxwellian molecules~ \cite{pareschi2013BOOK}. We introduce a uniform time grid $\tau^n:=n\Delta\tau$ with fixed step $\Delta\tau>0$ and we denote $g^n(w):=g(\tau^n,\,w)$. A forward discretisation of~\eqref{eq:boltzmann_strong} on such a mesh reads then
\begin{equation}
	g^{n+1}=\left(1-\frac{\Delta\tau}{\epsilon}\right)g^n+\frac{\Delta\tau}{\epsilon}Q^+(g^n,\,g^n).
	\label{eq:MCgeneral}
\end{equation}
From~\eqref{eq:boltzmann_strong}, owing to mass conservation, we see that $\int_{-1}^1Q^+(g,\,g)(\tau,\,w)\,dw=1$ for all $\tau>0$ if $\int_{-1}^1g(0,\,w)\,dw=1$, therefore $Q^+(g,\,g)(\tau,\,\cdot)$ can be regarded as a probability density function at all times. From~\eqref{eq:MCgeneral}, under the restriction $\Delta{\tau}\leq\epsilon$, we obtain therefore that $g^{n+1}$ is a convex combination of two probability density functions and is therefore in turn a probability density function. The probabilistic interpretation of~\eqref{eq:MCgeneral} is clear: with probability $\frac{\Delta{\tau}}{\epsilon}$ any two particles interact during the time step $\Delta{\tau}$; with complementary probability $1-\frac{\Delta{\tau}}{\epsilon}$ they do not. This is the basis on which to ground an MC-type numerical method for the approximate solution of~\eqref{eq:boltzmann_strong}.

However, it is in general numerically demanding to obtain from~\eqref{eq:MCgeneral} an accurate reconstruction of the asymptotic distribution $g_\infty$. To obviate this difficulty, one can take advantage of the fact that, for $\epsilon$ sufficiently small, the large time trend of~\eqref{eq:boltzmann_strong} is well approximated by the Fokker-Planck equation~\eqref{eq:FP.gen}. In~\cite{pareschi2018JSC}, an SP numerical scheme has been specifically designed to capture the large time behaviour of the solution to~\eqref{eq:FP.gen} with arbitrary accuracy and no restriction on the $w$-mesh size. Moreover, in the transient regime that scheme is second order accurate, preserves the non-negativity of the solution and is entropic for specific problems with gradient flow structure. See also~\cite{dimarco2018BOOKCH,herty2018SIAP,pareschi2018BOOKCH} for further applications.

To derive SP schemes for the Fokker-Planck equation~\eqref{eq:FP.gen}, we rewrite the latter in flux form:
\begin{equation}
	\partial_\tau g=\cF[g],
	\label{eq:FP.flux}
\end{equation}
where the flux is
$$ \cF[g](\tau,\,w):=\cC[g](\tau,\,w)g(\tau,\,w)+\frac{\sigma^2}{2}D^2(w)\partial_wg(\tau,w) $$
and
$$ \cC[g](\tau,\,w):=\gamma\int_{-1}^1P(w,\,w_\ast)(w-w_\ast)g(\tau,\,w_\ast)dw_\ast+\frac{\sigma^2}{2}{(D^2)}'(w). $$
Next, we introduce a uniform grid $\{w_i\}_{i=1}^N\subset [-1,\,1]$ such that $w_{i+1}-w_i=\Delta{w}>0$, we denote by $g_i(\tau)$ an approximation of the grid value $g(\tau,\,w_i)$ and we consider the conservative discretisation of~\eqref{eq:FP.flux}
\begin{equation}
	\frac{dg_i}{d\tau}=\frac{\cF_{i+1/2}-\cF_{i-1/2}}{\Delta w},
	\label{eq:FP.consdiscr}
\end{equation}
where $\cF_{i\pm1/2}$ is an approximation of $\cF$ at $w_{i\pm 1/2}:=w_i\pm\frac{\Delta{w}}{2}$. In particular, we choose a numerical flux of the form 
$$ \cF_{i+1/2}:=\tilde{\cC}_{i+1/2}\tilde{g}_{i+1/2}+\frac{\sigma^2}{2}D^2_{i+1/2}\frac{g_{i+1}-g_i}{\Delta w}, $$
with $\tilde g_{i+1/2}$ defined as a convex combination of $g_i$, $g_{i+1}$:
$$ \tilde{g}_{i+1/2}:=\left(1-\delta_{i+1/2}\right)g_{i+1}+\delta_{i+1/2}g_i. $$
The coefficient $\delta_{i+1/2}\in [0,\,1]$ has to be properly chosen. Setting in particular
\begin{equation}
	\tilde{\cC}_{i+1/2}:=\frac{\sigma^2D^2_{i+1/2}}{\Delta{w}}\left(\frac{\gamma}{\sigma^2}
		\int_{w_i}^{w_{i+1}}\frac{\cB[g](\tau,\,w)}{D^2(w)}\,dw+\log{\frac{D_{i+1}}{D_i}}\right),
	\label{eq:Ctilde}
 \end{equation}
where $\cB[g]$ is given by~\eqref{eq:FP.gen-B}, we obtain explicitly
$$ \delta_{i+1/2}:=\frac{1}{\lambda_{i+1/2}}+\frac{1}{1-\exp(\lambda_{i+1/2})} \quad \text{with} \quad
	\lambda_{i+1/2}:=\frac{2\Delta{w}\tilde{\cC}_{i+1/2}}{\sigma^2D^2_{i+1/2}}. $$

The order of this scheme for large times coincides with that of the quadrature formula employed for computing the integral contained in~\eqref{eq:Ctilde}. In particular, if a standard Gaussian quadrature rule is used then spectral accuracy is achieved in the $w$ variable. In the transient regime, instead, the scheme is always second order accurate. 

\subsubsection{Comparison of the numerical solutions for large times}
\label{sect:comp.homog}
We now compare the large time numerical solution of the Boltzmann-type equation~\eqref{eq:boltzmann_strong}, obtained by means of the MC scheme with the following specifications:
\begin{itemize}
\item $10^5$ particles;
\item quasi-invariant regime approximated by taking either $\epsilon=10^{-1}$ or $\epsilon=10^{-2}$,
\end{itemize}
with the numerical solution of the Fokker-Planck equation~\eqref{eq:FP.gen}, obtained by means of the SP scheme with the following specifications:
\begin{itemize}
\item $N=81$ grid points for the mesh $\{w_i\}_{i=1}^N\subset [-1,\,1]$, yielding a mesh step $\Delta{w}=2.5\cdot 10^{-2}$;
\item fourth order Runge-Kutta method for the time integration of~\eqref{eq:FP.consdiscr};
\item Gaussian quadrature rule, with $10$ quadrature points in each cell $[w_i,\,w_{i+1}]$, for the approximation of the integral in~\eqref{eq:Ctilde}.
\end{itemize}
We use the symmetric bounded-confidence-type compromise function $P$ given by~\eqref{eq:P_BC} with several choices of the confidence threshold $\Delta\in [0,\,2]$. At the initial time $\tau=0$, we prescribe the uniform distribution in $[-1,\,1]$, i.e.
$$ g(0,\,w)=\frac{1}{2}\chi(w\in[-1,\,1]). $$

\begin{figure}[!t]
\centering
\includegraphics[scale=0.5]{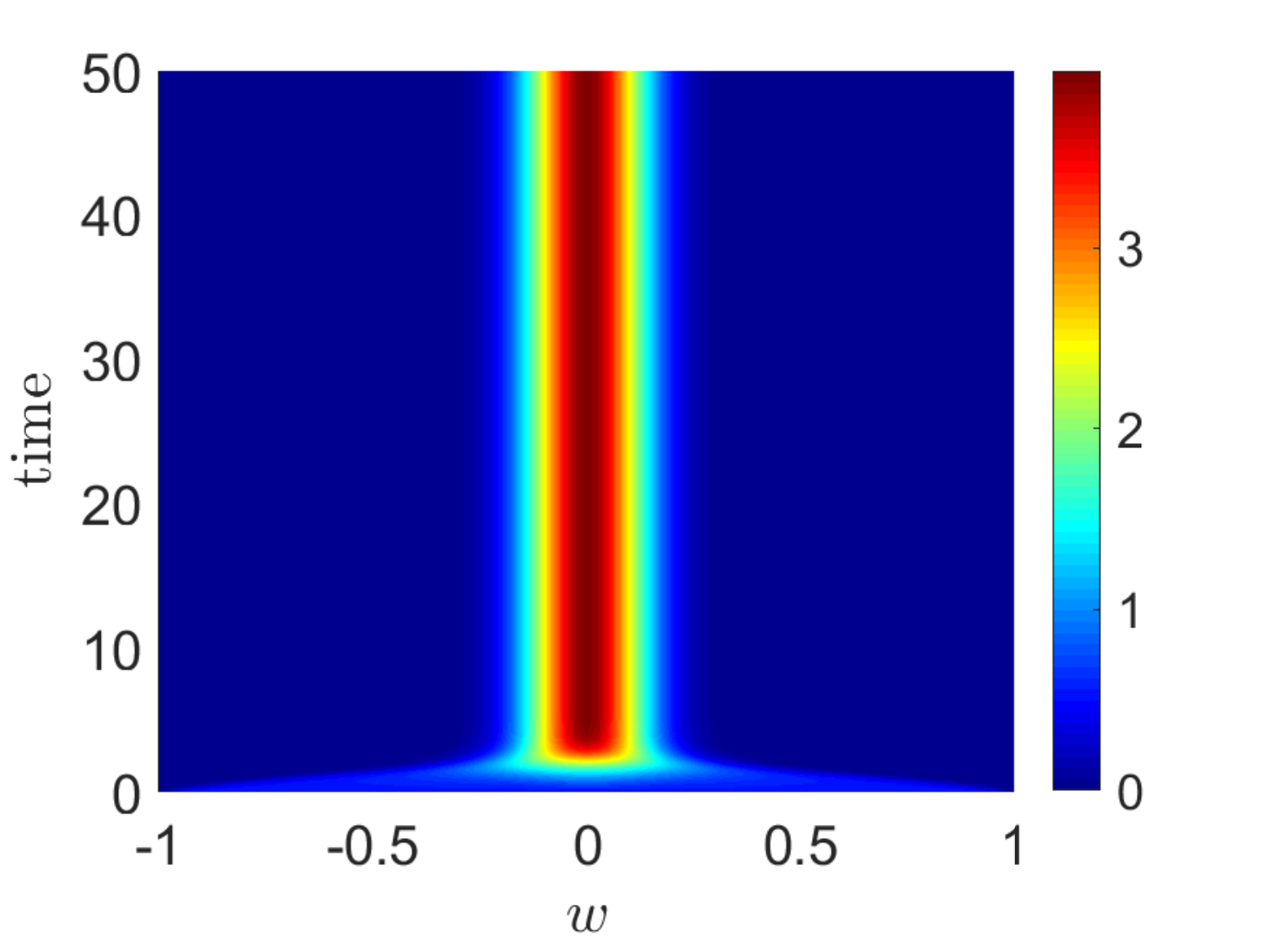}
\includegraphics[scale=0.5]{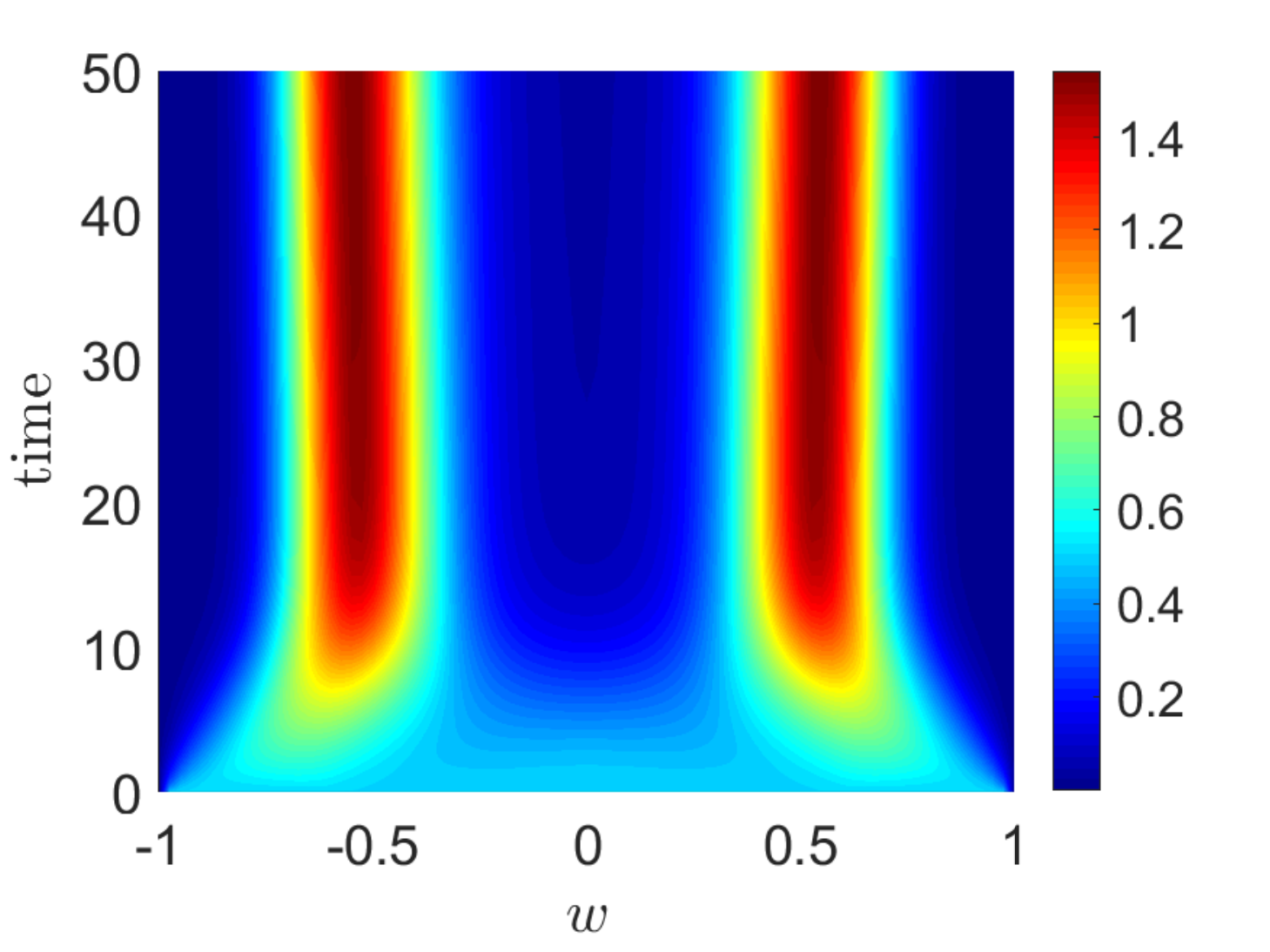} \\
\includegraphics[scale=0.5]{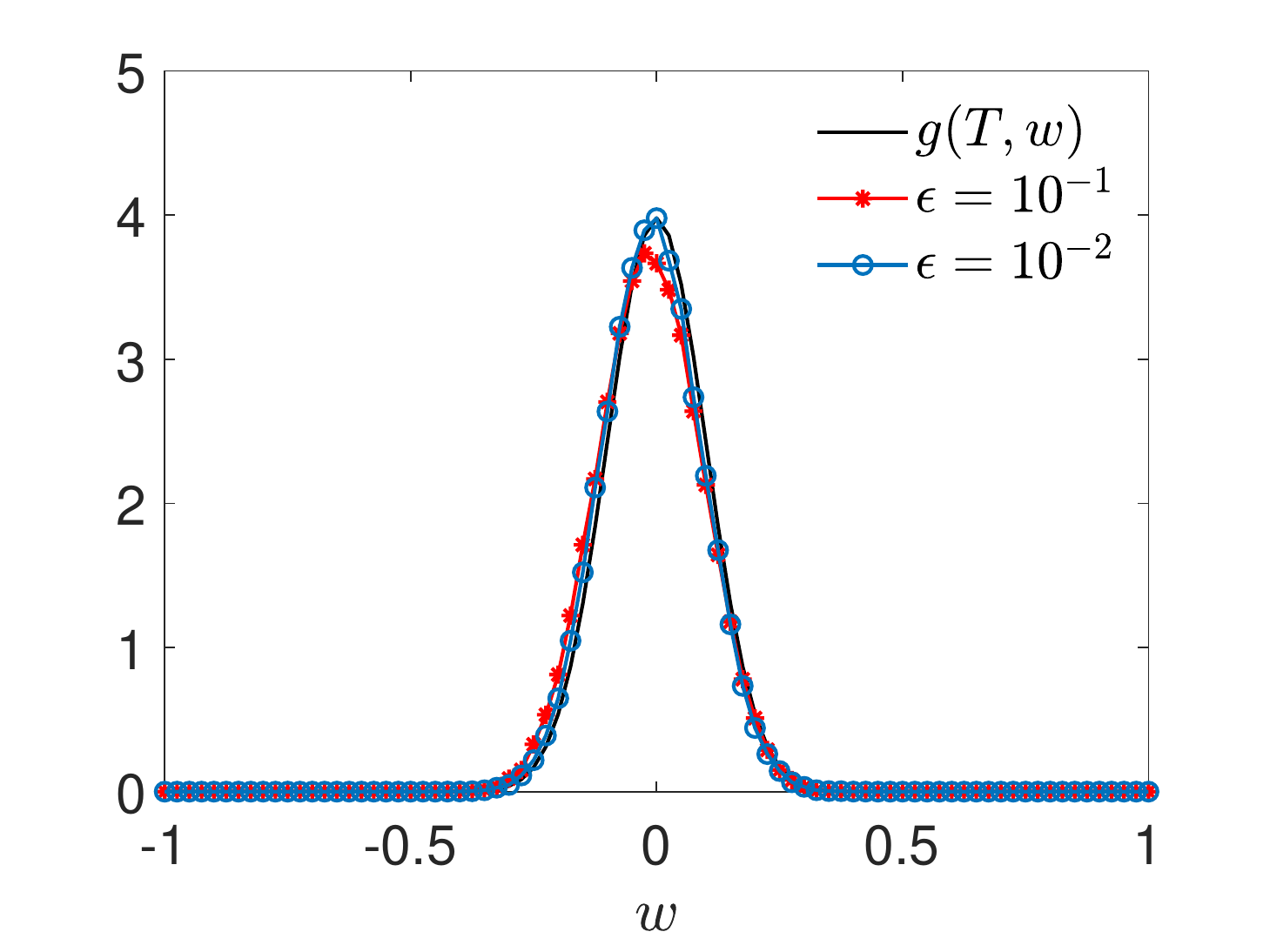}
\includegraphics[scale=0.5]{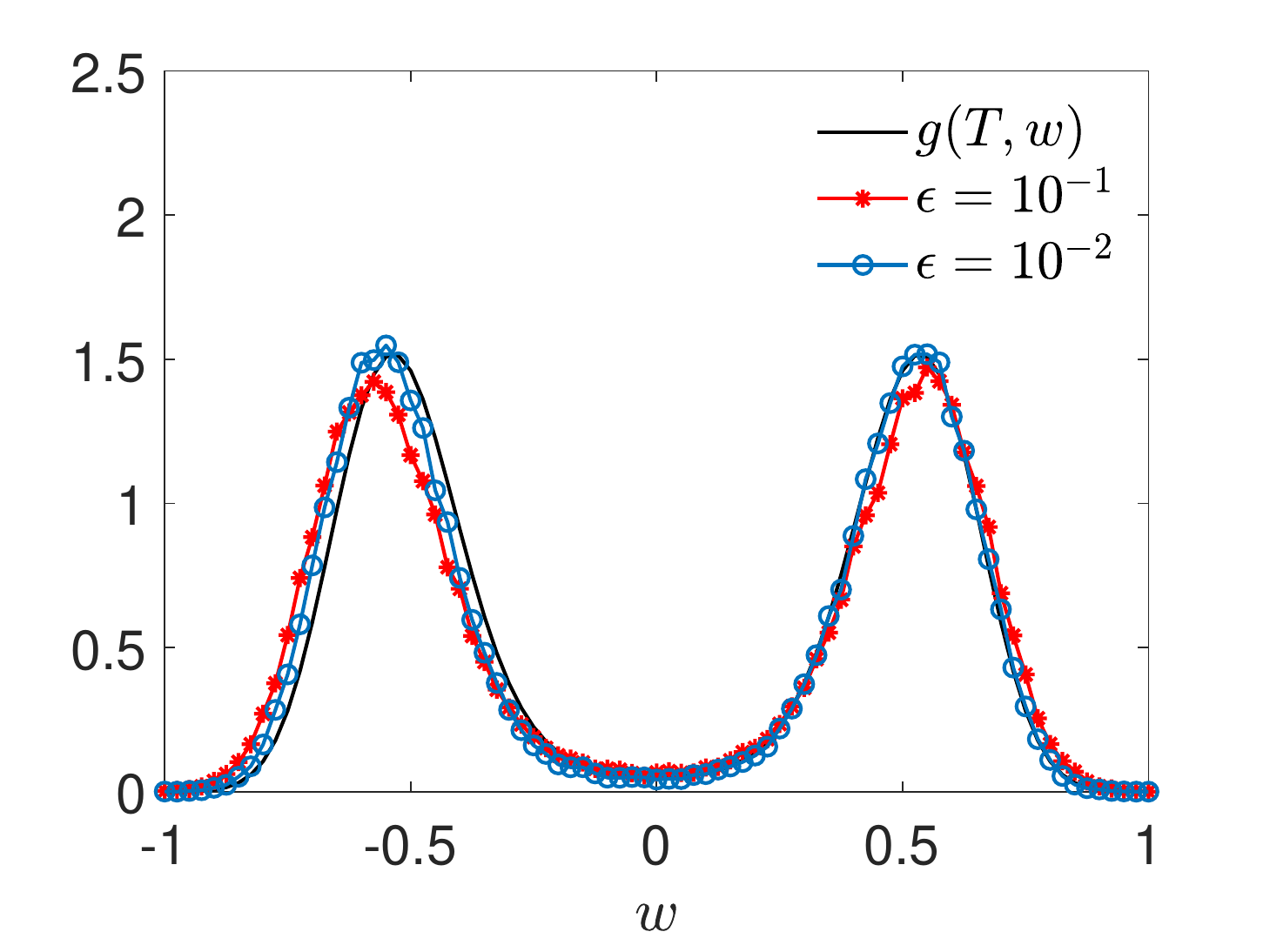}
\caption{\textbf{Top row}: contours of the distribution function $g$ computed numerically for $\tau\in (0,\,T]$, $T=50$, from the Fokker-Planck equation~\eqref{eq:FP.gen} with the SP scheme. \textbf{Bottom row}: comparison of the numerical approximations at $\tau=T$ of the large time distribution $g_\infty$ obtained with the previous SP scheme and with the MC scheme for the Boltzmann-type equation~\eqref{eq:boltzmann_strong} with two decreasing values of the parameter $\epsilon$ simulating the quasi-invariant regime. In both rows, the confidence thresholds are $\Delta=1$ (left) and $\Delta=0.4$ (right).}
\label{fig:BC1}
\end{figure}

In Figure~\ref{fig:BC1} we fix $\Delta=1$ (left panels) and $\Delta=0.4$ (right panels) and we take $\lambda=\sigma^2/\gamma=5\cdot 10^{-3}$. We observe that, as expected, the smaller $\epsilon$ the more the MC solution coincides with the SP solution of the Fokker-Planck equation for either value of the confidence threshold $\Delta$. Furthermore, the asymptotic profiles compare qualitatively well with those obtained with the deterministic microscopic model~\eqref{eq:micro_w}, cf. Figure~\ref{fig:BCmicro}(a, b), in terms of number and location of the opinion clusters.

\begin{figure}[!t]
\centering
\includegraphics[scale=0.5]{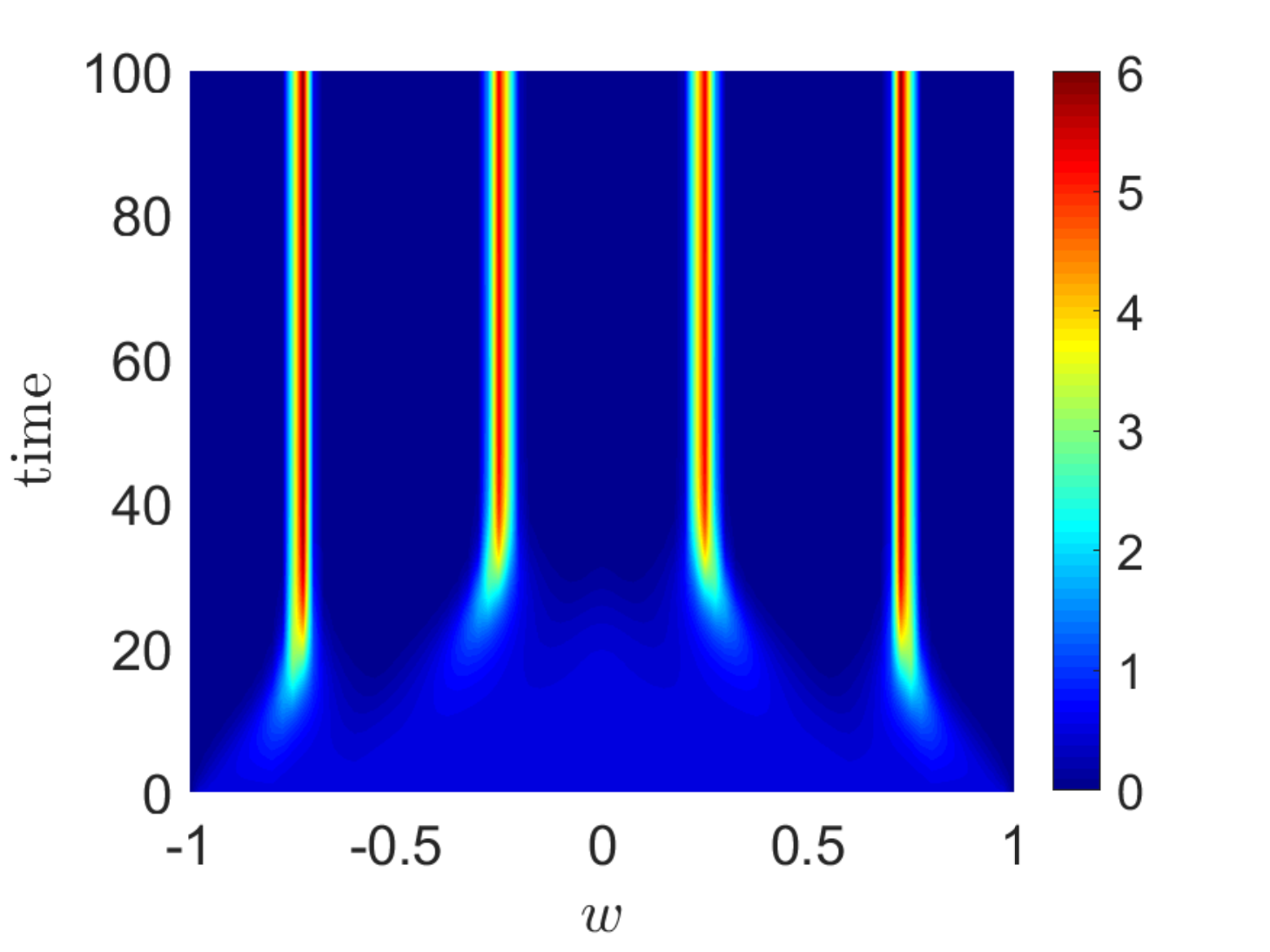} 
\includegraphics[scale=0.5]{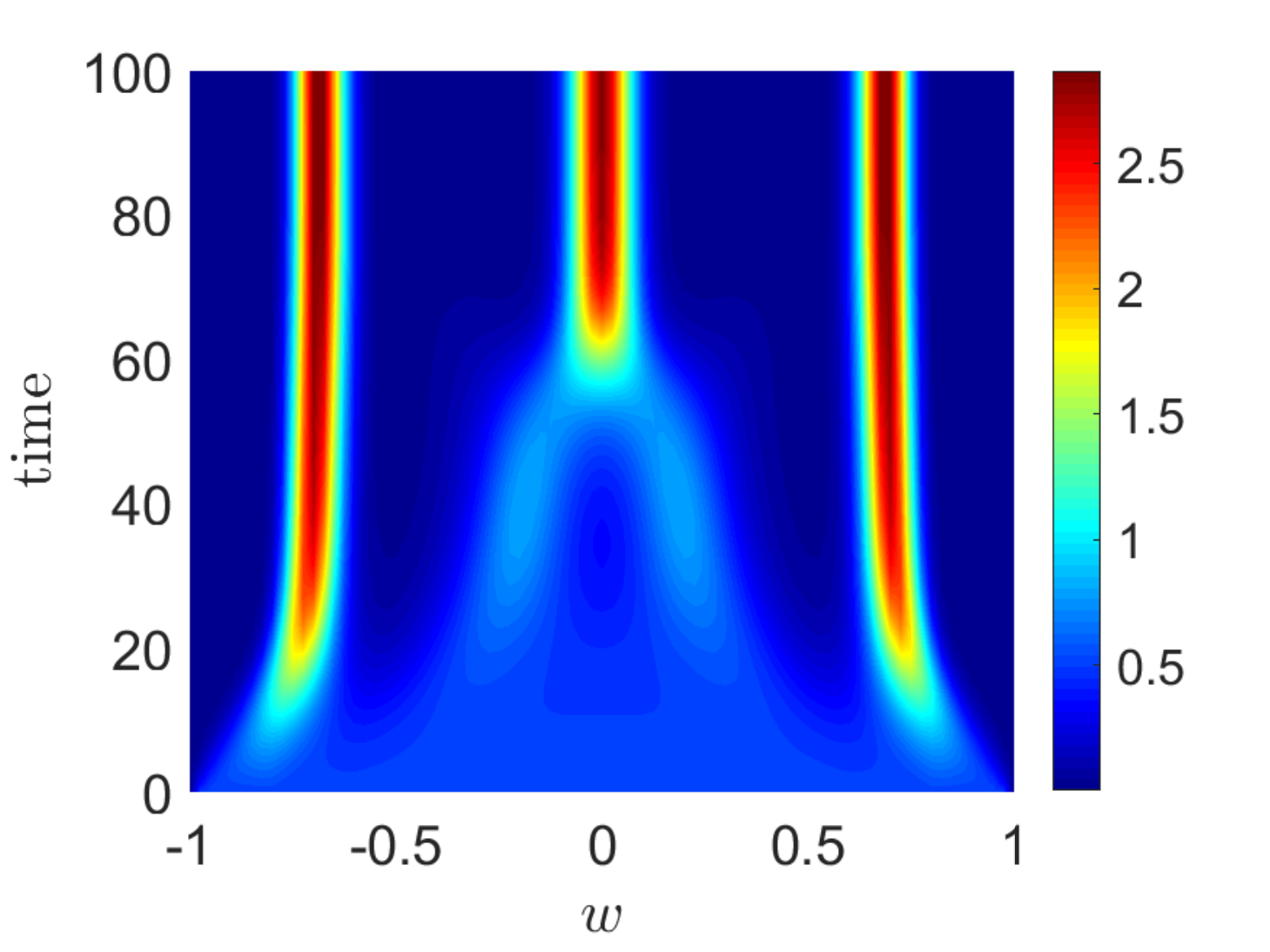} \\
\includegraphics[scale=0.5]{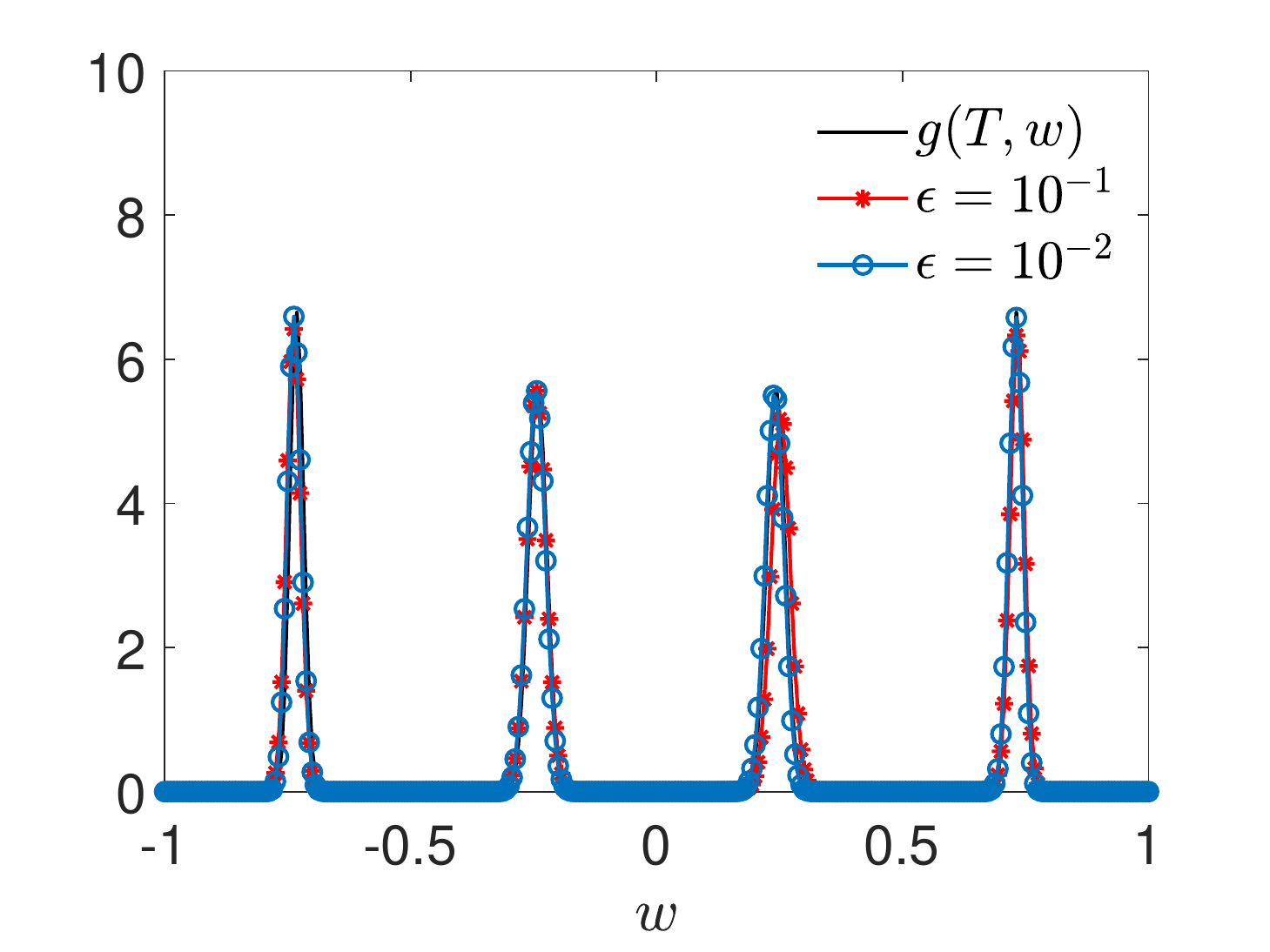}
\includegraphics[scale=0.5]{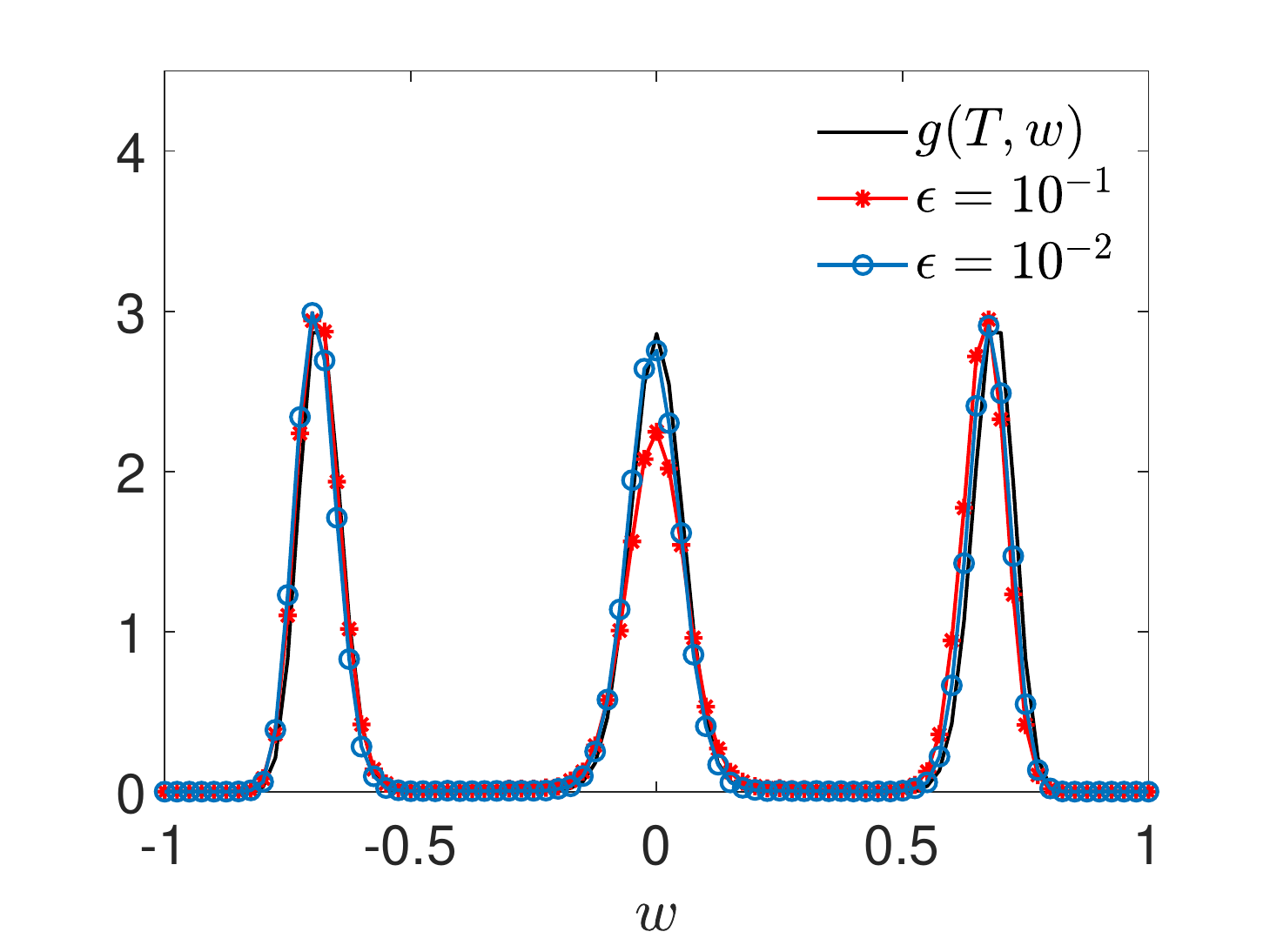}
\caption{The same as Figure~\ref{fig:BC1} but with $\Delta=0.2$. In the left panels we use $\lambda=10^{-4}$, in the right panels $\lambda=10^{-3}$. In the latter case, the asymptotic distribution features only three opinion clusters, well reproduced by both the SP Fokker-Planck solution and the MC Boltzmann solution (especially with $\epsilon=10^{-2}$), because the two central clusters merge during the transient due to a higher relevance of the self-thinking (diffusion) with respect to the tendency to compromise (transport).}
\label{fig:BC2}
\end{figure}

In Figure~\ref{fig:BC2} we repeat the same comparisons between the MC and SP numerical solutions but with $\Delta=0.2$. For $\lambda=10^{-4}$ (left panels) we recover both a transient behaviour and an asymptotic trend of the solution fully consistent with those already observed with the deterministic microscopic model~\eqref{eq:micro_w}. In particular, four opinion clusters emerge in the long run. Interestingly, for a slightly larger parameter $\lambda=10^{-3}$, indicating a higher relevance of the self-thinking (stochastic fluctuation) in the behaviour of the individuals, two opinion clusters merge, thereby giving rise to just three clusters in the long run. This aggregate phenomenon can only be observed if some microscopic randomness is duly taken into account in the model.

\subsection{Inhomogeneous kinetic equation~\texorpdfstring{\eqref{eq:boltzmann.inhomog.g}}{}}
\label{subsect:inhom}
We now pass to the inhomogeneous kinetic model, in which the formation of the preferences is driven by an interplay with the opinion dynamics studied before.

We start by outlining the procedure by which we solve the inhomogeneous Boltzmann-type equation~\eqref{eq:boltzmann.inhomog.g}. Since the Knudsen-like number $\delta$ is assumed to be small, at each time step we adopt the very same splitting procedure already discussed in Section~\ref{sect:boltzmann.inhomog}. Therefore, upon introducing a time discretisation $\tau^n:=n\Delta{\tau}$, with $\Delta{\tau}>0$ constant, we proceed as follows.
\begin{description}
\item[Interaction step.] At time $\tau=\tau^n$, we solve the interactions towards the equilibrium during half a time step:
\begin{equation}
	\begin{cases}
		\partial_\tau G(\tau,\,\xi,\,w)=\dfrac{1}{\delta}Q(G,\,G)(\tau,\,\xi,\,w), & \tau\in (\tau^{n},\,\tau^{n+1/2}] \\[2mm]
		G(\tau^n,\,\xi,\,w)=g(\tau^n,\,\xi,\,w)
	\end{cases}
	\label{eq:int.step}
\end{equation}
for all $\xi=\xi_i$ belonging to a suitable mesh $\{\xi_i\}_i\subset [-1,\,1]$. In this step, we take advantage of the MC scheme introduced in Section~\ref{sect:MC-SP}, which has proved to give asymptotic solutions comparable to those of the more accurate SP scheme, provided the parameter $\delta$ is sufficiently small. In particular, we use a sample of $10^6$ particles and we fix $\delta=10^{-2}$.

In~\eqref{eq:int.step}, $Q$ denotes the collision operator that appears on the right-hand side of~\eqref{eq:boltzmann.inhomog.g} once this equation has been written in strong form.
\item[Transport step.] Next, we take the asymptotic distribution obtained in the interaction step as the input of a pure transport towards the next time step $\tau^{n+1}$:
\begin{equation*}
	\begin{cases}
		\partial_{\tau}g(\tau,\,\xi,\,w)+(w-\alpha)\partial_{\xi}\left(\Phi(\xi)g(\tau,\,\xi,\,w)\right)=0, 
			& \tau\in (\tau^{n+1/2},\,\tau^{n+1}] \\[2mm]
		g(\tau^{n+1/2},\,\xi,\,w)=G(\tau^{n+1/2},\,\xi,\,w).
	\end{cases}
\end{equation*}
\end{description}

In the tests of this section, unless otherwise specified, we prescribe the uniform distribution in the variables $\xi$, $w$ as initial datum:
\begin{equation}
	g(0,\,\xi,\,w):=\frac{1}{4}\chi((\xi,w)\in [-1,\,1]^2)
	\label{eq:initial_inhom}
\end{equation}
we fix $\lambda=10^{-3}$ and we take the function $\Phi$ given in~\eqref{eq:Phi_3poles}.

\subsubsection{Symmetric~\texorpdfstring{$\boldsymbol{P}$}{}}
\begin{figure}[!t]
\centering
\subfigure[$\tau=1$]{\includegraphics[scale=0.33]{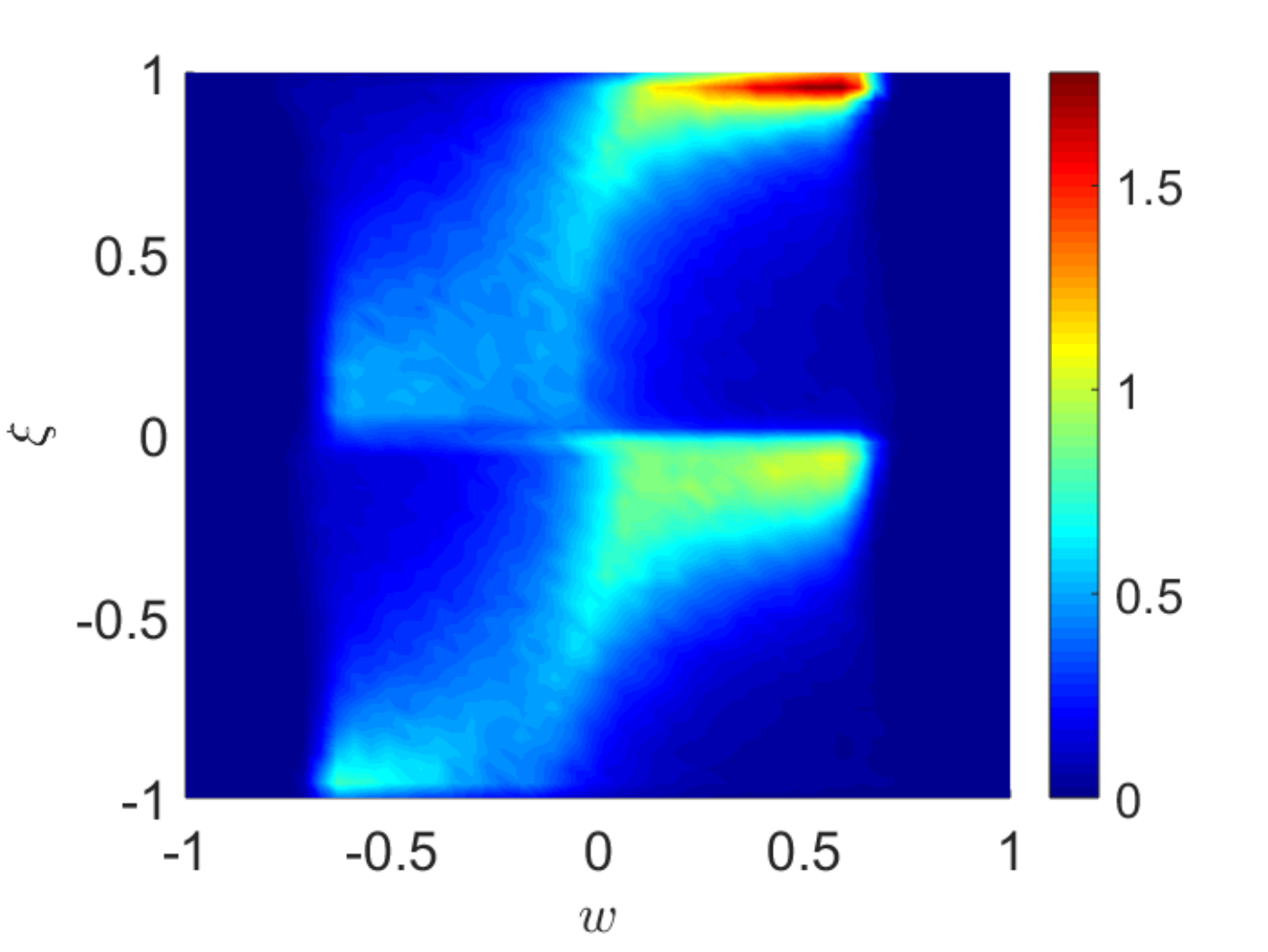}}
\subfigure[$\tau=3$]{\includegraphics[scale=0.33]{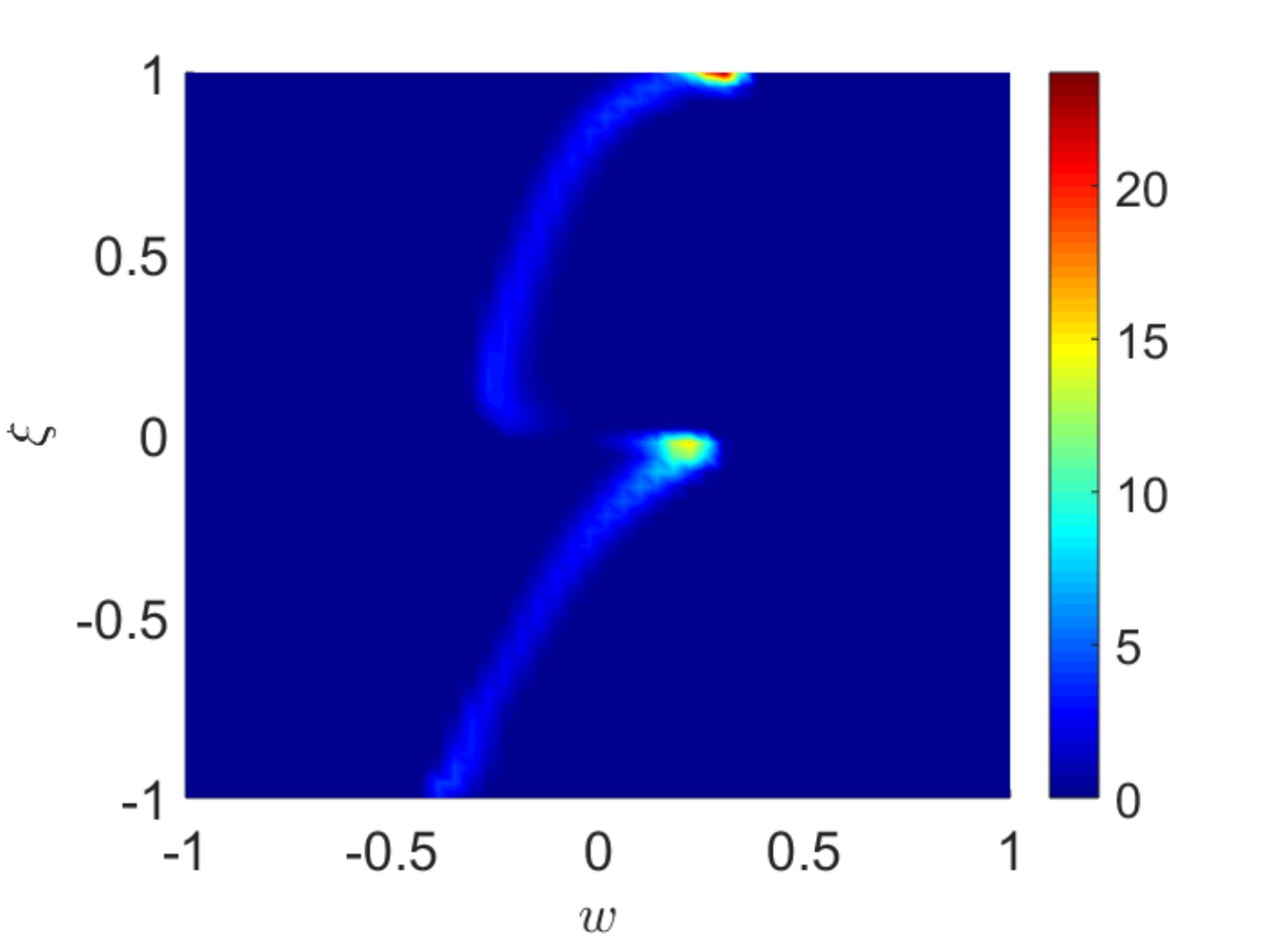}}
\subfigure[$\tau=5$]{\includegraphics[scale=0.33]{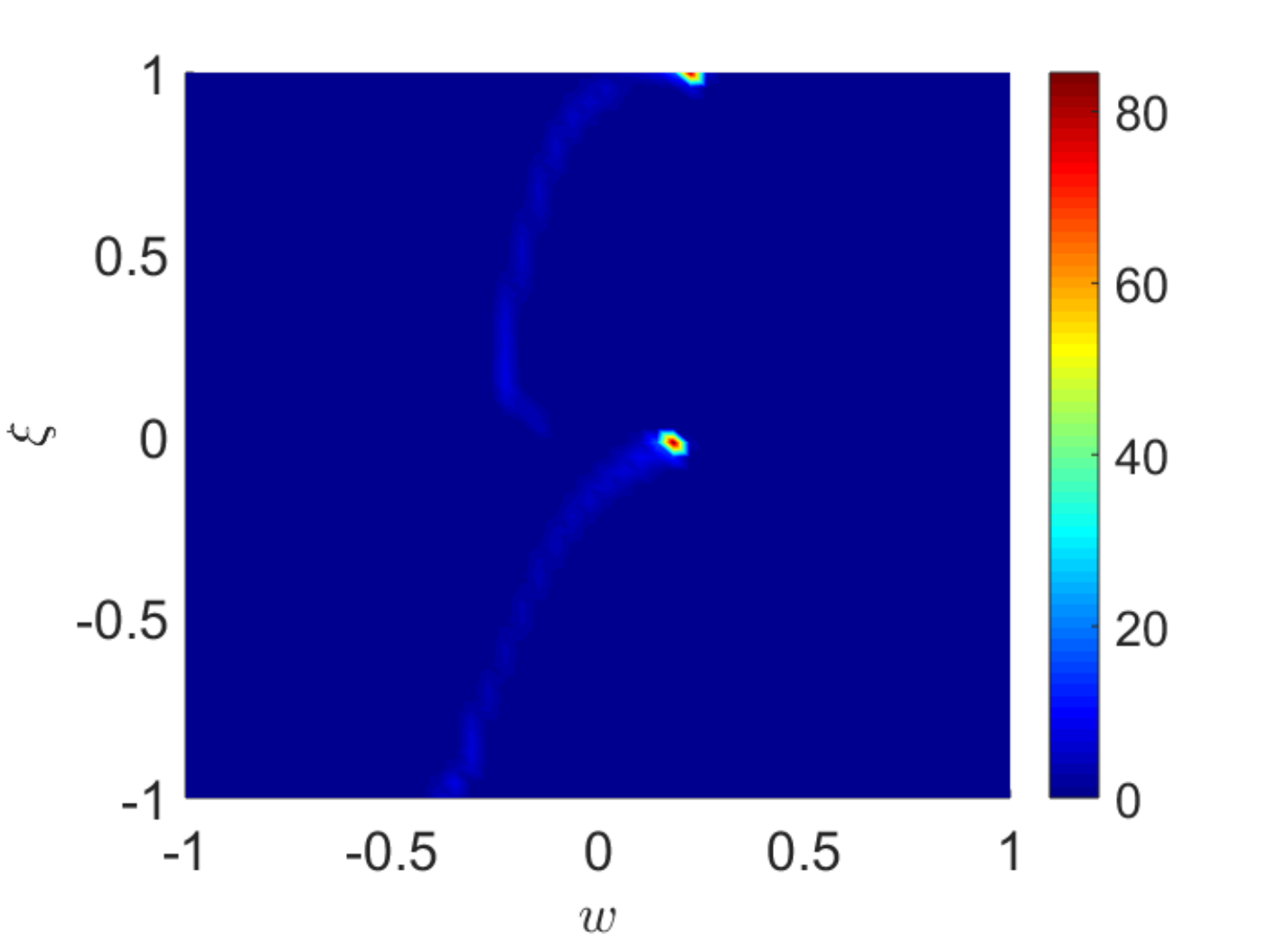}}
\caption{Contours of the inhomogeneous kinetic distribution $g(\tau,\,\xi,\,w)$ at different times with $\Delta=1$ and $\alpha=-0.3$.}
\label{fig:inhom1}
\end{figure}
\begin{figure}[!t]
\centering
\subfigure[$\tau=1$]{\includegraphics[scale=0.33]{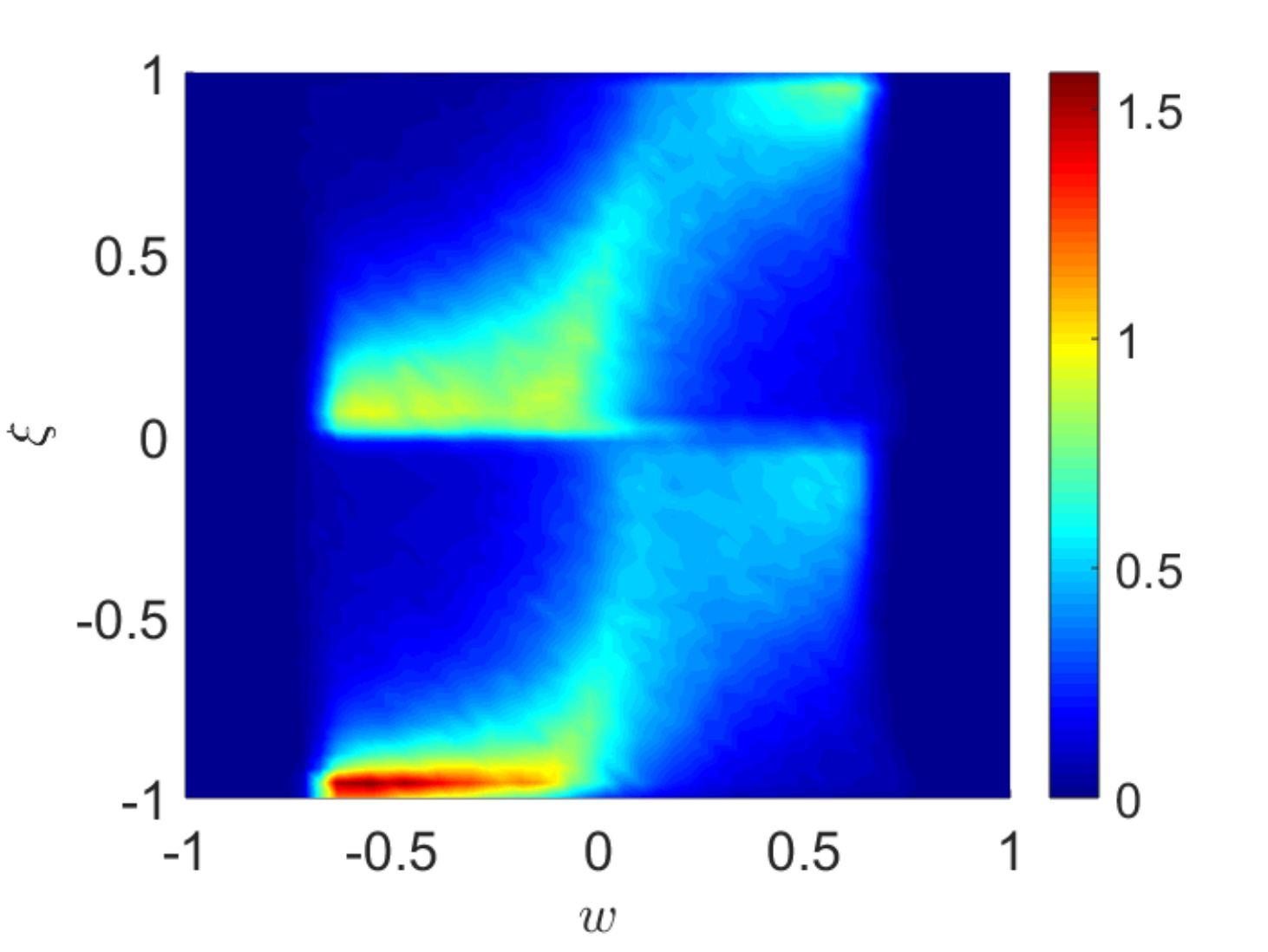}}
\subfigure[$\tau=3$]{\includegraphics[scale=0.33]{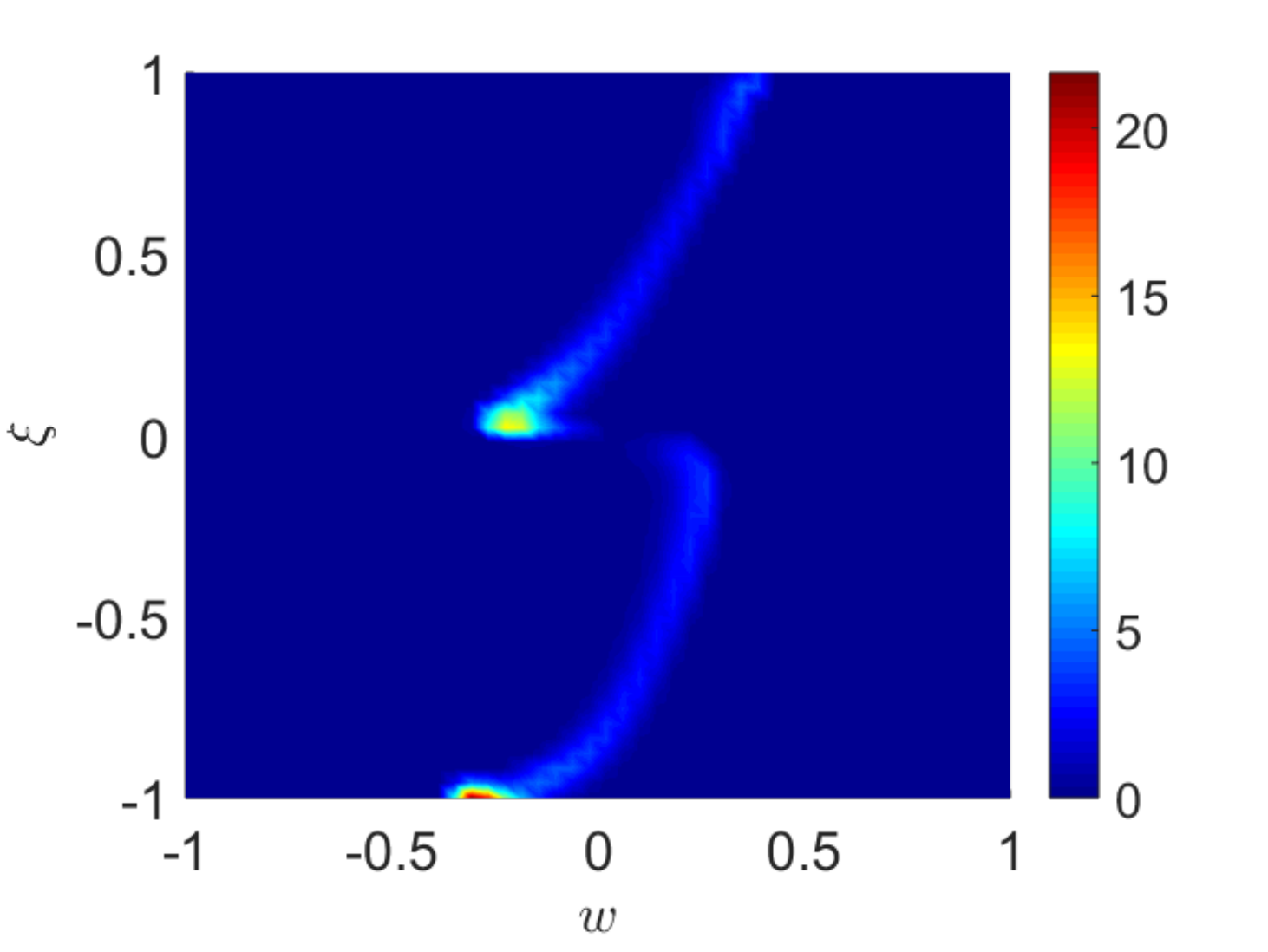}}
\subfigure[$\tau=5$]{\includegraphics[scale=0.33]{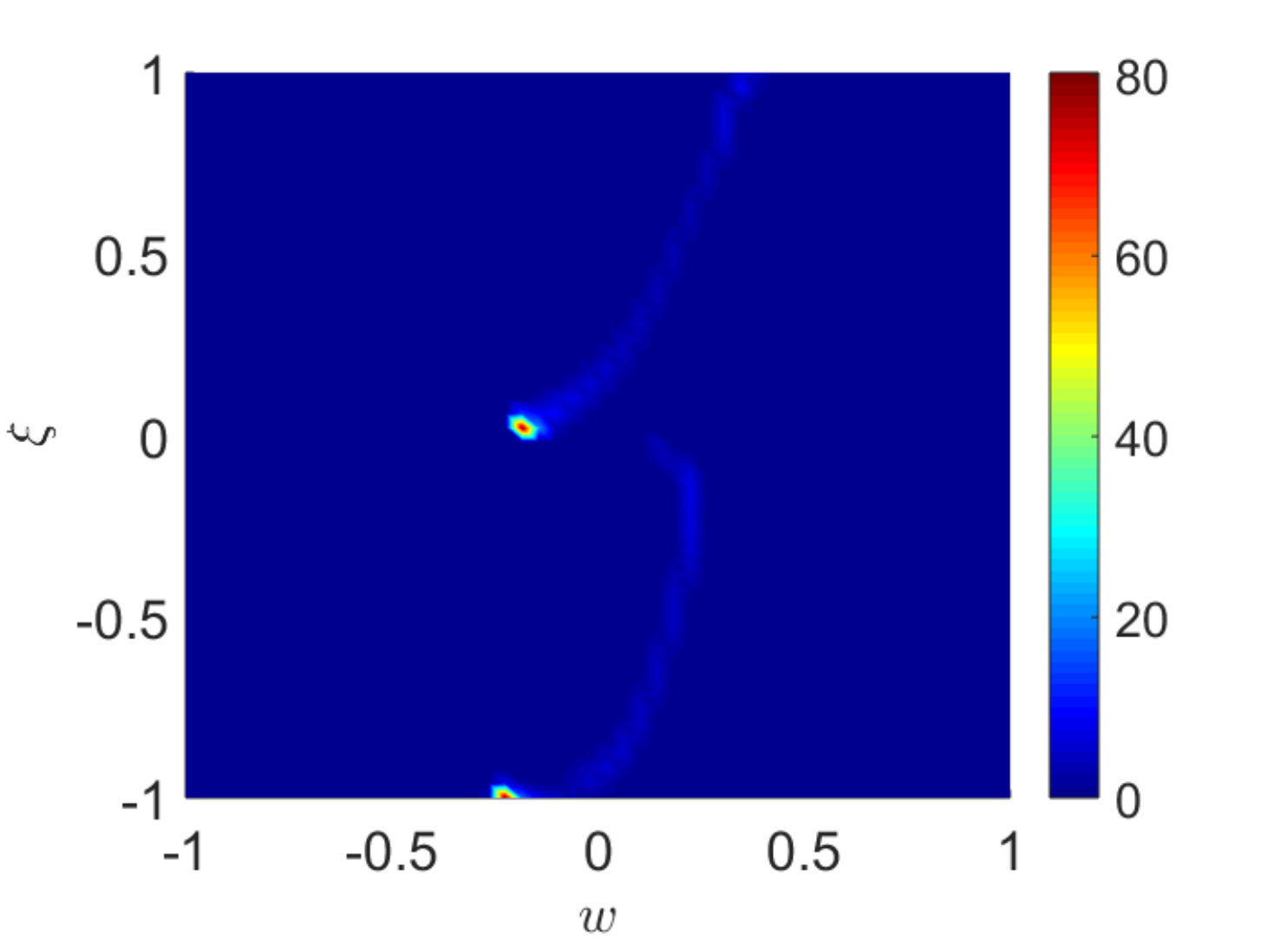}}
\caption{The same as Figure~\ref{fig:inhom1} but with $\alpha=0.3$.}
\label{fig:inhom2}
\end{figure}

First, we consider symmetric interactions described again by the bounded confidence compromise function~\eqref{eq:P_BC} with $\Delta=1$. In Figures~\ref{fig:inhom1},~\ref{fig:inhom2} we show the evolution of the inhomogeneous kinetic model for two different choices of the perceived social opinion, $\alpha=\pm 0.3$ respectively. We clearly observe that while the opinions distribute around the conserved mean opinion $m=0$, as expected, the preferences polarise in two possible ways. For $\alpha=-0.3$, cf. Figure~\ref{fig:inhom1}, polarisations emerge in $\xi = 0$ and $\xi=1$. Specifically, individuals with an initial preference in $[-1,\,0]$ tend to polarise in $\xi=0$, whereas individuals with an initial preference in $(0,\,1]$ tend to polarize in $\xi=1$. For $\alpha=0.3$, cf. Figure~\ref{fig:inhom2}, the mirror trends emerge. These polarisation patterns of the preferences are very much consistent with those observed in Section~\ref{sect:micro_xi_w} with the deterministic microscopic model~\eqref{eq:micro_xi}-\eqref{eq:micro_w}, cf. Figure~\ref{fig:xi1}.

\begin{figure}[!t]
\centering
\subfigure[$\tau=1$]{\includegraphics[scale=0.33]{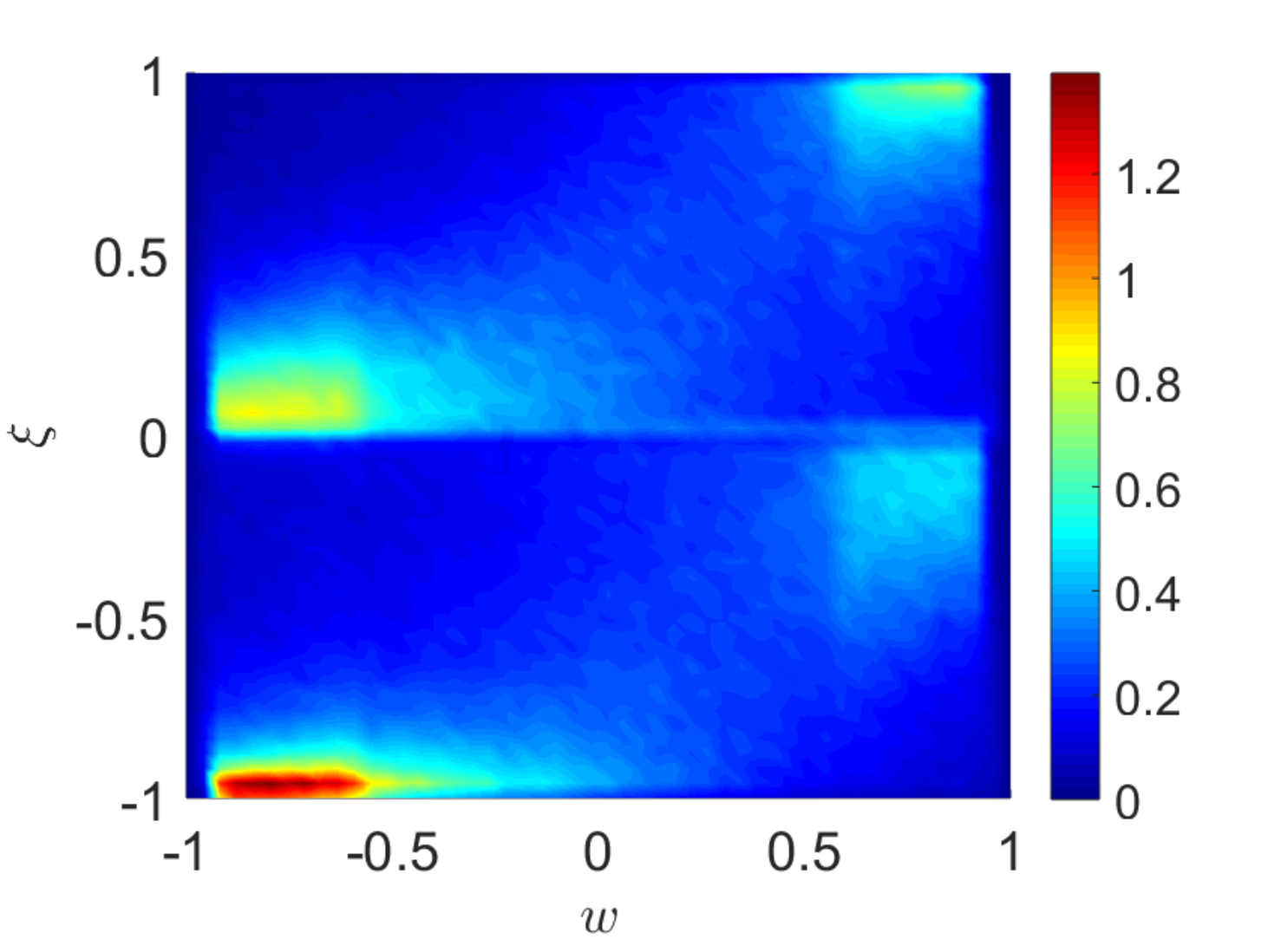}}
\subfigure[$\tau=3$]{\includegraphics[scale=0.33]{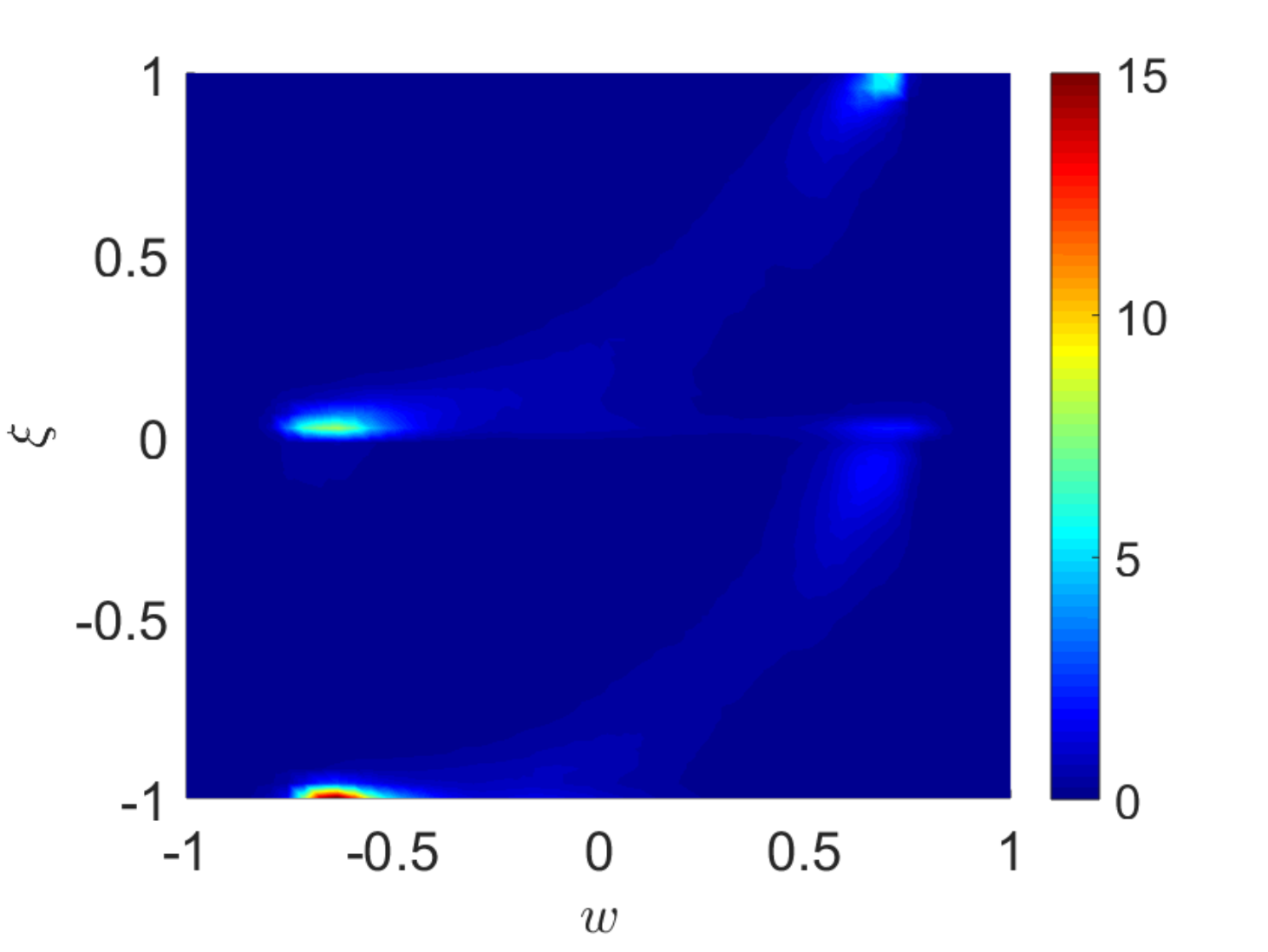}}
\subfigure[$\tau=5$]{\includegraphics[scale=0.33]{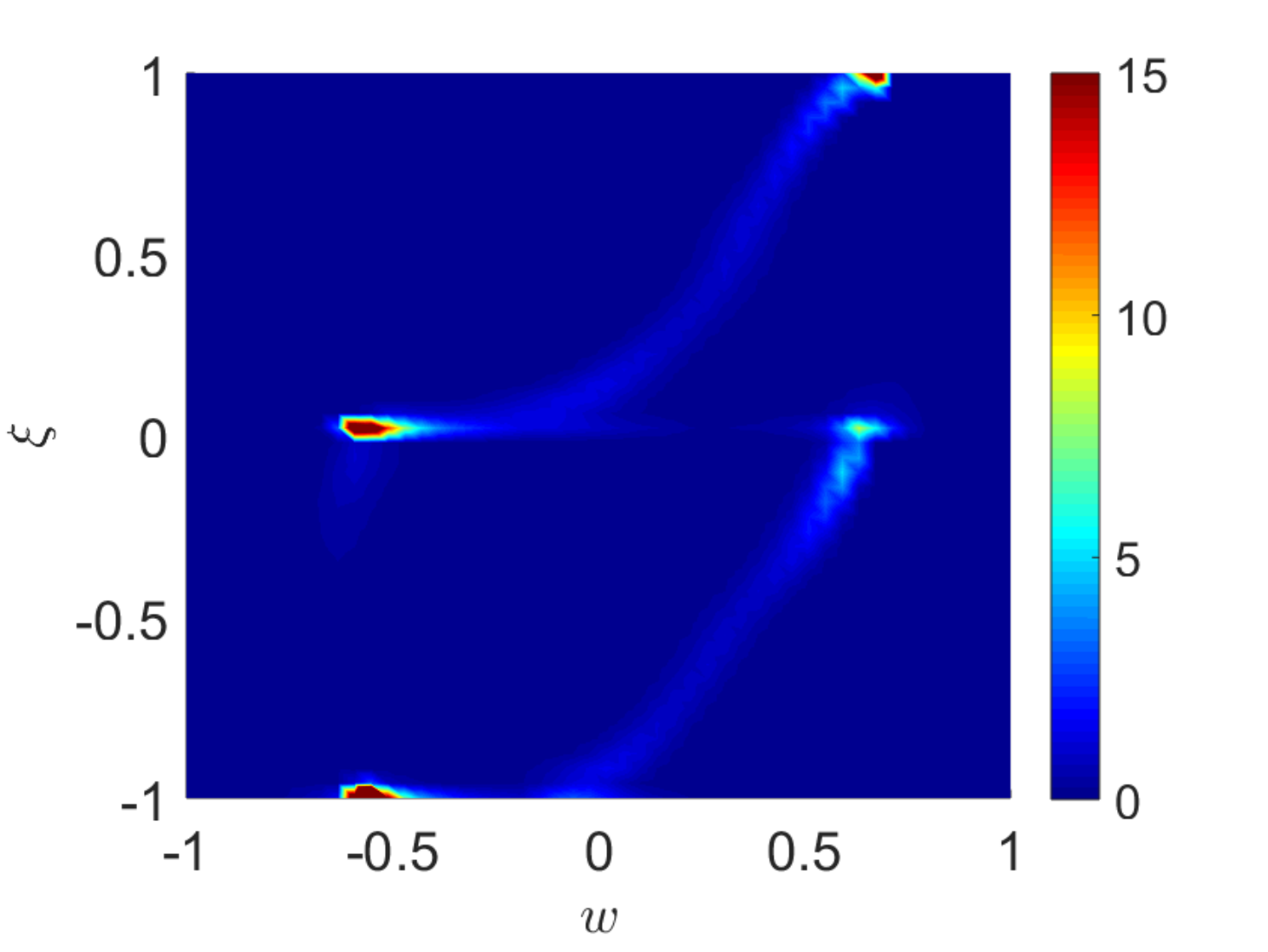}} \\
\caption{The same as Figure~\ref{fig:inhom1} but with $\Delta=0.4$ and $\alpha=0.3$.}
\label{fig:inhom3}
\end{figure}

\begin{figure}[!t]
\centering
\subfigure[$w$-marginal at $T=10$]{\includegraphics[scale=0.45]{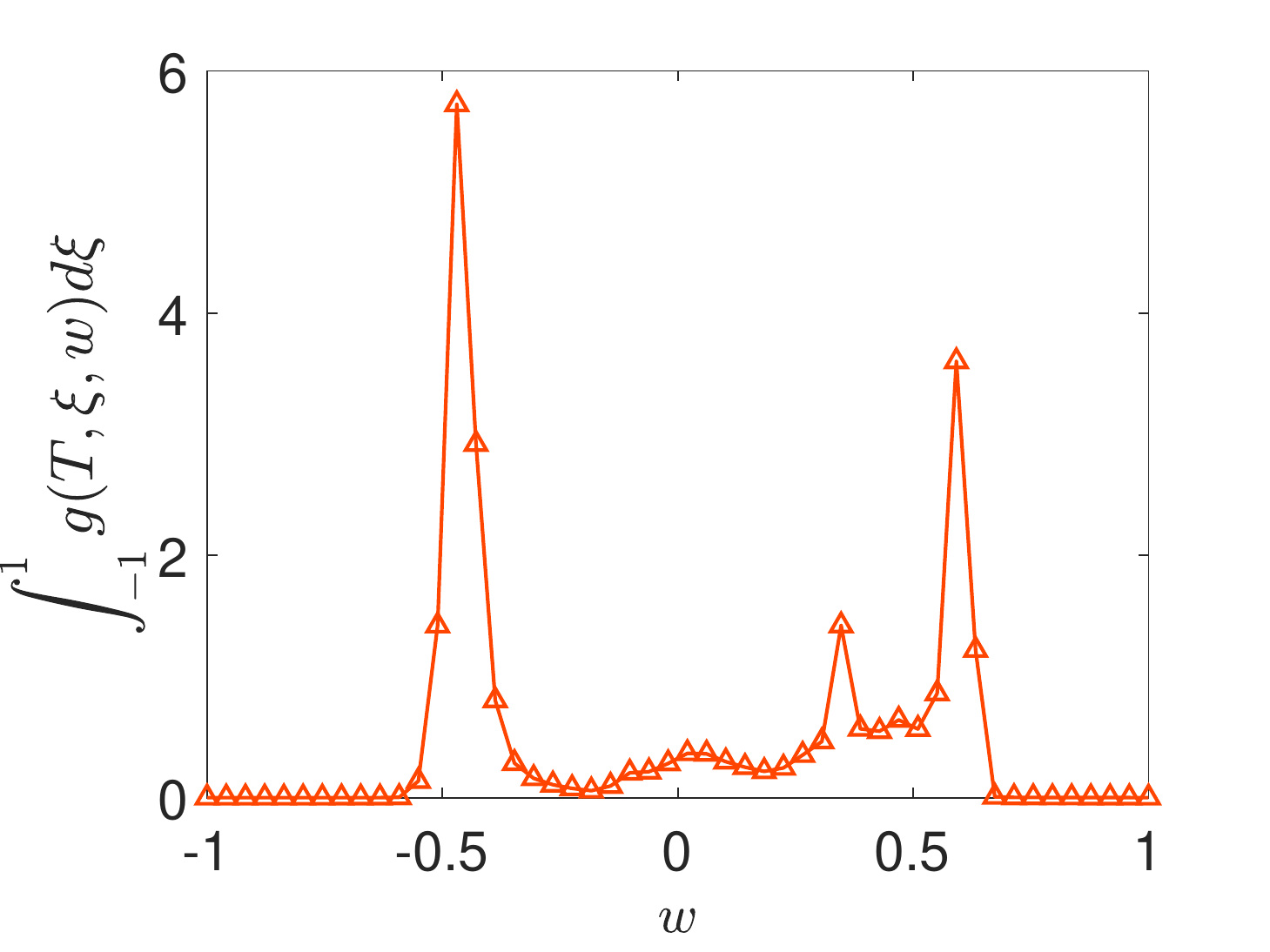}}
\subfigure[$\xi$-marginal at $T=10$]{\includegraphics[scale=0.45]{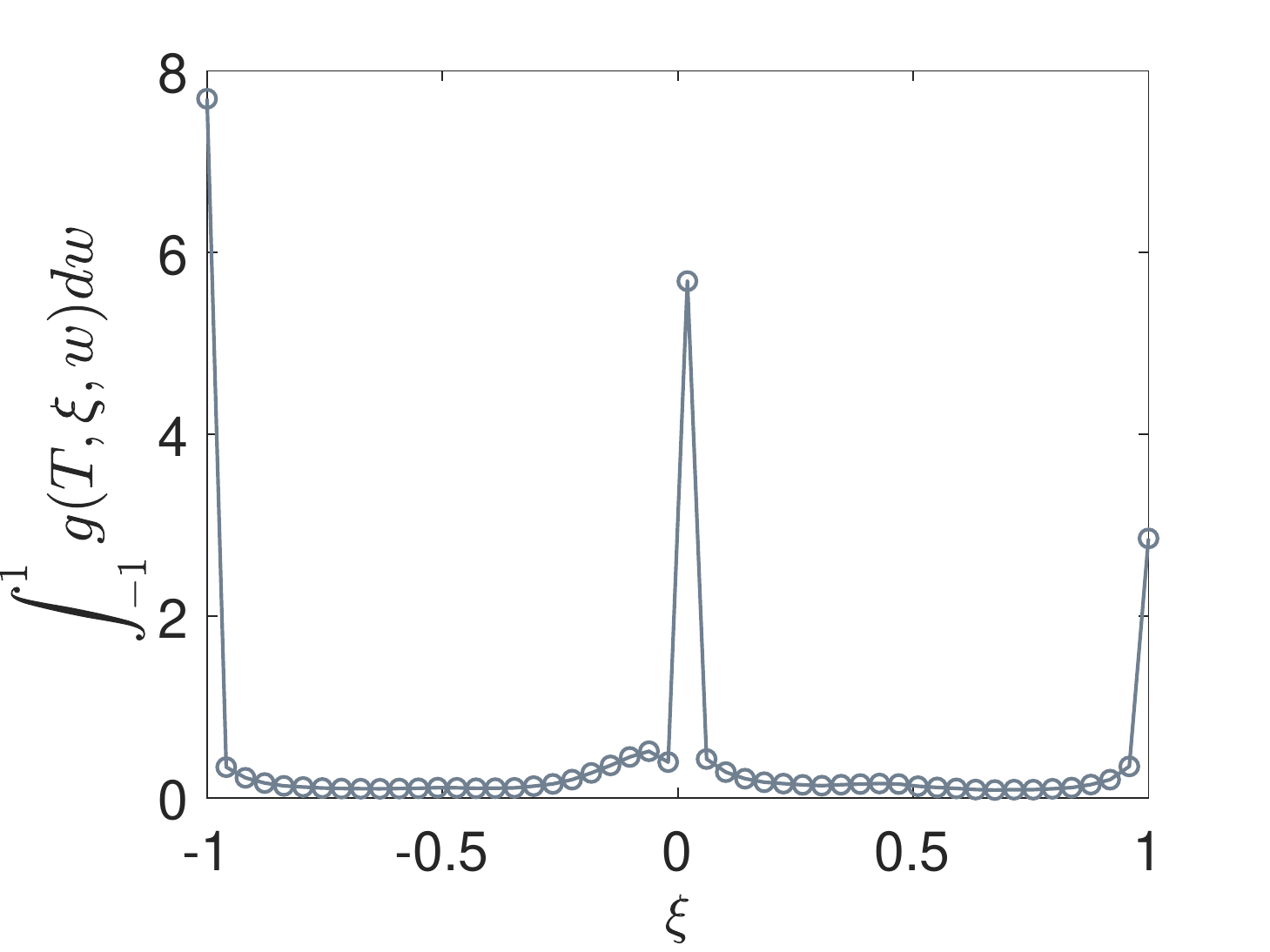}}
\caption{Marginal distributions of the opinions (a) and of the preferences (b) for the numerical test of Figure~\ref{fig:inhom3}.}
\label{fig:inhom3.marginals}
\end{figure}

Next, we consider the same symmetric compromise function $P$ as before but now we fix $\Delta=0.4$. In Figure~\ref{fig:inhom3} we depict the evolution of the inhomogeneous kinetic model for $\alpha=0.3$. As far as the opinion dynamics are concerned, we recognise that individuals tend to cluster in two well distinct positions, see Figure~\ref{fig:inhom3.marginals}(a), directly comparable with the emerging clusters shown in Figure~\ref{fig:BC1}(right) and also, up to diffusion, in Figure~\ref{fig:BCmicro}(b). Nevertheless, the social detail is now higher, because we clearly distinguish that individuals with the same asymptotic opinion may actually polarise in different preferences. More specifically, the opinion cluster near $w=-0.5$ is formed by individuals with preferences polarised in either $\xi=-1$ or $\xi=0$, while the opinion cluster near $w=0.5$ is formed by individuals with preference polarised in either $\xi=0$ or $\xi=1$, see Figure~\ref{fig:inhom3}(c). Remarkably, three polarisations of the preference emerge on the whole in the long run, see Figure~\ref{fig:inhom3.marginals}(b), because the opinions do not reach a global consensus.

Also these polarisation patterns of the preferences are consistent with those discussed in Section~\ref{sect:micro_xi_w}, indeed the deterministic microscopic model can account in principle for three preference poles. The fact that Figure~\ref{fig:xi2}(b) shows asymptotically only two of them depends essentially on the choice of the initial conditions, which in a particle model hardly allow one to observe the representative average trend in a single realisation.

The case $\alpha=-0.3$ is qualitatively analogous to the one just discussed, therefore we do not report it in detail.

\subsubsection{Non-symmetric~\texorpdfstring{$\boldsymbol{P}$}{}}
Finally, we investigate the effect of a non-symmetric compromise function $P$. As already discussed in Section~\ref{sect:P.nonsymm}, we recall that the asymmetry of $P$ can be understood as a systematic bias of the individuals, who for some reason are more prone to change opinion in a specific direction. In this numerical example, we remain in the class of the bounded confidence models and, taking inspiration from~\cite{hegselmann2002JASSS}, we consider
\begin{equation}
	P(w,\,w_\ast)=\chi(-\Delta_L\leq w_\ast-w\leq\Delta_R),
	\label{eq:P_BC.nonsymm}
\end{equation}
where $\Delta_L,\,\Delta_R\in [0,\,2]$ are two confidence thresholds.

In order to understand the effect of function~\eqref{eq:P_BC.nonsymm}, we observe that if $w\leq w_\ast$ then interactions are allowed provided $\abs{w_\ast-w}=w_\ast-w\leq\Delta_R$. Otherwise, if $w\geq w_\ast$ then interactions are allowed provided $\abs{w_\ast-w}=w-w_\ast\leq\Delta_L$. Thus, if e.g. $\Delta_R>\Delta_L$ then an individual with opinion $w$ is more incline to interact with other individuals with opinion $w_\ast\geq w$. The converse holds if instead $\Delta_R<\Delta_L$.

\begin{remark}
If $\Delta_L=\Delta_R$ then~\eqref{eq:P_BC.nonsymm} actually reduces to~\eqref{eq:P_BC} with $\Delta=\Delta_R$.
\end{remark}

\begin{figure}[!t]
\centering
\subfigure[$\tau=1$]{\includegraphics[scale=0.33]{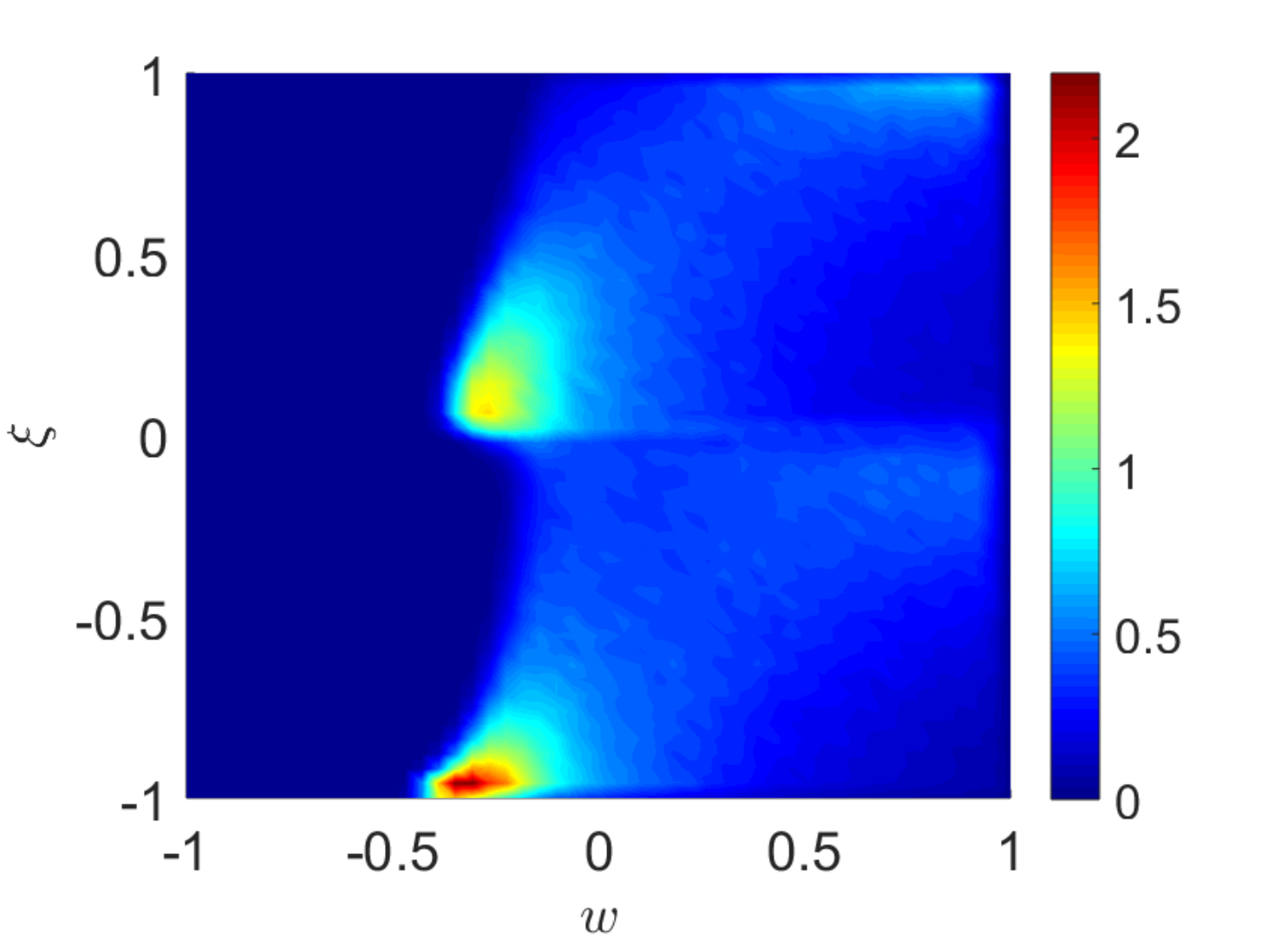}}
\subfigure[$\tau=5$]{\includegraphics[scale=0.33]{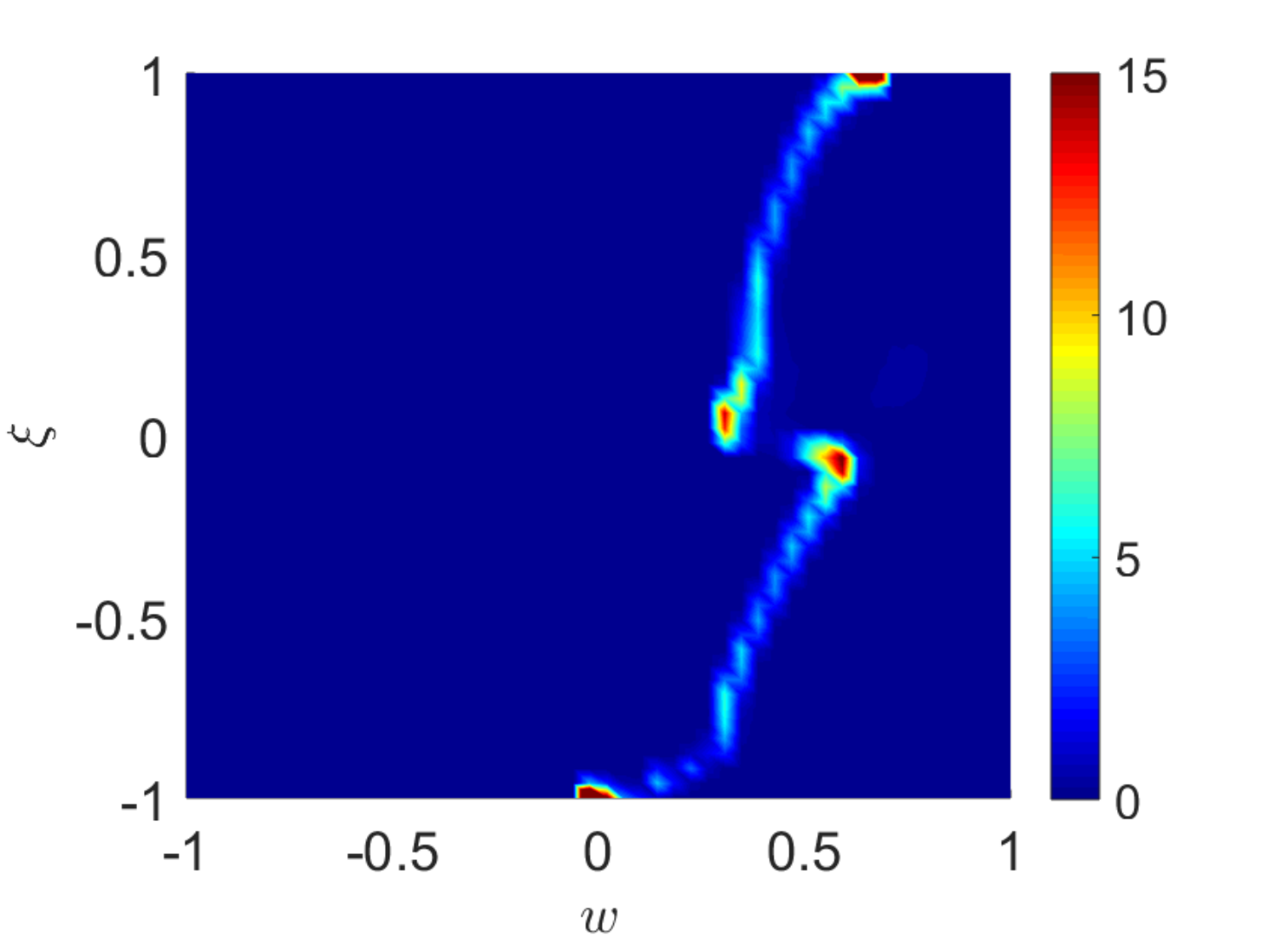}}
\subfigure[$\tau=10$]{\includegraphics[scale=0.33]{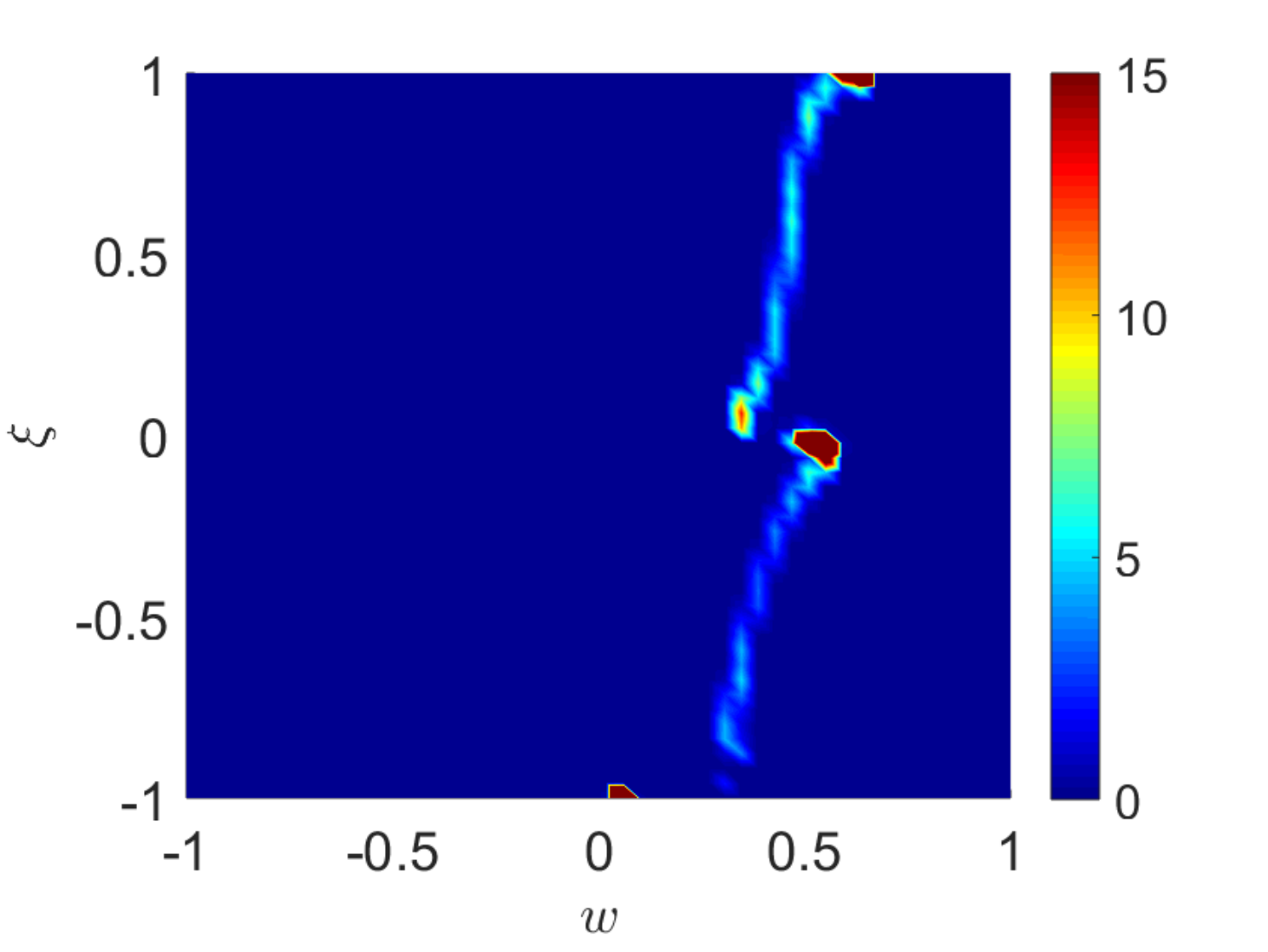}} \\
\subfigure[Mean opinion trend]{\includegraphics[scale=0.33]{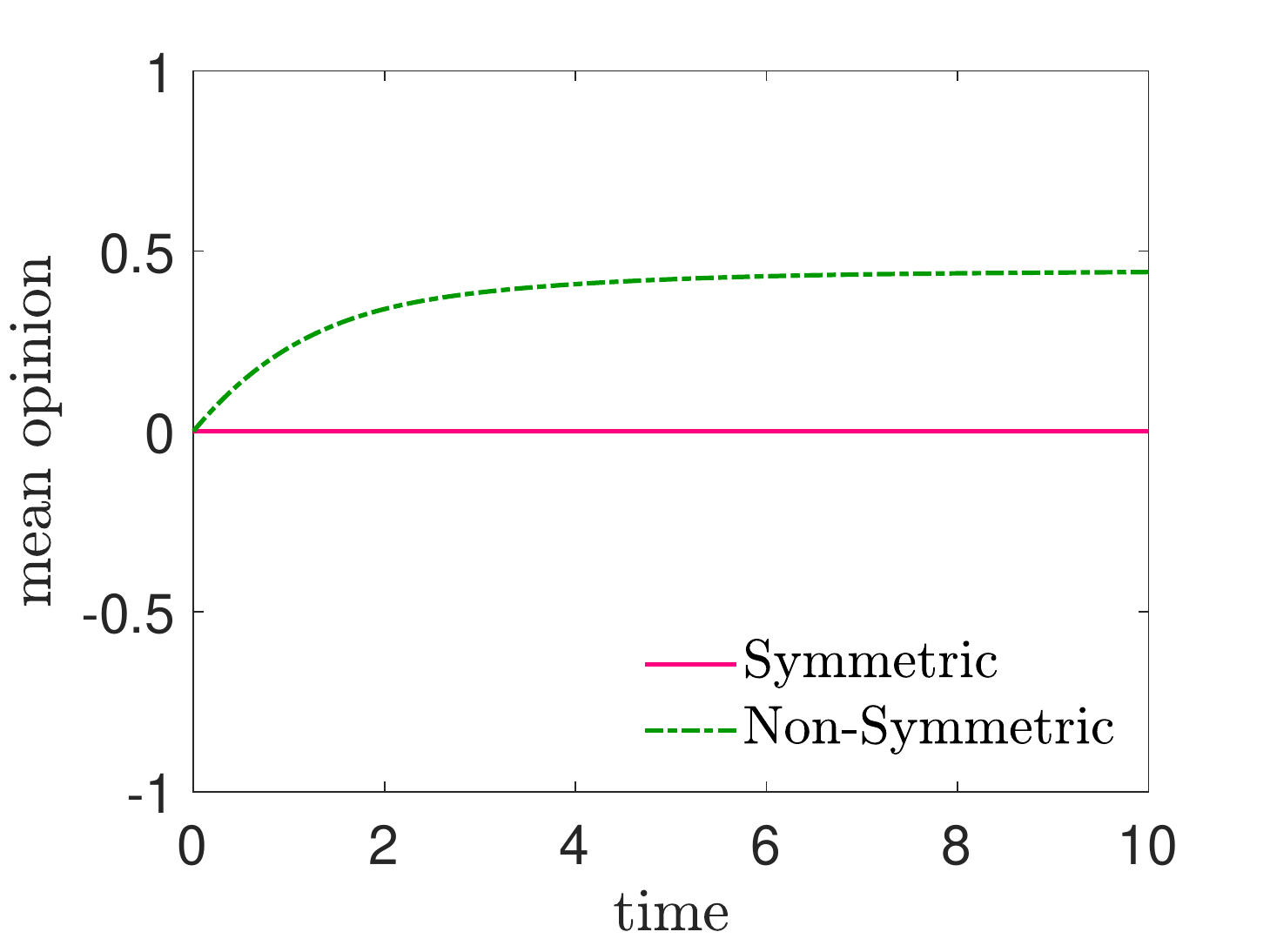}}
\subfigure[$w$-marginal at $T=10$]{\includegraphics[scale=0.33]{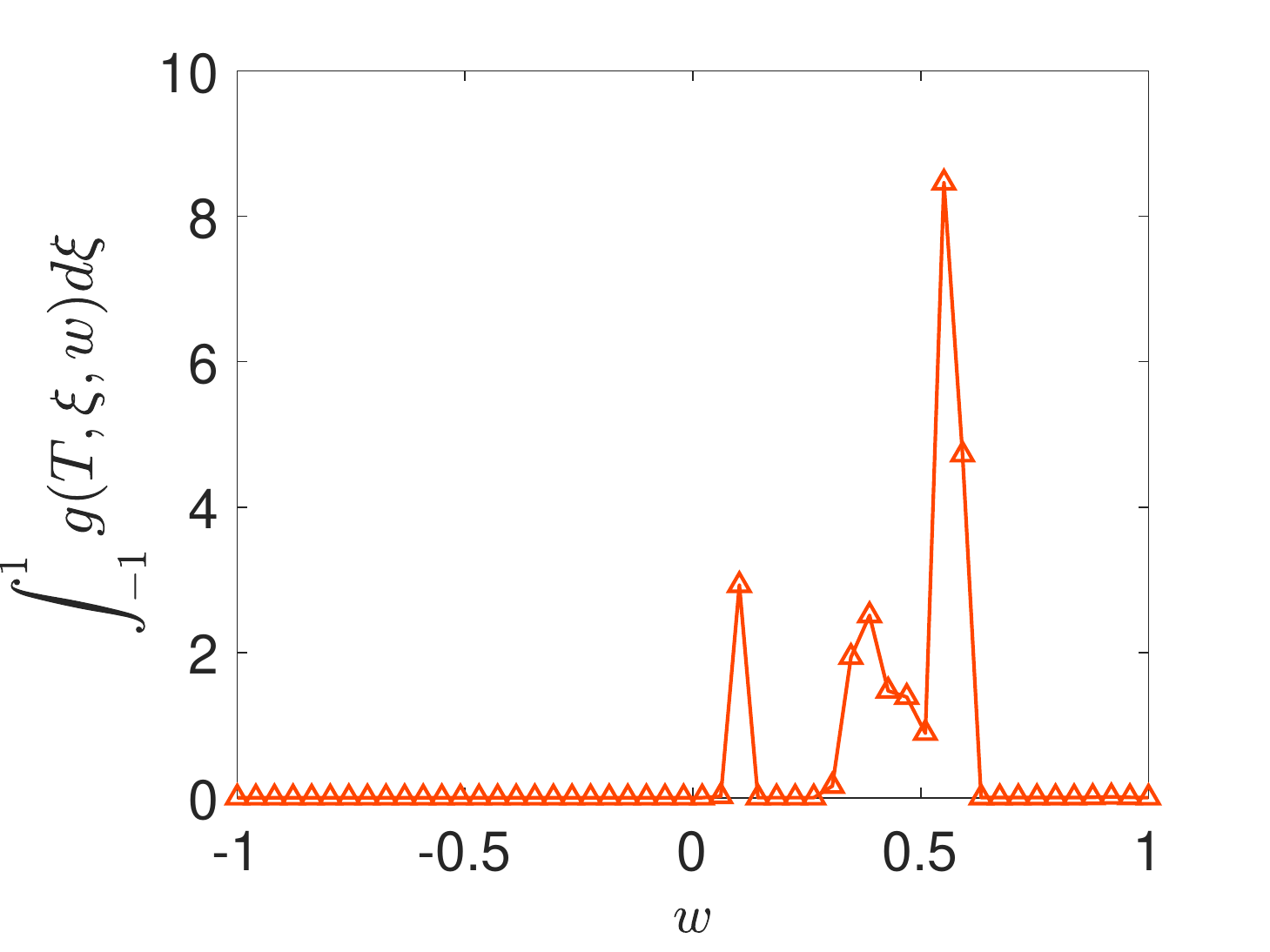}}
\subfigure[$\xi$-marginal at $T=10$]{\includegraphics[scale=0.33]{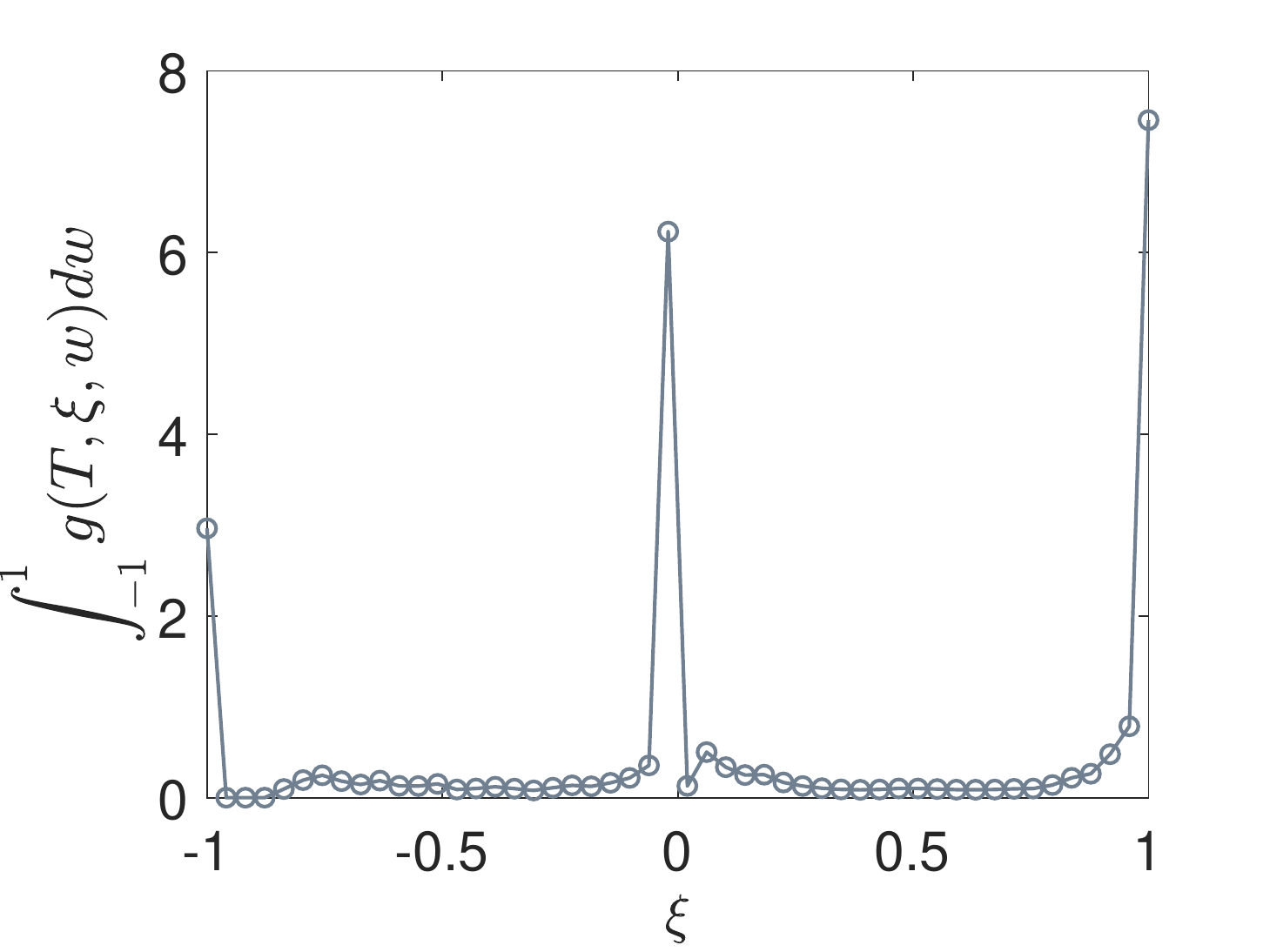}}
\caption{\textbf{Top row}: Contours of the inhomogeneous kinetic distribution $g(\tau,\,\xi,\,w)$ at different times with the non-symmetric compromise function~\eqref{eq:P_BC.nonsymm} featuring $\Delta_L=0.3$, $\Delta_R=0.7$. \textbf{Bottom row}: Time trend of the mean opinion (the symmetric case is plotted for duly comparison) and marginal distributions of opinions and preferences at time $T=10$.}
\label{fig:asym}
\end{figure}

We choose $\Delta_L=0.3$ and $\Delta_R=0.7$, meaning that individuals compromise preferentially with other individuals with an opinion located on the right of their own. Moreover, we consider the perceived social opinion $\alpha=0.3$. In Figure~\ref{fig:asym} we show the evolution of the inhomogeneous kinetic model starting from the uniform distribution~\eqref{eq:initial_inhom}. 

We observe that initially the mean opinion is neutral at any preference, indeed
$$ \int_{-1}^1wg(0,\,\xi,\,w)\,dw=0, \quad \forall\,\xi\in [-1,\,1]. $$
Nevertheless, due to the non-symmetric interactions, the mean opinion is not conserved in time, cf. Figure~\ref{fig:asym}(d). In particular, owing to the bias induced by $\Delta_R>\Delta_L$, the opinions tend to shift on the whole rightwards, cf. Figure~\ref{fig:asym}(e), while the preferences polarise in the three poles $\xi=-1,\,0,\,1$, cf. Figure~\ref{fig:asym}(f). Again, we notice that the joint picture preference-opinion is a lot more informative than the sole opinion dynamics, because it allows us to observe e.g. that two clusters with nearly the same asymptotic opinion about $w\approx 0.5$ actually include individuals expressing strongly different preferences ($\xi=0,\,1$), cf. Figures~\ref{fig:asym}(b, c).

\subsection{Hydrodynamic model}
Now we test the hydrodynamic model of preference formation derived in Section~\ref{sect:macro}. In particular, since the dynamics predicted by the first order models of Section~\ref{sect:first_order.hydro} are quite well understood analytically, we focus on the second order model presented in Section~\ref{sect:second_order.hydro}, cf.~\eqref{eq:SO}.

To discretise the system of conservation laws~\eqref{eq:SO}, we introduce a uniform mesh in the preference domain $[-1,\,1]$ made of $300$ grid points. Furthermore, we choose a time step $\Delta\tau>0$ such that the following CFL condition is met at each computational time:
$$ \frac{\Delta\tau}{\Delta\xi}\max_{\xi\in [-1,\,1]}\left[\Phi(\xi)\max\{\abs{\mu_1(\tau,\,\xi)},\,\abs{\mu_2(\tau,\,\xi)}\}\right]\leq 1, $$
where $\mu_1$, $\mu_2$ are the eigenvalues of the matrix~\eqref{eq:A}. Then we use a WENO reconstruction in the variable $\xi$ with a Godunov-type numerical flux, coupled with a third order Runge-Kutta integration in time. See~\cite{shu2009SIREV} for a detailed description of the numerical scheme.

In all the numerical tests of this section we consider the following initial condition:
\begin{equation}
	\rho(0,\,\xi)=\frac{1}{2}\chi(\xi\in [-1,\,1]), \quad m(0,\,\xi)=0,
	\label{eq:initial_num}
\end{equation}
which represents a uniform distribution of the density over the whole range of preferences with a null mean opinion, denoting initial indecisiveness, at all preferences.

\begin{figure}[!t]
\subfigure[Density]{\includegraphics[scale=0.5]{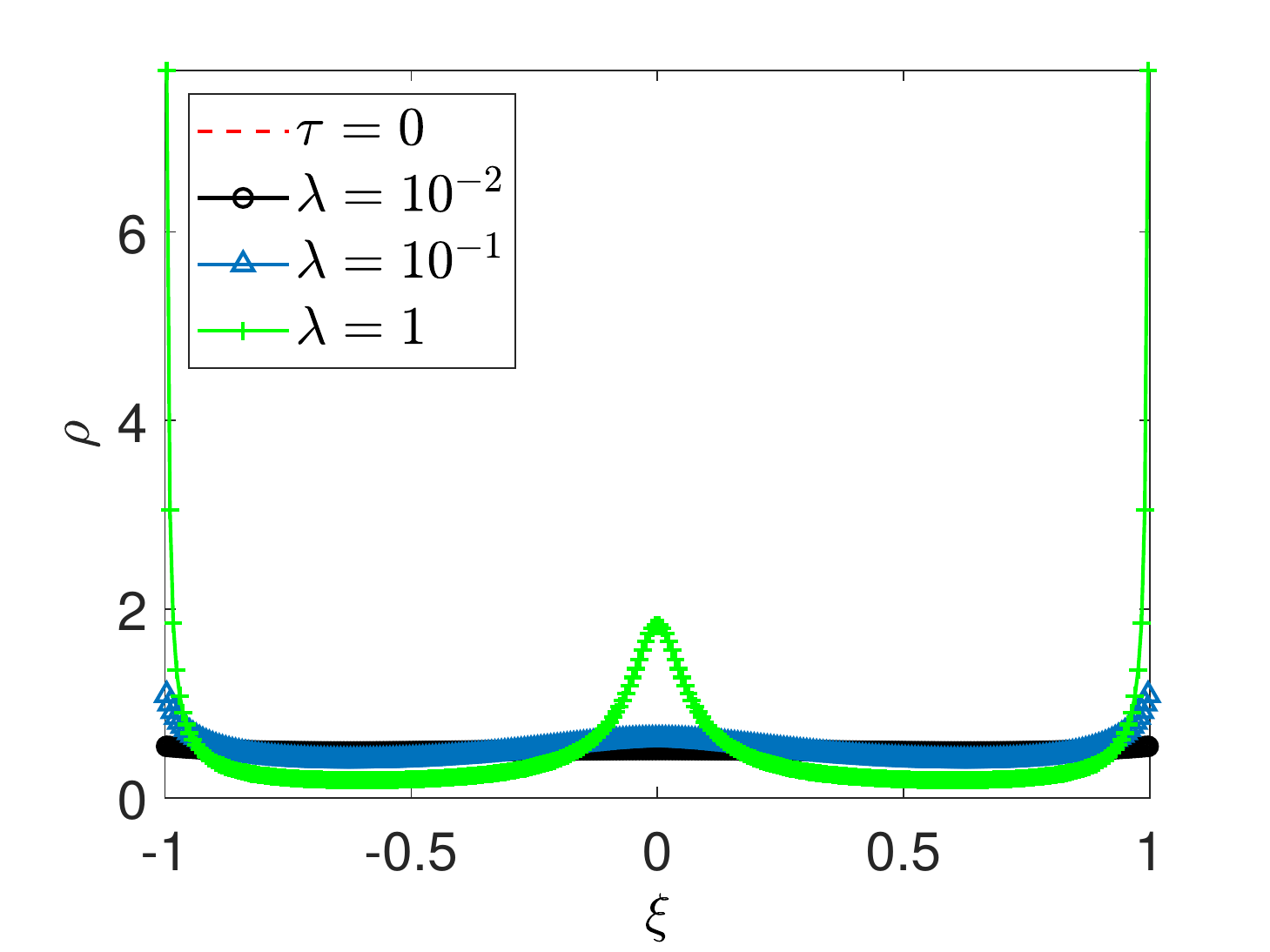}}
\subfigure[Mean opinion]{\includegraphics[scale=0.5]{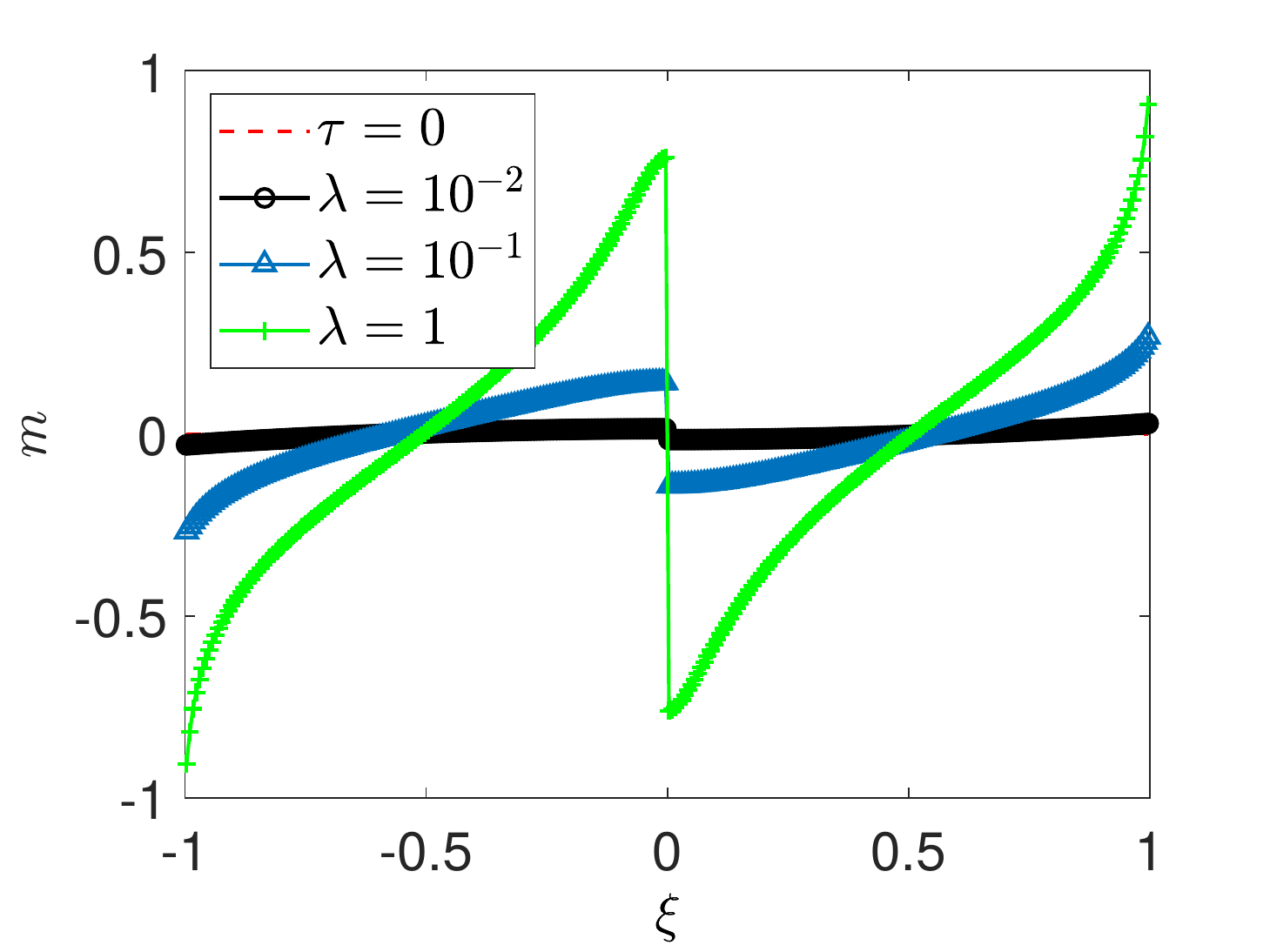}} \\
\subfigure[$\lambda=10^{-2}$]{\includegraphics[scale=0.5]{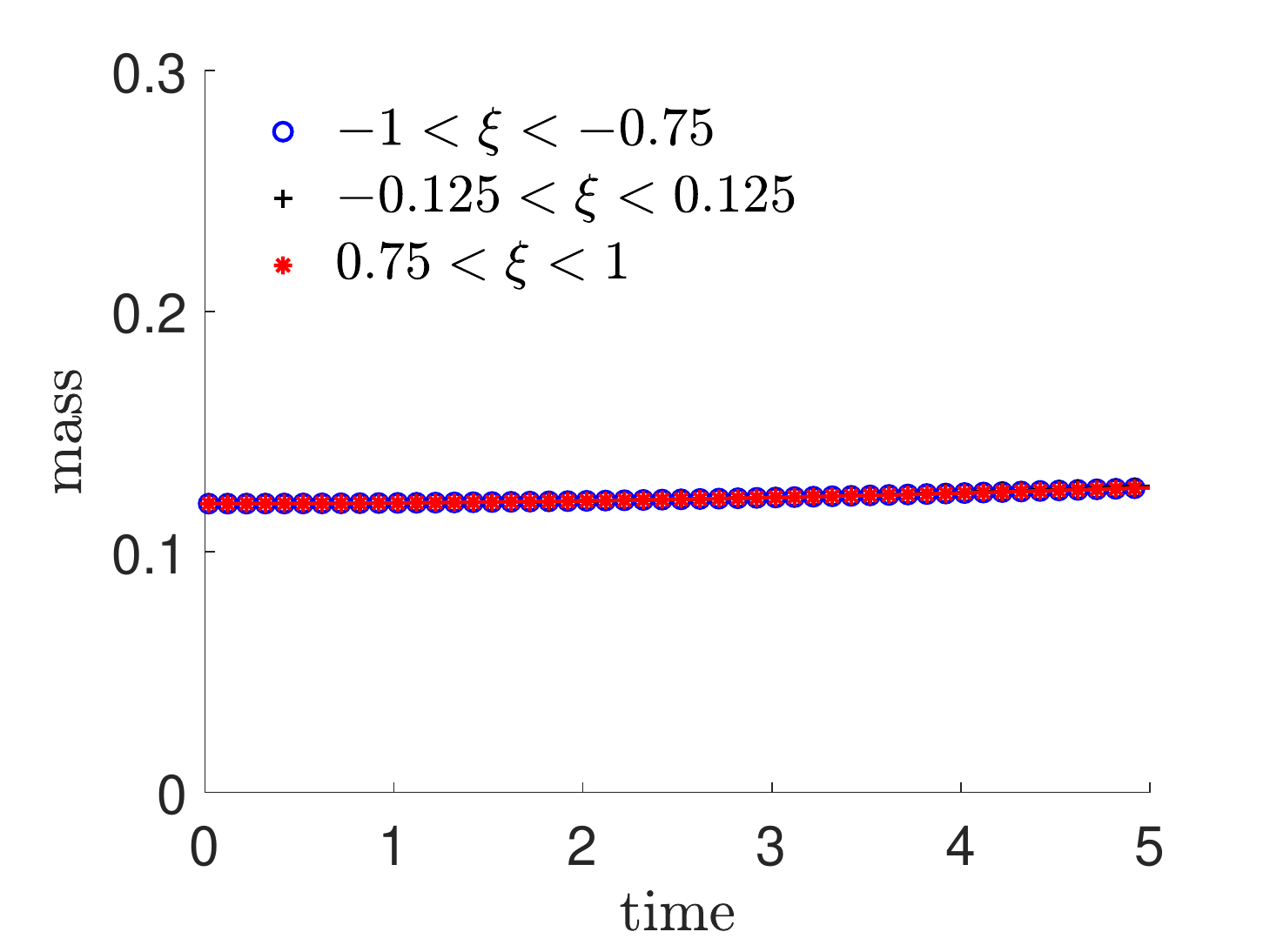}}
\subfigure[$\lambda=1$]{\includegraphics[scale=0.5]{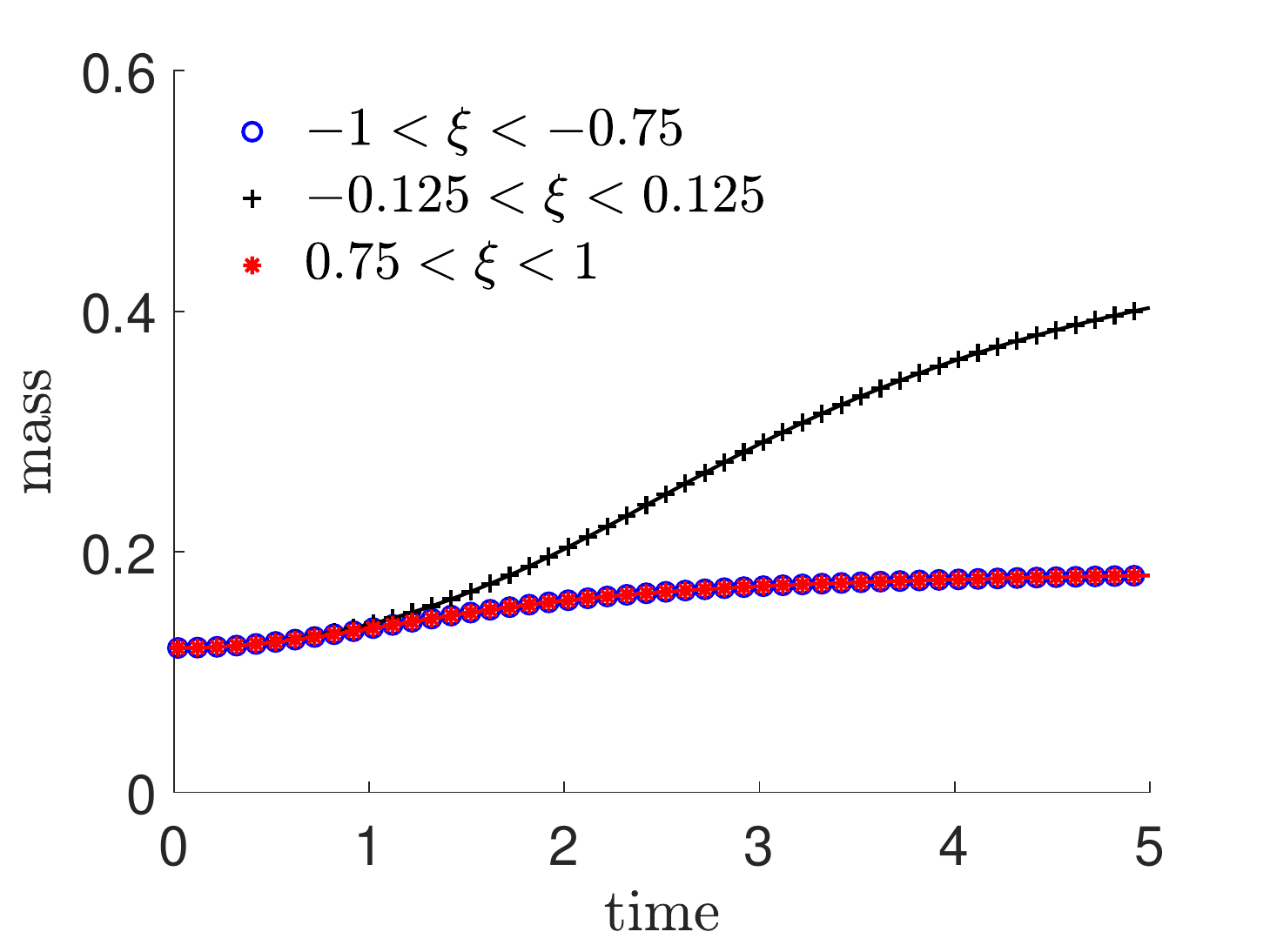}}
\caption{\textbf{Top row}: density (a) and mean opinion (b) at time $\tau=3$ computed from the second order hydrodynamic model~\eqref{eq:SO} with $\alpha=0$ and increasing values of $\lambda$, starting from the initial condition~\eqref{eq:initial_num}. \textbf{Bottom row}: time evolution of the mass of agents concentrating about the three preference poles $\xi=-1,\,0,\,1$ for (c) small and (d) large $\lambda$.}
\label{fig:alpha0}
\end{figure}

In the first test, whose results are displayed in Figure~\ref{fig:alpha0} at time $\tau=3$, we fix the perceived social opinion $\alpha=0$ and we consider three increasing values of the parameter $\lambda$, in particular $\lambda=10^{-2},\,10^{-1},\,1$, denoting a progressively stronger influence of the self-thinking. The numerical solution clearly shows that, due to $\alpha=0$, the symmetry of the initial density is preserved, cf. Figure~\ref{fig:alpha0}(a). In other words, the system is not driven spontaneously towards a specific preference, hence three symmetric polarisations emerge in $\xi=-1,\,0,\,1$. The inspection of the trend of the mean opinion, cf. Figure~\ref{fig:alpha0}(b), reveals that individuals ending in one of the two choices $\xi=\pm 1$ tend to develop opinions in agreement with their preference, while individuals ending in the neighbourhood of the choice $\xi=0$ tend to develop opinions markedly opposite to their preference. Such shapes of the solution are more and more evident for increasing $\lambda$. The role of the parameter $\lambda$ is further stressed by Figures~\ref{fig:alpha0}(c),~(d), which show the time trend of the mass of agents concentrating in suitable neighbourhoods of the preference poles $\xi=-1,\,0,\,1$, specifically the intervals $[-1,\,-\frac{3}{4}]$, $[-\frac{1}{8},\,\frac{1}{8}]$, $[\frac{3}{4},\,1]$. For small $\lambda$, cf. Figure~\ref{fig:alpha0}(c), there is a perfect equilibrium among the concentrations in the three poles. For relatively larger $\lambda$, cf. Figure~\ref{fig:alpha0}(d), denoting an increased relevance of the self-thinking, the individuals who concentrate about $\xi=0$ raise considerably with respect to those who concentrate instead in $\xi=\pm 1$. On the other hand, the latter remain symmetric.

\begin{figure}[!t]
\subfigure[Density]{\includegraphics[scale=0.5]{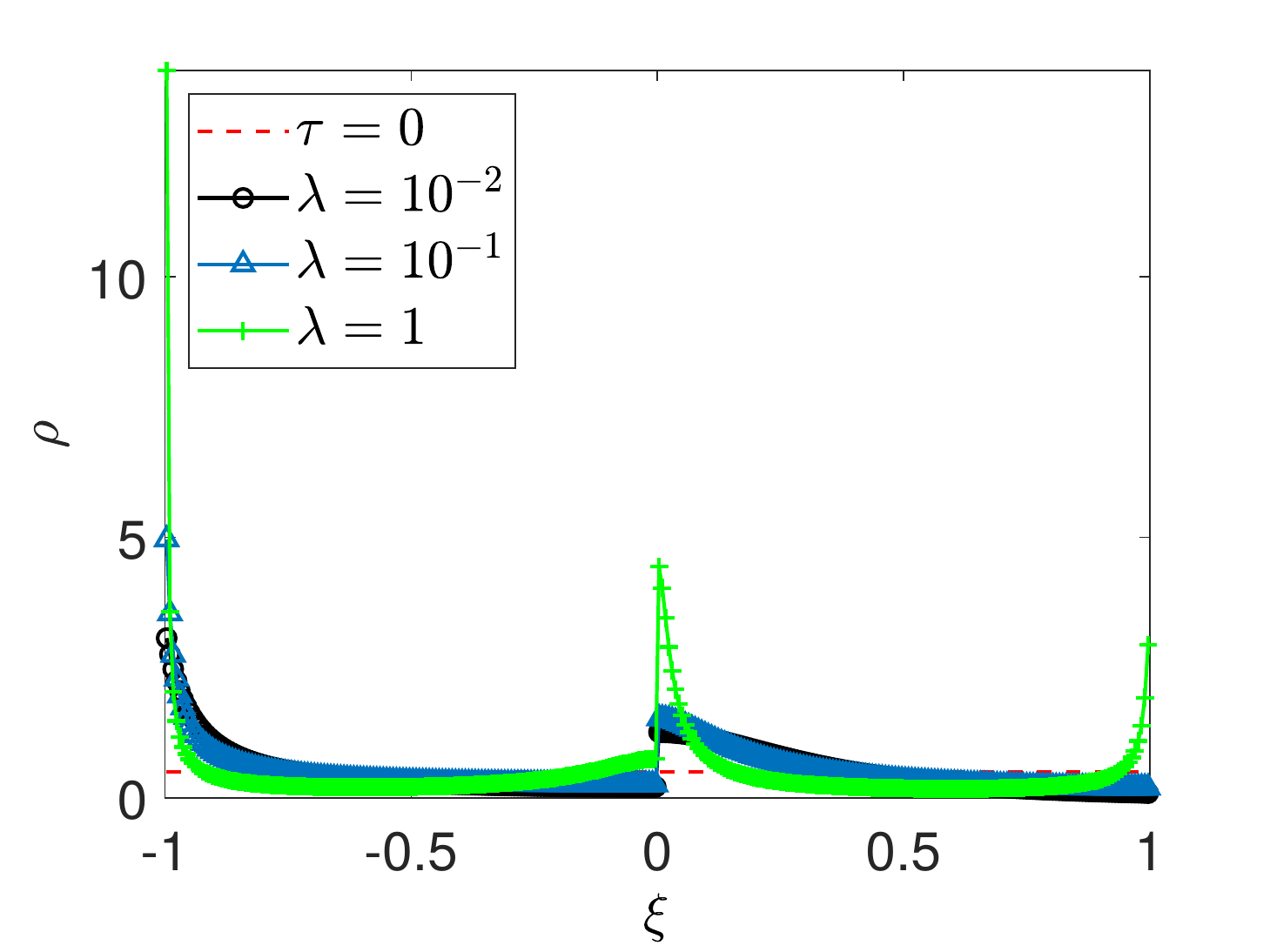}}
\subfigure[Mean opinion]{\includegraphics[scale=0.5]{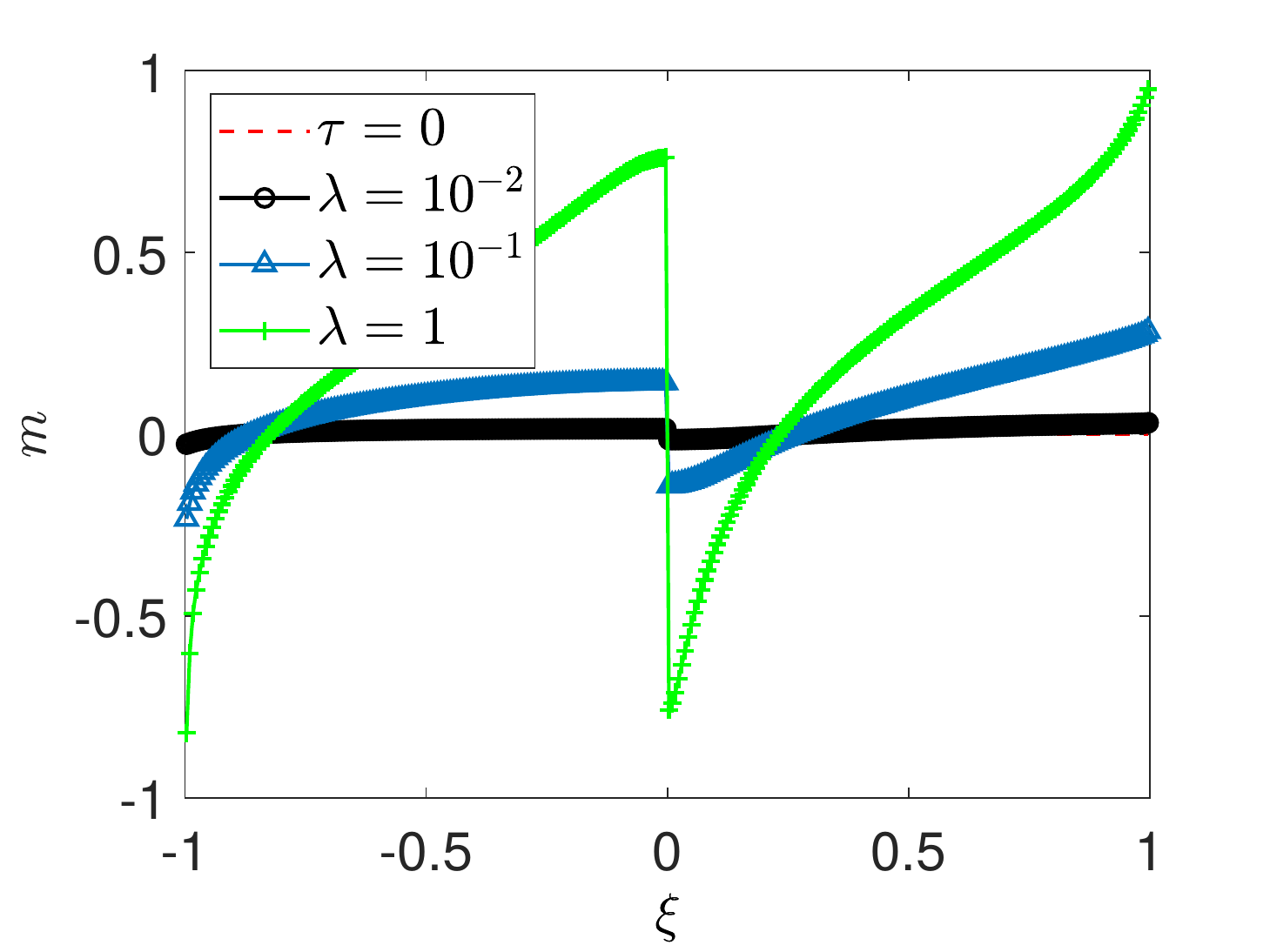}} \\
\subfigure[$\lambda=10^{-2}$]{\includegraphics[scale=0.5]{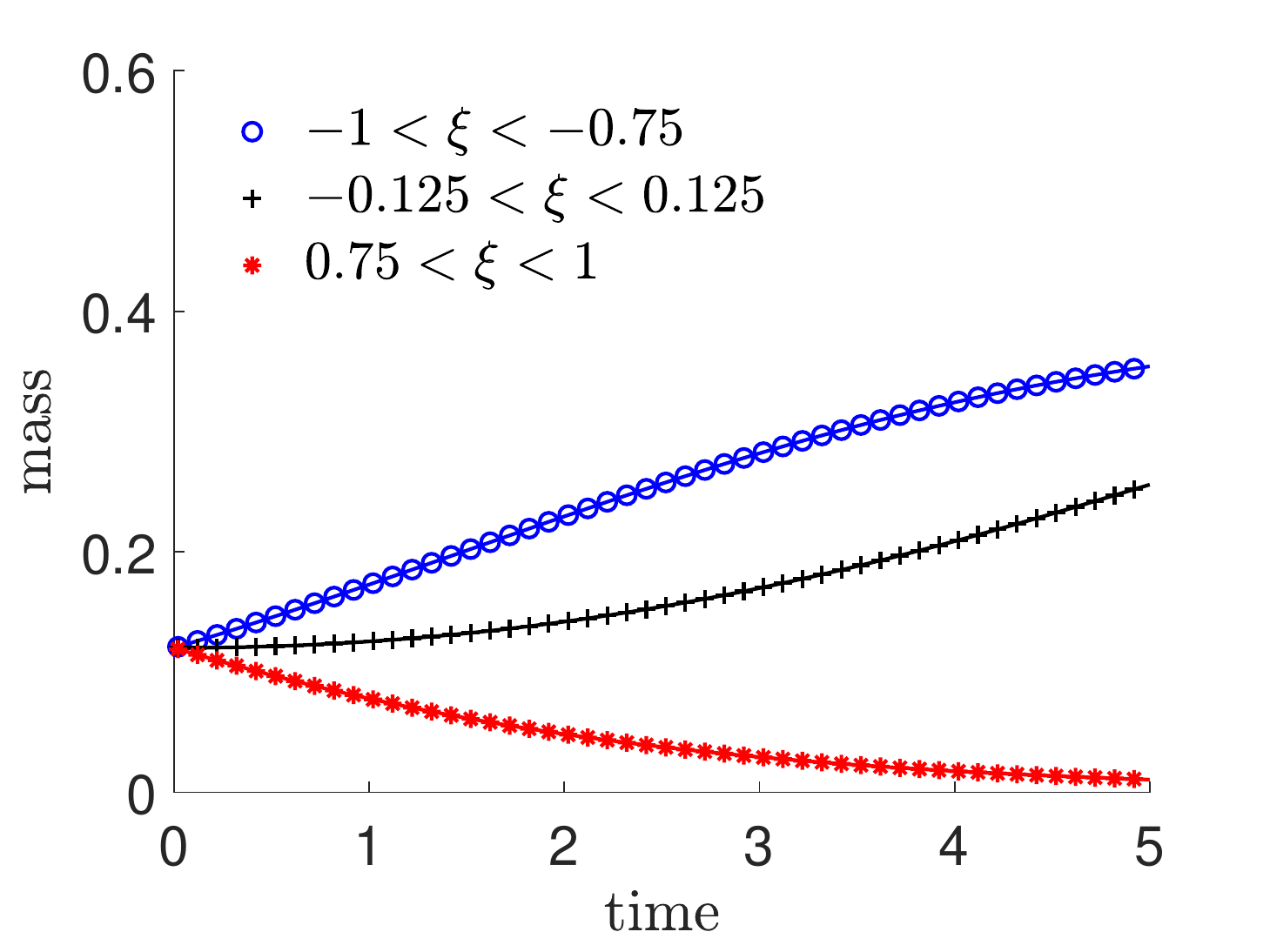}}
\subfigure[$\lambda=1$]{\includegraphics[scale=0.5]{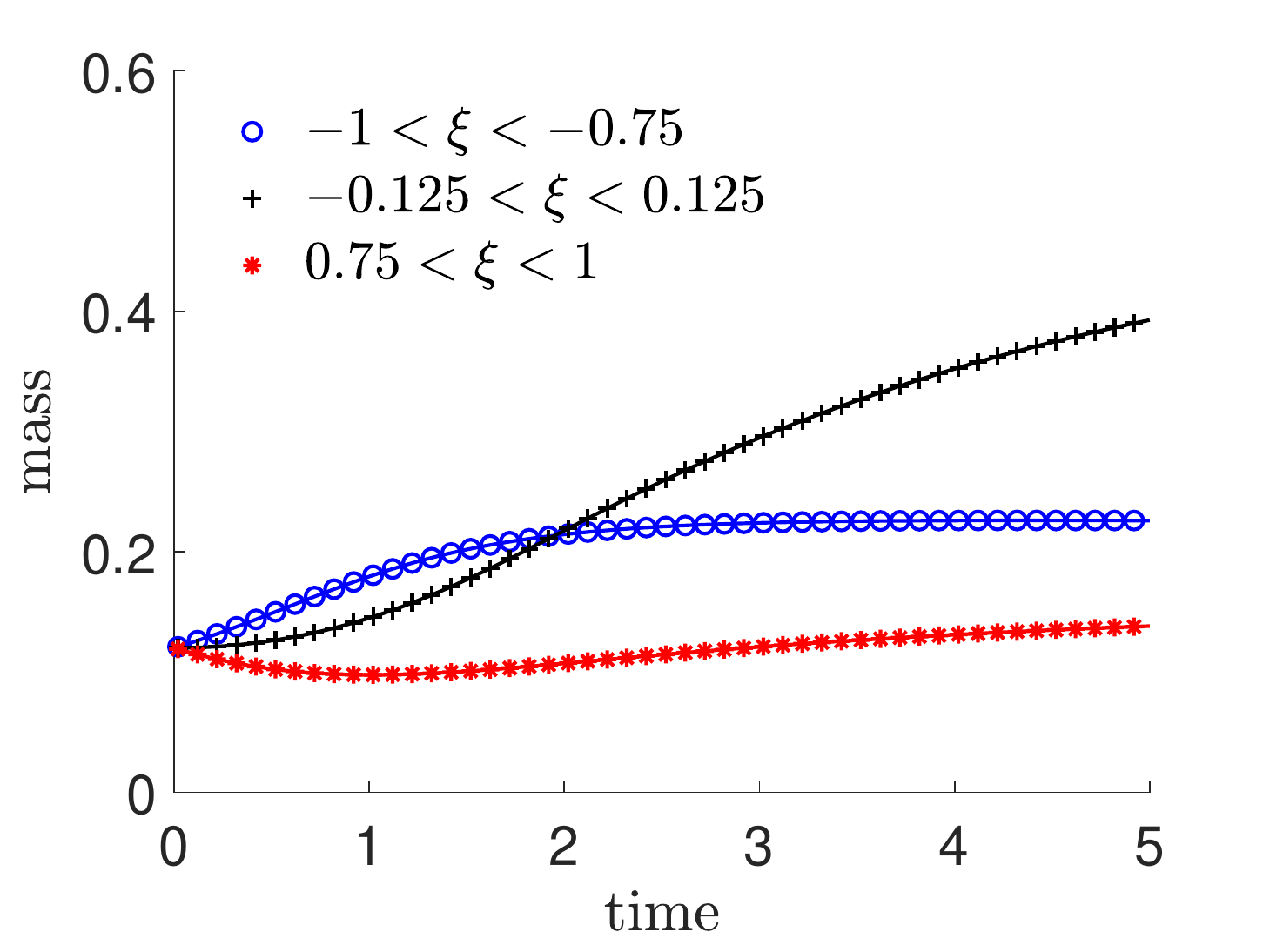}}
\caption{The same as Figure~\ref{fig:alpha0} but with $\alpha=0.3$.}
\label{fig:alpha_p03}
\end{figure}

In the second test, cf. Figure~\ref{fig:alpha_p03}, we consider the same situation described before but we set the perceived social opinion to $\alpha=0.3$, which induces a bias in the transport of the preference. The system exhibits again three polarisations, but now with an overall trend towards $\xi=-1,\,0$ and a residual polarisation in $\xi=1$, cf. Figure~\ref{fig:alpha_p03}(a). The reason is that now the perceived social opinion is higher than the initial mean opinion of the individuals, therefore initially the dominant drift is leftwards. During time, mean opinions higher than $\alpha$ emerge, cf. Figure~\ref{fig:alpha_p03}(b), which give rise to the polarisation in $\xi=1$ and to a further contribution to the polarisation in $\xi=0$. The latter subtracts mass to the polarisation in $\xi=-1$, especially for large $\lambda$, cf. Figures~\ref{fig:alpha_p03}(c),~(d).

\begin{figure}[!t]
\centering
\includegraphics[scale=0.45]{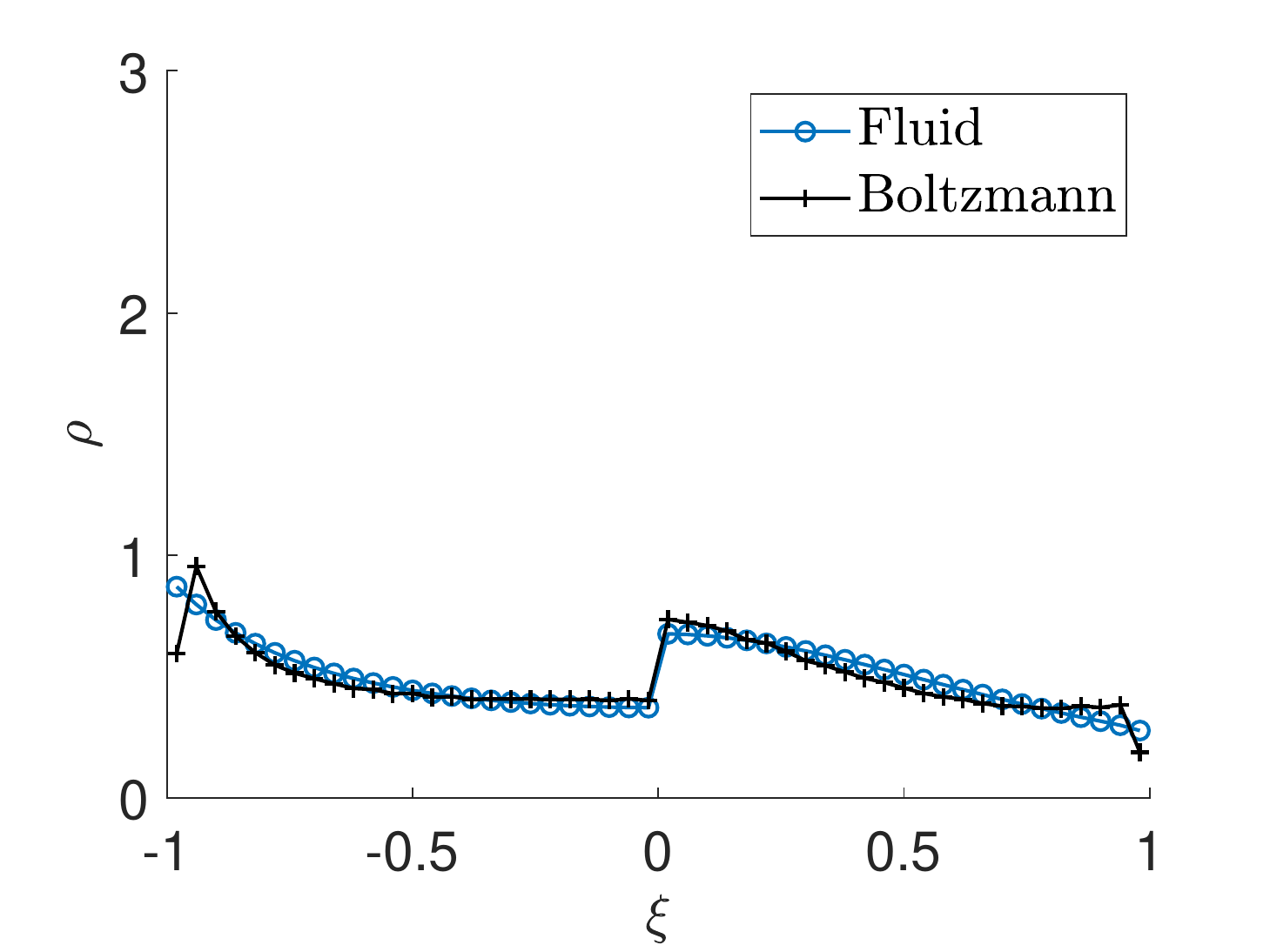}
\includegraphics[scale=0.45]{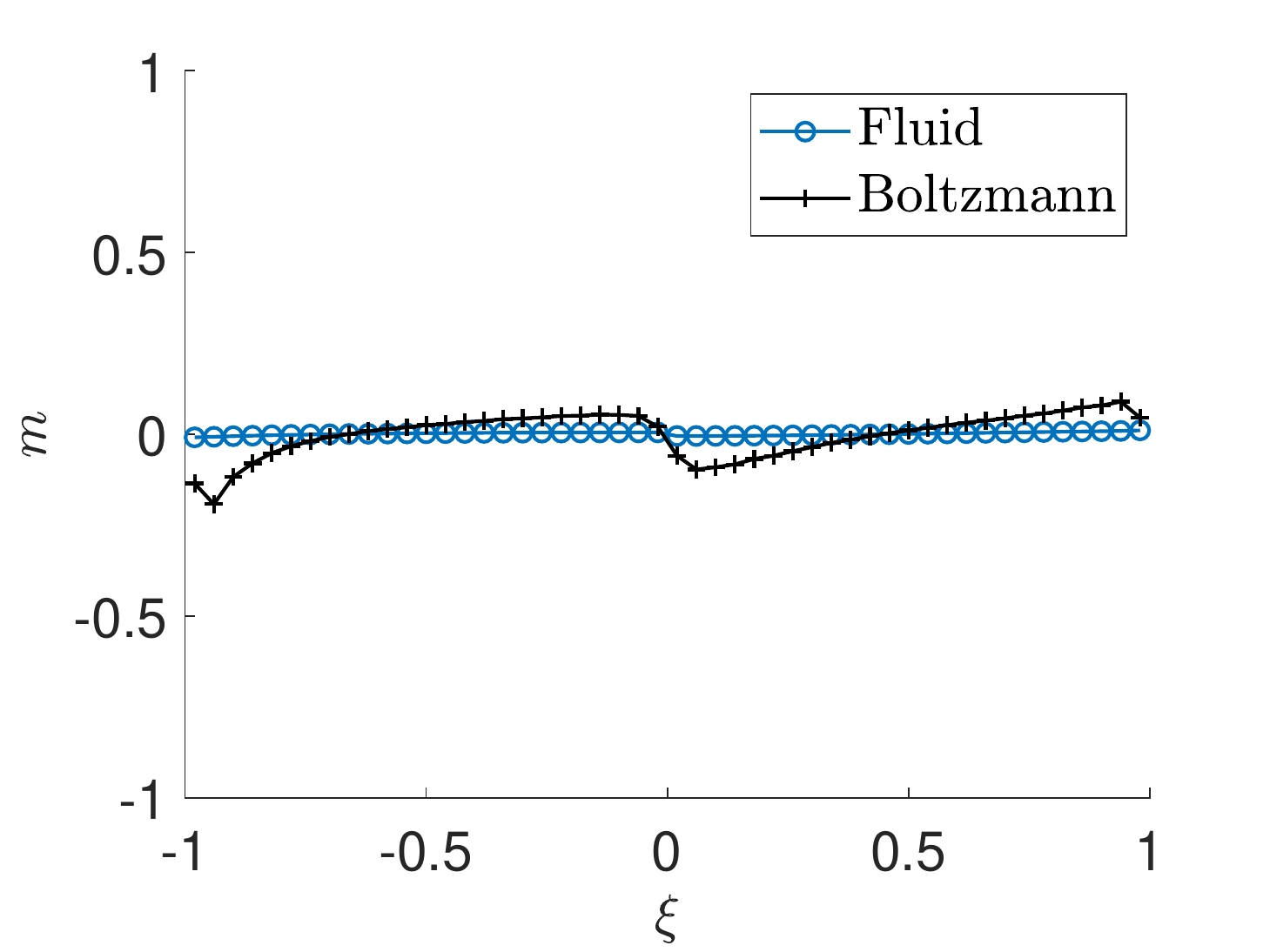} \\
\includegraphics[scale=0.45]{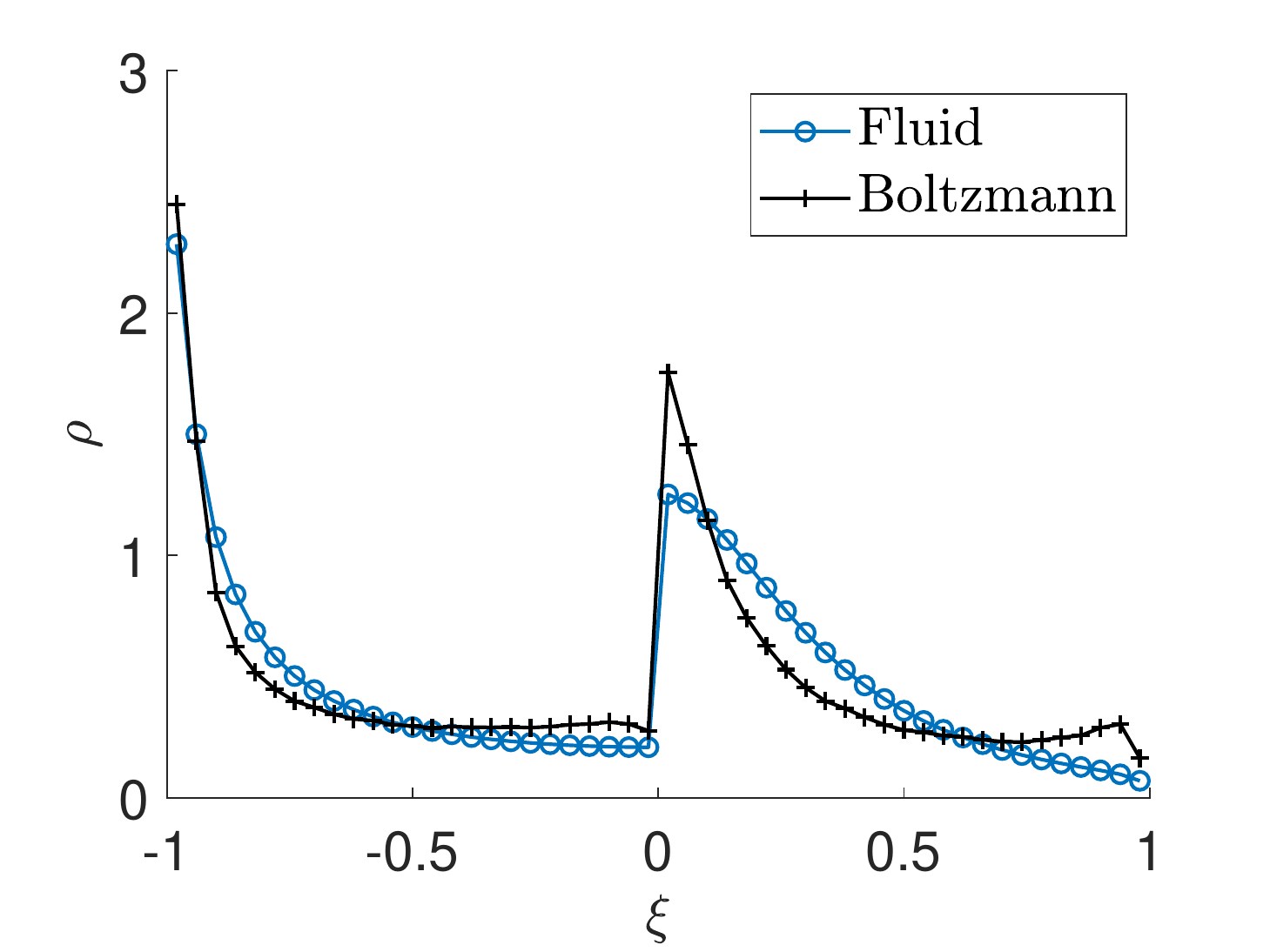}
\includegraphics[scale=0.45]{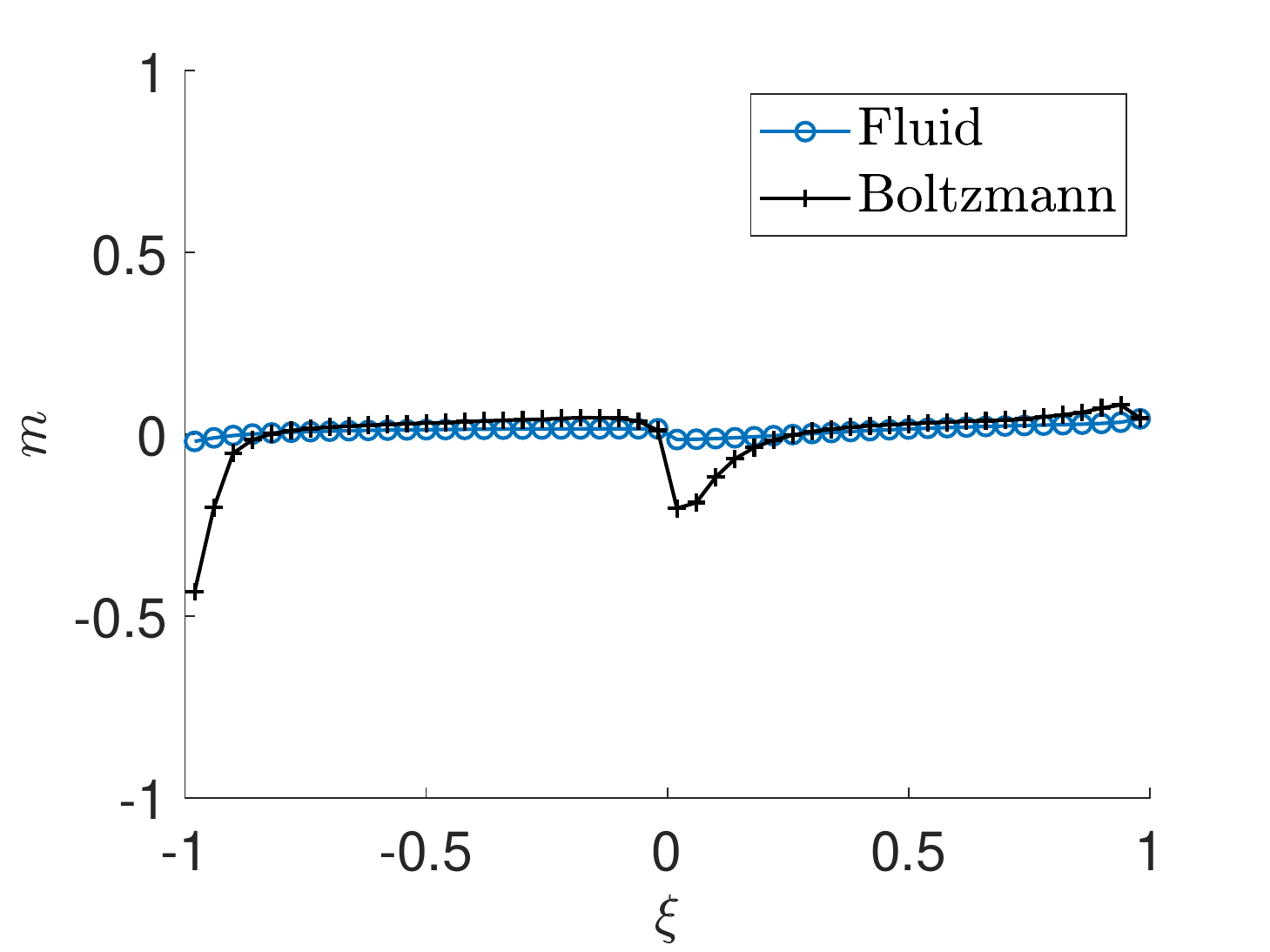}
\caption{Comparison of the numerical solution obtained in Figure~\ref{fig:alpha_p03} in the case $\lambda=10^{-2}$ with the numerical solution of the Boltzmann-type equation~\eqref{eq:boltzmann.inhomog.g} in the quasi-invariant scaling with $\delta=10^{-2}$ at two successive times: $\tau=1$ (\textbf{top row}) and $\tau=3$ (\textbf{bottom row}).}
\label{fig:fluid_bol}
\end{figure}

In order to validate the macroscopic model, in Figure~\ref{fig:fluid_bol} we compare the hydrodynamic quantities $\rho$, $m$ with the numerical marginals $\int_{-1}^1g(\tau,\,\xi,\,w)\,dw$, $\int_{-1}^1wg(\tau,\,\xi,\,w)\,dw$ computed from the solution $g$ to the inhomogeneous Boltzmann model~\eqref{eq:boltzmann.inhomog.g}. In particular, we solve the kinetic equation via the numerical procedure outlined in Section~\ref{subsect:inhom} with $50$ grid points in the $\xi$-mesh. Starting from the initial condition~\eqref{eq:initial_inhom} for the Boltzmann model, to which there corresponds the initial condition~\eqref{eq:initial_num} at the hydrodynamic level, we observe that the macroscopic model describes consistently the time evolution of the density of agents and of their mean opinion, as expected.

\begin{figure}[!t]
\subfigure[Density]{\includegraphics[scale=0.5]{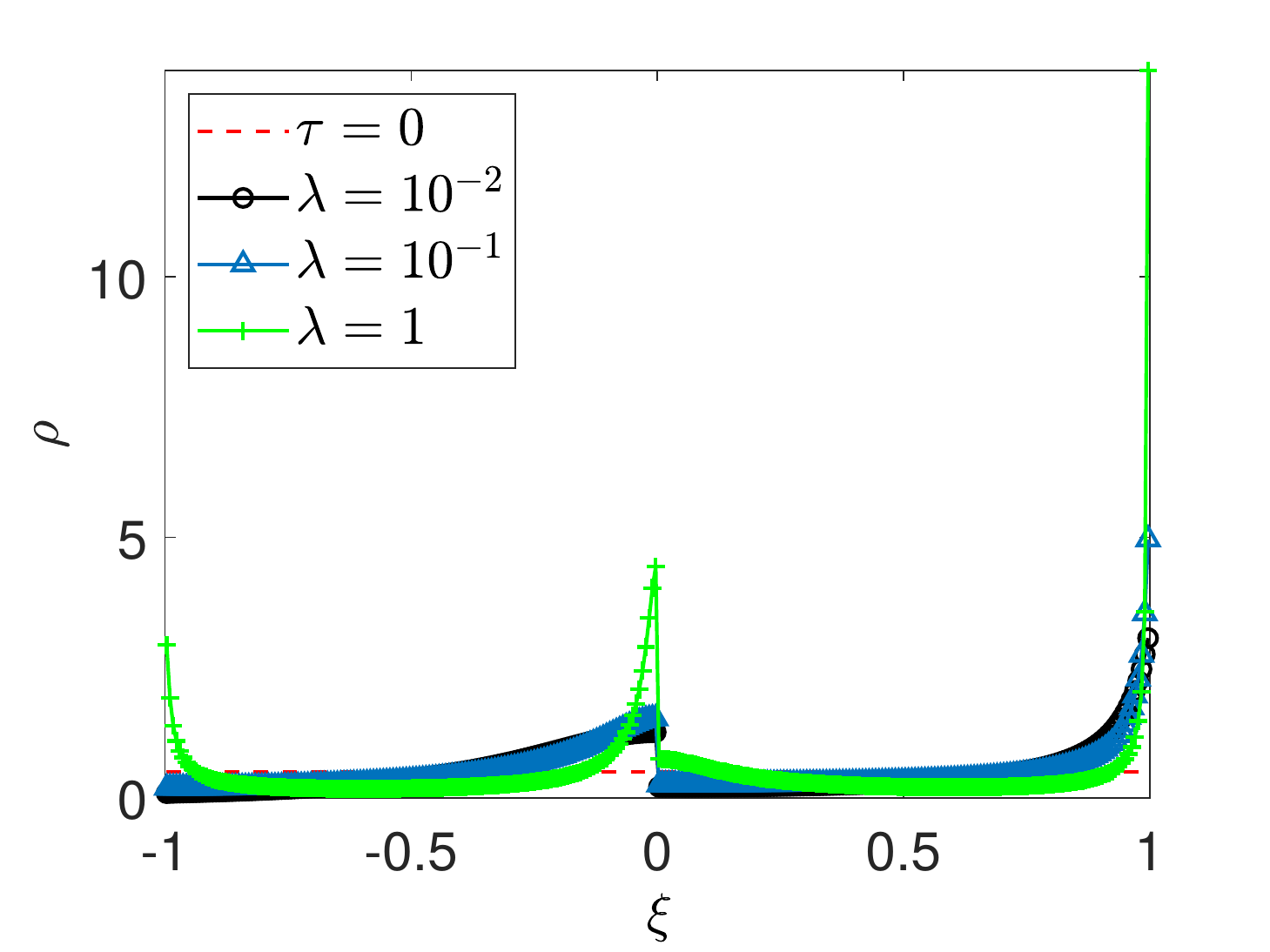}}
\subfigure[Mean opinion]{\includegraphics[scale=0.5]{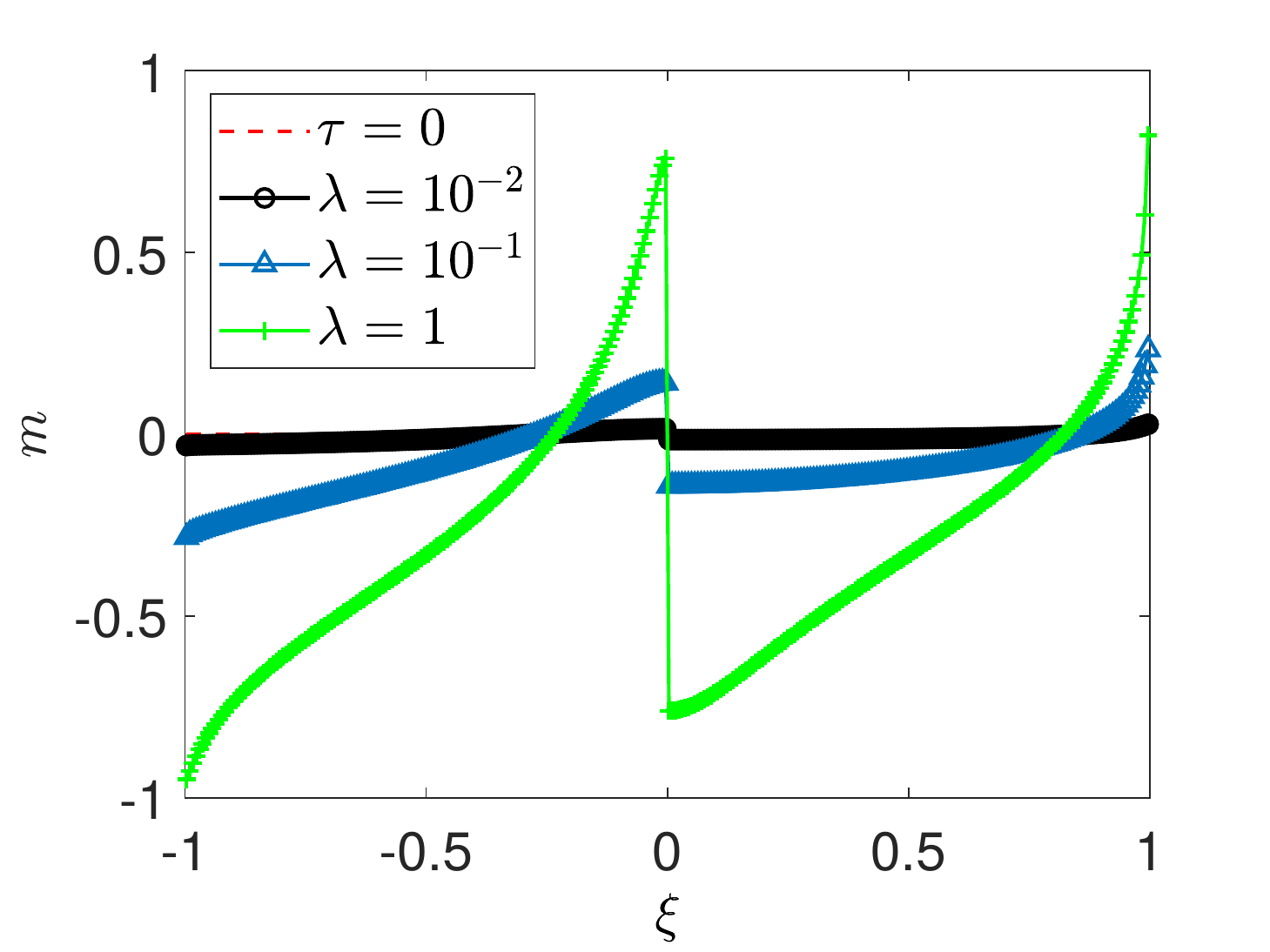}}
\caption{The same as Figure~\ref{fig:alpha0} but with $\alpha=-0.3$.}
\label{fig:alpha_m03}
\end{figure}

Finally, in Figure~\ref{fig:alpha_m03} we show that the mirror behaviour is observed with $\alpha=-0.3$, with an overall trend towards $\xi=0,\,1$ and a residual polarisation in $\xi=-1$.

\section{Summary and outlook}
\label{sect:summary}
In this paper we have proposed a development of classical consensus-based opinion formation models by including an additional variable, that we have called preference, which is transported in time by the evolving opinions of the agents. Although commensurable with an opinion, the preference does not simply replicate the opinion dynamics. It is rather the expression of a final choice, which is often necessarily much sharper, i.e. more polarised, than the opinion, like for instance in case of polls, referendums, elections.

At the agent-based level, the inspiration to model the interplay between opinion and preference is brought from the classical kinematic relation between position and velocity in mechanics: roughly speaking, the time variation of the preference $\xi$ is dictated by the sign of the instantaneous opinion $w$. However, here we introduce two main differences:
\begin{enumerate*}[label=(\roman*)]
\item a polarisation mechanism of the preference on some values denoting the choices available to the agents;
\item a bias linked to the general feeling perceived in the society, that we identify with a perceived social opinion $\alpha$.
\end{enumerate*}
Hence, ultimately, the time variation of $\xi$ depends on the signed distance $w-\alpha$. The meaning is clear: an agent will tend to move his/her preference, hence to orient his/her choice, according to the relative collocation in the society that s/he perceives for him/herself. For instance, if there are three possible choices, viz. preference poles, say $\xi=-1,\,0,\,1$, and the perceived social opinion is $\alpha>0$, an agent with preference $\xi>0$ and opinion $w>\alpha$ will move his/her preference towards $\xi=1$, because $w-\alpha>0$; conversely, if the opinion is $w<\alpha$, the agent will move his/her preference towards $\xi=0$, because $w-\alpha<0$. Such a distinction between opinion and preference is crucial to explain how the choice processes, although originating from the opinion dynamics, produce finally outcomes different from those of the opinion formation processes.

Taking advantage of the methods of statistical physics, in particular of the kinetic theory, we have given an aggregate description of the interplay between opinion and preference by means of an inhomogeneous Boltzmann-type equation, in which $w$ plays morally the role of the ``velocity'' and $\xi$ that of the ``position''. Specifically, the collision term of this equation accounts for the opinion formation due to binary interactions among the agents, while the transport term accounts for the opinion-driven preference formation. From here, we have finally derived macroscopic hydrodynamic models of preference formation with the technique of the local equilibrium closure, thanks to the possibility to identify precisely, from the kinetic model, the local equilibrium distribution of the opinions.

The analytical investigation of our models, and the related numerical results further extending the scope of the qualitative analysis, have shown the soundness and consistency of our hierarchical approach across all the considered scales. Moreover, they have highlighted the importance of the parameter $\alpha$ in shaping the collective distribution of the preferences. In this work, we have considered $\alpha$ simply as a given constant. Future research may instead address a variable perceived social opinion, which changes in time driven by various social forces. For instance, the control of $\alpha$ could be the goal of antithetical communication strategies commonly seen in contemporary political scenarios. Majority parties might try to weaken the opposition parties by creating the possibly exaggerated perception of a social opinion strongly oriented in their favour. At the same time, opposition parties might try to exaggerate the social bias towards the majority parties, in order to guard the electorate against the risks of too extremist positions. On the other hand, truly extremist parties might prefer to falsely soften the perceived social opinion, in order to avoid scaring the electorate and losing consensus. All these competitive strategies amount to controlling $\alpha$ with the aim of optimising a certain cost functional. This implies controlling the flux of the macroscopic models of preference formation proposed in this paper, taking however into account that the control strategy depends necessarily on the instantaneous opinion distribution, because $\alpha$ is itself an opinion. From the mathematical point of view, this requires to control the inhomogeneous Boltzmann-type kinetic equation and then to study the passage from the controlled kinetic model to hydrodynamic equations. Suitable techniques need to be explored, in order to make such a passage feasible from both the analytical and the numerical point of view.

\section*{Acknowledgements}
This work has been written within the activities of the Excellence Project \textrm{CUP: E11G18000350001} of the Department of Mathematical Sciences "G. L. Lagrange" of Politecnico di Torino funded by MIUR (Italian Ministry for Education, University and Research), and within the activities of GNFM (Gruppo Nazionale per la Fisica Matematica) and GNCS (Gruppo Nazionale per il Calcolo Scientifico) of INdAM (Istituto Nazionale di Alta Matematica), Italy. This work is also part of the activities of the Starting Grant ``Attracting Excellent Professors'' funded by ``Compagnia di San Paolo'' (Torino) and promoted by Politecnico di Torino.


\end{document}